\documentstyle[amssymb,12pt,epsfig]{article}

\begin{document}

\onecolumn

\begin{titlepage}
\begin{center}
{\LARGE \bf Chaos in an Exact Relativistic 3-body Self-Gravitating System}\\ 
\vspace{2cm}
F. Burnell\footnotemark\footnotetext{email: 
fburnell@physics.ubc.ca}
J.J. Malecki\footnotemark\footnotetext{email: 
jjmaleck@uwaterloo.ca}
R.B. Mann\footnotemark\footnotetext{email: 
mann@avatar.uwaterloo.ca},\\

\vspace{0.5cm} 
Dept. of Physics,
University of Waterloo
Waterloo, ONT N2L 3G1, Canada\\
\vspace{1cm}
T. Ohta \footnotemark\footnotetext{email:
t-oo1@ipc.miyakyo-u.ac.jp}\\
\vspace{0.5cm} 
Department of Physics, Miyagi University of Education,
Aoba-Aramaki, Sendai 980, Japan\\
\vspace{2cm}
\vspace{2cm}
\today\\
\end{center}

\begin{abstract}
We consider the problem of three body motion for a relativistic one-dimensional 
self-gravitating system.  After describing the canonical decomposition of the action, we
find an exact expression for the 3-body Hamiltonian, implicitly determined in terms of
the four coordinate and momentum degrees of freedom in the system.  
Non-relativistically these degrees of freedom can be rewritten in terms of a 
single particle moving in a two-dimensional hexagonal well.  We find the exact
relativistic generalization of  this potential, along with its post-Newtonian 
approximation. We then specialize to the equal mass case and numerically solve the 
equations of motion that follow from the Hamiltonian. Working in hexagonal-well 
coordinates, we obtaining orbits in both the hexagonal and 3-body representations
of the system, and plot the Poincare sections as a function of the relativistic
energy parameter $\eta $.  We find two broad categories of periodic and 
quasi-periodic motions that we refer to as the annulus and pretzel patterns, as well as
a set of chaotic motions that appear in the region of phase-space between 
these two types. Despite the high degree of non-linearity in the relativistic system,
we find that the the global structure of its phase space remains qualitatively the
same as its non-relativisitic counterpart for all values of $\eta $ that we could study. 
However the relativistic system has a weaker symmetry and so its Poincare section 
develops an asymmetric distortion that increases with increasing $\eta $. For the
post-Newtonian system we find that it experiences a KAM breakdown for
$\eta \simeq 0.26$: above which the near integrable regions degenerate 
into chaos.

\end{abstract}
\end{titlepage}\onecolumn

\section{INTRODUCTION}

The $N$-body problem, that of determining the motion of a system of $N$
particles mutually interacting through specified forces, is one of the
oldest problems in physics. It continues to remain of key importance over a
variety of distinct subfields, including nuclear physics, atomic physics,
stellar dynamics, and cosmology. When the interactions are purely
gravitational the problem is particularly challenging: while an exact
solution is known for pure Newtonian gravity in three spatial dimensions in
the $N=2$ case, there is no corresponding solution in the
general-relativistic case. This is due to dissipation of energy in the form
of gravitational radiation, which so far has necessitated recourse to
various approximation schemes.

Considerable progress has been made in recent years by reducing the number
of spatial dimensions. Indeed, non-relativistic one-dimensional
self-gravitating systems (OGS) of $N$ particles have played an important
role in astrophysics and cosmology for more than 30 years \cite{Rybicki}. \
While used primarily as prototypes for studying the behaviour of gravity in
higher dimensions, they also approximate the behaviour of some physical
systems in $3$ spatial dimensions. For example, very long-lived core-halo
configurations, reminiscent of structures observed in globular clusters, are
known to exist in the OGS phase space \cite{yawn}. These model a dense
massive core in near-equilibrium, surrounded by a halo of high kinetic
energy stars that interact only weakly with the core . Further examples
include collisions of flat parallel domain walls moving in directions
perpendicular to their surfaces and the dynamics of stars in a direction
orthogonal to the plane of a highly flattened galaxy. In addition to this, a
number of open questions remain concerning the statistical properties of the
OGS, including its ergodic behaviour, the circumstances (if any) under which
equipartition of energy can be attained, whether or not it can reach a true
equilibrium state from arbitrary initial conditions, and the appearance of
fractal behaviour \cite{fractal}.

In a relativistic context, reduction of the number of spatial dimensions
results in an absence of gravitational radiation whilst retaining most (if
not all) of the remaining conceptual features of relativistic gravity.
Consequently one might hope to obtain insight into the nature of
relativistic classical and quantum gravitation in a wide variety of physical
situations by studying the relativistic OGS, or ROGS.

Comparatively little has been known about the ROGS (even for $N=2$) until
quite recently, when a prescription for obtaining its Hamiltonian from a
generally covariant minimally-coupled action was obtained \cite{OR}. In the
non-relativistic limit ($c\rightarrow \infty $), the Hamiltonian reduces to
that of the OGS. This opened up the possibility of extending the insights of
the OGS into the relativistic regime, and indeed, considerable progress has
been made. Exact closed-form solutions to the $2$-body problem have been
obtained \cite{2bd}. These have been extended to include both a cosmological
constant \cite{2bdcossh,2bdcoslo}\ and electromagnetic interactions \cite%
{2bdchglo}, and a new exact solution to the static-balance problem has been
obtained \cite{statbal}. In the $N$-body case the Hamiltonian can be
obtained as a series expansion in inverse powers of the speed of light $\ c$
to arbitrary order, and a complete derivation of the partition and
single-particle distribution functions in both the canonical and
microcanonical ensembles \cite{pchak}, providing interesting information
concerning the influence of relativistic effects on self-gravitating
systems. Very recently formulation of the ROGS has been extended to circular
topologies \cite{circle} (forbidden for the OGS), and a new $N$-body dynamic
equilibrium solution has been found \cite{ryan}.

In this paper we consider the $3$-body problem for a relativistic
self-gravitating system in lineal gravity. Its non-relativistic counterpart
models several interesting physical systems, including perfectly elastic
collisions of a particle with a wedge in a uniform gravitational field \cite%
{LMiller}, two elastically colliding billiard balls in a uniform
gravitational field \cite{Goodings}\ and a bound state of three quarks to
form a ``linear baryon'' \cite{Butka}. These systems have recently been
shown to be subject to experimental test \cite{optbill}. Ours is the first
study of 3-body motion in a fully relativistic context.

We work with a 2D theory of gravity on a line (lineal gravity) that models
4D general relativity in that it sets the Ricci scalar $R$ equal to the
trace of the stress-energy of prescribed matter fields and sources. Hence,
as in $(3+1)$ dimensions, the evolution of spacetime curvature is governed
by the matter distribution, which in turn is governed by the dynamics of
spacetime \cite{r3}. Sometimes referred to as $R=T$ theory, it is a
particular member of a broad class of dilaton gravity theories formulated on
a line. What singles it out for consideration is its consistent
non-relativistic (i.e. $c\rightarrow \infty $ ) limit \cite{r3}, in general
a problematic limit for a generic $(1+1)$-dimensional theory of gravity \cite%
{jchan}. Consequently it contains each of the aforementioned
non-relativistic self-gravitating systems as special cases. Furthermore, it
reduces to Jackiw-Teitelboim (JT) theory \cite{JT} when the stress-energy is
that of a cosmological constant.

We have found that the most effective means by which to extract and study
the dynamics of the ROGS is to work in the canonical formalism \cite{OR}. We
formulate the $3$-body problem in relativistic gravity by taking the matter
action to be that of $3$ point-particles minimally coupled to gravity. We
obtain an exact expression for the Hamiltonian in terms of the four physical
degrees of freedom of the system (the two proper separations and their
conjugate momenta), given as a transcendental equation. Under a simple
coordinate transformation the non-relativistic system is equivalent to that
of a single particle moving in a hexagonal-well potential in 2 spatial
dimensions. The system we study is an exact relativistic generalization of
the hexagonal-well problem, affording insight into intrinsically
non-perturbative relativistic effects, as well allowing a controlled study
into its slow motion, weak field limit so as to determine its relativistic
corrections to leading order. When the masses of all particles are equal the
cross-sectional shape of the well in the non-relativistic case is that of a
regular hexagon; unequal masses distort this symmetry to that of a hexagon
with sides of differing length. Relativistic effects maintain this symmetry
in both cases, but curve the sides of the hexagon outward.

The action principle underlying the dynamics of the system must include a
scalar (dilaton) field \cite{BanksMann} since the Einstein action is a
topological invariant in (1+1) dimensions. We find upon canonical reduction
that the Hamiltonian is given in terms of a spatial integral of the second
derivative of the dilaton field, regarded as a function of the canonical
variables of the particles (coordinates and momenta) and is determined from
the constraint equations. Solving these equations matched across the
particles yields a transcendental equation that determines the Hamiltonian
in terms of the remaining degrees of freedom of the system when $N=3$. Since
we can determine from it the Hamiltonian in terms of the relative proper
separations of the bodies and their conjugate momenta, we refer to this
transcendental equation as the determining equation. From the determining
equation we can derive the canonical equations of motion. The equations are
considerably more complicated than their non-relativistic counterparts, and
we solve them numerically. We find an extremely rich and interesting
dynamics dependent upon the initial conditions imposed on the system.

\bigskip

In order to have a controlled investigation and comparison of the
relativistic effects, we consider three distinct physical systems: the
non-relativistic (N) system, whose Hamiltonian has been considered
previously \cite{LMiller,Goodings,Butka} in a variety of contexts, its exact
relativistic (R) counterpart Hamiltonian system, and the post-Newtonian (pN)
expansion of the R-system, truncated to leading order in $c^{-2}$, where $c$
is the speed of light. \ The $c\rightarrow \infty $ limit of both the R and
pN systems is the N system; consequently we have both an exact relativistic
generalization of the OGS and a well-defined relativistic approximation to
it. \ We find intriguing relationships and striking differences between all
three systems. \ For example, tightly bound states of two bodies undergoing
a low-frequency oscillation with the third take place in both the N and R
systems; however the motion in the R system for the bound-pair and the third
body take on features similar to that of $2$-body ROGS motion studied
previously \cite{2bd,2bdchglo,2bdcossh} whereas the corresponding motions in
the N system have the expected parabolic behaviour. \ In general bound-state
oscillations in the R system at a given energy have a higher frequency and
cover a smaller region of the position part of the phase space than its N
and pN counterparts do at the same energy.

\bigskip

The global structure of phase space can be probed using Poincare sections.
Remarkably, the Poincare plots of the R system are qualitatively similar to
those of the N system, but distorted toward the lower-right of the phase
plane. This is because there is a component to the gravitational momentum in
the R case that is absent in the N case, continuously transforming the basic
structure of the Poincare plot. On the other hand the pN system develops
additional regions of chaos in phase-space that neither the N nor R systems
have. \ This suggests that there are limits to the reliability of a pN
approximation to an R system.

\bigskip

In section II we review the formalism of the $N$-body problem in lineal
gravity, discussing the canonical decomposition of the action and
Hamiltonian, and the formulation of the equations of motion. We then go on
in section III to solve these equations in the $3$-body case, finding the
determining equation of the Hamiltonian and deriving the equations of motion
which follow from it. \ Before solving this system of equations we first
consider some of its general properties in section IV. We find its
post-Newtonian expansion and use this to study how the hexagonal-well
potential is modified by relativistic corrections. In section V we describe
our methods for numerically solving the 3-body system. \ Working in
hexagonal-well coordinates, we describe our methods for obtaining orbits,
Poincare maps, and graphs that illustrate the oscillation patterns of the
three particles. We then go on to numerically solve the equations of motion
of the system in section VI\ in the equal-mass case. \ We find two broad
categories of periodic and quasi-periodic motions that we refer to as the
annulus and pretzel patterns. We also find a set of chaotic motions that
appear in the region of phase-space between these two other types. To
complete our investigation we present various Poincare maps in section VII.
Here we discuss the striking similarities and differences in the global
structure of phase space between the three systems.\ In section VIII we
discuss the salient features of our solutions and make some conjectures
regarding their general properties. We close our paper with some concluding
remarks and directions for further work, including an appendix containing
the transformation to hexagonal coordinates.

\section{Canonical Reduction of the $N$-body Problem in lineal gravity}

The general procedure for the derivation of the Hamiltonian via canonical
reduction \cite{OK} has been given previously \cite{2bd,2bdcoslo}, and so
here we briefly review this work, highlighting those aspects that are
peculiar to the $3$-body case.

We begin with an action that describes the minimal coupling of $N$ point
masses to gravity%
\begin{eqnarray}
I &=&\int d^{2}x\left[ \frac{1}{2\kappa }\sqrt{-g}g^{\mu \nu }\left\{ \Psi
R_{\mu \nu }+\frac{1}{2}\nabla _{\mu }\Psi \nabla _{\nu }\Psi \right\}
\right.  \label{act1} \\
&&\makebox[2em]{}\left. +\sum_{a=1}^{N}\int d\tau _{a}\left\{ -m_{a}\left(
-g_{\mu \nu }(x)\frac{dz_{a}^{\mu }}{d\tau _{a}}\frac{dz_{a}^{\nu }}{d\tau
_{a}}\right) ^{1/2}\right\} \delta ^{2}(x-z_{a}(\tau _{a}))\right] \;, 
\nonumber
\end{eqnarray}%
where $\Psi $ is the dilaton field, $g_{\mu \nu }$ and $g$ are the metric
and its determinant, $R$ is the Ricci scalar. and $\tau _{a}$ is the proper
time of $a$-th particle, respectively, with $\kappa =8\pi G/c^{4}$. We
denote by $\nabla _{\mu }$ the covariant derivative associated with $g_{\mu
\nu }$.

\bigskip From the action (\ref{act1}) the field equations are 
\begin{eqnarray}
&&R-g^{\mu \nu }\nabla _{\mu }\nabla _{\nu }\Psi =0\;,  \label{eq-R} \\
&&\frac{1}{2}\nabla _{\mu }\Psi \nabla _{\nu }\Psi -\frac{1}{4}g_{\mu \nu
}\nabla ^{\lambda }\Psi \nabla _{\lambda }\Psi +g_{\mu \nu }\nabla ^{\lambda
}\nabla _{\lambda }\Psi -\nabla _{\mu }\nabla _{\nu }\Psi =\kappa T_{\mu \nu
}  \label{Psieq} \\
&&m_{a}\left[ \frac{d}{d\tau _{a}}\left\{ g_{\mu \nu }(z_{a})\frac{%
dz_{a}^{\nu }}{d\tau _{a}}\right\} -\frac{1}{2}g_{\nu \lambda ,\mu }(z_{a})%
\frac{dz_{a}^{\nu }}{d\tau _{a}}\frac{dz_{a}^{\lambda }}{d\tau _{a}}\right]
=0\;,  \label{eq-z}
\end{eqnarray}%
where 
\begin{equation}
T_{\mu \nu }=\sum_{a}m_{a}\int d\tau _{a}\frac{1}{\sqrt{-g}}g_{\mu \sigma
}g_{\nu \rho }\frac{dz_{a}^{\sigma }}{d\tau _{a}}\frac{dz_{a}^{\rho }}{d\tau
_{a}}\delta ^{2}(x-z_{a}(\tau _{a}))\;,  \label{stressenergy}
\end{equation}%
is the stress-energy of the $N$-body system. Eq.(\ref{Psieq}) guarantees the
conservation of $T_{\mu \nu }$. By inserting the trace of Eq.(\ref{Psieq})
into Eq.(\ref{eq-R}) we obtain 
\begin{equation}
R=\kappa T_{\;\;\mu }^{\mu }\;.  \label{RT}
\end{equation}%
Eqs. (\ref{eq-z}) and (\ref{RT}) form a closed system of equations for the $%
N $-body system coupled to gravity.

\bigskip In the canonical formalism the action (\ref{act1}) is written in
the form 
\begin{equation}
I=\int d^{2}x\left\{ \sum_{a=1}^{N}p_{a}\dot{z}_{a}\delta
(x-z_{a}(x^{0}))+\pi \dot{\gamma}+\Pi \dot{\Psi}+N_{0}R^{0}+N_{1}R^{1}\right%
\}  \label{actcan}
\end{equation}%
where the metric is 
\begin{equation}
ds^{2}=-N_{0}^{2}\left( x,t\right) dt^{2}+\gamma \left( dx+\frac{N_{1}}{%
\gamma }dt\right) ^{2}  \label{metric}
\end{equation}%
and $\pi $ and $\Pi $ are conjugate momenta to $\gamma $ and $\Psi $
respectively. \ The quantities $R^{0}$ and $R^{1}$ are given by%
\begin{eqnarray}
R^{0} &=&-\kappa \sqrt{\gamma }\gamma \pi ^{2}+2\kappa \sqrt{\gamma }\pi \Pi
+\frac{1}{4\kappa \sqrt{\gamma }}(\Psi ^{\prime })^{2}-\frac{1}{\kappa }%
\left( \frac{\Psi ^{\prime }}{\sqrt{\gamma }}\right) ^{\prime
}-\sum_{a=1}^{N}\sqrt{\frac{p_{a}^{2}}{\gamma }+m_{a}^{2}}\;\delta
(x-z_{a}(x^{0}))  \label{R0con} \\
R^{1} &=&\frac{\gamma ^{\prime }}{\gamma }\pi -\frac{1}{\gamma }\Pi \Psi
^{\prime }+2\pi ^{\prime }+\sum_{a=1}^{N}\frac{p_{a}}{\gamma }\delta
(x-z_{a}(x^{0}))\;\;.  \label{R1con}
\end{eqnarray}%
and describe the constraints of the system, with the symbols dot and prime
denoting $\partial _{0}$ and $\partial _{1}$, respectively. Setting $R^{0}=0$
yields an energy-balance equation, in which the total energy of the
particles is offset by the energy of the gravitational field. Setting $%
R^{1}=0$ yields an equation in which the total momenta of the particles is
balanced by the momentum of the gravitational field. \ 

The transformation from (\ref{act1}) to (\ref{actcan}) is carried out by
rewriting the particle Lagrangian into first order form using the
decomposition of the scalar curvature in terms of the extrinsic curvature $K$
via 
\[
\sqrt{-g}R=-2\partial _{0}(\sqrt{\gamma }K)+2\partial _{1}[\sqrt{\gamma }%
(N^{1}K-\gamma ^{-1}\partial _{1}N_{0})] 
\]%
where $K=(2N_{0}\gamma )^{-1}(2\partial _{1}N_{1}-\gamma ^{-1}N_{1}\partial
_{1}\gamma -\partial _{0}\gamma )$.

The action (\ref{actcan}) leads to the system of field equations : 
\begin{eqnarray}
\dot{\pi} &+&N_{0}\left\{ \frac{3\kappa }{2}\sqrt{\gamma }\pi ^{2}-\frac{%
\kappa }{\sqrt{\gamma }}\pi \Pi +\frac{1}{8\kappa \sqrt{\gamma }\gamma }%
(\Psi ^{\prime })^{2}-\sum_{a}\frac{p_{a}^{2}}{2\gamma ^{2}\sqrt{\frac{%
p_{a}^{2}}{\gamma }+m_{a}^{2}}}\;\delta (x-z_{a}(x^{0}))\right\}  \nonumber
\\
&+&N_{1}\left\{ -\frac{1}{\gamma ^{2}}\Pi \Psi ^{\prime }+\frac{\pi ^{\prime
}}{\gamma }+\sum_{a}\frac{p_{a}}{\gamma ^{2}}\;\delta
(x-z_{a}(x^{0}))\right\} +N_{0}^{\prime }\frac{1}{2\kappa \sqrt{\gamma }%
\gamma }\Psi ^{\prime }+N_{1}^{\prime }\frac{\pi }{\gamma }=0  \label{feqpi}
\\
&&\makebox[5em]{}\dot{\gamma}-N_{0}(2\kappa \sqrt{\gamma }\gamma \pi
-2\kappa \sqrt{\gamma }\Pi )+N_{1}\frac{\gamma ^{\prime }}{\gamma }%
-2N_{1}^{\prime }=0  \label{feq2} \\
&&\makebox[5em]{}R^{0}=0  \label{feq3} \\
&&\makebox[5em]{}R^{1}=0  \label{feq4} \\
&&\makebox[5em]{}\dot{\Pi}+\partial _{1}(-\frac{1}{\gamma }N_{1}\Pi +\frac{1%
}{2\kappa \sqrt{\gamma }}N_{0}\Psi ^{\prime }+\frac{1}{\kappa \sqrt{\gamma }}%
N_{0}^{\prime })=0  \label{feq5} \\
&&\makebox[5em]{}\dot{\Psi}+N_{0}(2\kappa \sqrt{\gamma }\pi )-N_{1}(\frac{1}{%
\gamma }\Psi ^{\prime })=0  \label{feq6} \\
&&\dot{p}_{a}+\frac{\partial N_{0}}{\partial z_{a}}\sqrt{\frac{p_{a}^{2}}{%
\gamma }+m_{a}^{2}}-\frac{N_{0}}{2\sqrt{\frac{p_{a}^{2}}{\gamma }+m_{a}^{2}}}%
\frac{p_{a}^{2}}{\gamma ^{2}}\frac{\partial \gamma }{\partial z_{a}}-\frac{%
\partial N_{1}}{\partial z_{a}}\frac{p_{a}}{\gamma }+N_{1}\frac{p_{a}}{%
\gamma ^{2}}\frac{\partial \gamma }{\partial z_{a}}=0  \label{feq7} \\
&&\makebox[5em]{}\dot{z_{a}}-N_{0}\frac{\frac{p_{a}}{\gamma }}{\sqrt{\frac{%
p_{a}^{2}}{\gamma }+m_{a}^{2}}}+\frac{N_{1}}{\gamma }=0\;\;.  \label{feq8}
\end{eqnarray}%
where all metric components ($N_{0}$, $N_{1}$, $\gamma $) are evaluated at
the point $x=z_{a}$ in eqs.(\ref{feq7},\ref{feq8}), with 
\[
\frac{\partial f}{\partial z_{a}}\equiv \left. \frac{\partial f(x)}{\partial
x}\right| _{x=z_{a}} 
\]%
{} The quantities $N_{0}$ and $N_{1}$ are Lagrange multipliers which yield
the constraint equations (\ref{feq3}) and (\ref{feq4}). It is
straightforward to show \cite{OR} that this system of equations is
equivalent to the set of equations (\ref{Psieq}), (\ref{eq-z}) and (\ref{RT}%
).\\


Full canonical reduction of the action (\ref{act1}) involves elimination of
the redundant variables by employing the constraint equations to fix the
coordinate conditions. The constraint equations (\ref{feq3}) and (\ref{feq4}%
) may be solved for the quantities $\left( \Psi ^{\prime }/\sqrt{\gamma }%
\right) ^{\prime }$ and $\pi ^{\prime }$, since they are the only linear
terms present. We then transform the total generator obtained from the end
point variation into an appropriate form to fix the coordinate conditions.
These conditions can consistently be chosen to be \cite{OR,2bd} {\bf \ } 
\begin{equation}
\gamma =1\makebox[2em]{}\mbox{and}\makebox[2em]{}\Pi =0\;.  \label{cc}
\end{equation}%
and, upon elimination of the constraints, yields 
\begin{equation}
I=\int d^{2}x\left\{ \sum_{a}p_{a}\dot{z}_{a}\delta (x-z_{a})+\frac{1}{%
\kappa }\triangle \Psi \right\} \;,  \label{actred}
\end{equation}%
for the action (\ref{actcan}). From this we read off the reduced Hamiltonian
for the system of $N$\ particles 
\begin{equation}
H=\int dx{\cal H}=-\frac{1}{\kappa }\int dx\left( \triangle \Psi \right) \;
\label{ham1}
\end{equation}%
where $\Psi $ is understood to be a function of $z_{a}$ and $p_{a}$
determined by solving the constraints (\ref{feq3}) and (\ref{feq4}). Under
the coordinate conditions (\ref{cc}) these become 
\begin{equation}
\triangle \Psi -\frac{1}{4}(\Psi ^{\prime })^{2}+\kappa ^{2}\pi ^{2}+\kappa
\sum_{a}\sqrt{p_{a}^{2}+m_{a}^{2}}\delta (x-z_{a})=0  \label{cst1}
\end{equation}%
\begin{equation}
2\pi ^{\prime }+\sum_{a}p_{a}\delta (x-z_{a})=0\;\;.  \label{cst2}
\end{equation}

The consistency of this canonical reduction can be demonstrated by showing
that the canonical equations of motion derived from the reduced Hamiltonian (%
\ref{ham1}) are identical with the equations eqs.(\ref{feq7},\ref{feq8}) %
\cite{2bd,2bdcoslo}.

\bigskip \bigskip

\section{Solving the 3-body Constraint Equations}

The standard approach for investigating the dynamics of particles is to get
first an explicit expression of the Hamiltonian and to derive the equations
of motion, from which the solution of trajectories are obtained. In this
section we show how to derive the Hamiltonian from the solution to the
constraint equations (\ref{cst1}) and (\ref{cst2}) and get an exact equation
expressing the Hamiltonian as a function of the phase-space degrees of
freedom for a system of three particles.

Defining $\phi $ and $\chi $ by 
\begin{equation}
\Psi =-4\mbox{log}|\phi |\qquad \pi =\chi ^{\prime }  \label{phichi}
\end{equation}%
the constraints (\ref{cst1}) and (\ref{cst2}) for a three-body system become 
\begin{eqnarray}
\triangle \phi -\frac{\kappa ^{2}}{4}(\chi ^{\prime })^{2}\phi &=&\frac{%
\kappa }{4}\left\{ \sqrt{p_{1}^{2}+m_{1}^{2}}\;\phi (z_{1})\delta (x-z_{1})+%
\sqrt{p_{2}^{2}+m_{2}^{2}}\;\phi (z_{2})\delta (x-z_{2})\right.  \nonumber \\
&&\left. \makebox[2em]{}+\sqrt{p_{3}^{2}+m_{3}^{2}}\;\phi (z_{3})\delta
(x-z_{3})\right\}  \label{e-phi} \\
\triangle \chi &=&-\frac{1}{2}\left\{ p_{1}\delta (x-z_{1})+p_{2}\delta
(x-z_{2})+p_{3}\delta (x-z_{3})\right\} \;\;.  \label{e-chi}
\end{eqnarray}%
The general solution to (\ref{e-chi}) is 
\begin{equation}
\chi =-\frac{1}{4}\left\{
p_{1}|x-z_{1}|+p_{2}|x-z_{2}|+p_{3}|x-z_{3}|\right\} -\epsilon Xx+\epsilon
C_{\chi }\;\;.  \label{chisol}
\end{equation}%
The factor $\epsilon $ ($\epsilon ^{2}=1$) flips sign under time-reversal $%
{\Bbb T}$, and has been introduced in the constants $X$ and $C_{\chi }$ so
that this property of $\chi $ is explicitly manifest.

Our next task is to solve (\ref{e-phi}).\ Consider first the case $%
z_{3}<z_{2}<z_{1}$, for which we may divide spacetime into four regions%
\[
z_{1}<x\makebox[3em]{}\mbox{(+) region} 
\]%
\[
z_{2}<x<z_{1}\makebox[1.5em]{}\mbox{(1) region} 
\]%
\[
z_{3}<x<z_{2}\makebox[1.5em]{}\mbox{(2) region} 
\]%
\[
x<z_{3}\makebox[3em]{}\mbox{(-) region} 
\]%
within each of which $\chi ^{\prime }$ is constant: 
\begin{equation}
\chi ^{\prime }=\left\{ 
\begin{array}{ll}
-\epsilon X-\frac{1}{4}(p_{1}+p_{2}+p_{3}) & \makebox[3em]{}\mbox{(+) region}
\\ 
-\epsilon X+\frac{1}{4}(p_{1}-p_{2}-p_{3}) & \makebox[3em]{}\mbox{(1) region}
\\ 
-\epsilon X+\frac{1}{4}(p_{1}+p_{2}-p_{3}) & \makebox[3em]{}\mbox{(2) region}
\\ 
-\epsilon X+\frac{1}{4}(p_{1}+p_{2}+p_{3}) & \makebox[3em]{}\mbox{(-) region}%
\end{array}%
\right.
\end{equation}%
It is straightforward to solve the homogeneous equation $\triangle \phi
-(\kappa ^{2}/4)(\chi ^{\prime })^{2}\phi =0$ in each region:%
\begin{equation}
\left\{ 
\begin{array}{l}
\phi _{+}(x)=A_{+}e^{\frac{\kappa }{2}K_{+}x}+B_{+}e^{-\frac{\kappa }{2}%
K_{+}x} \\ 
\phi _{1}(x)=A_{1}e^{\frac{\kappa }{2}K_{1}x}+B_{1}e^{-\frac{\kappa }{2}%
K_{1}x} \\ 
\phi _{2}(x)=A_{2}e^{\frac{\kappa }{2}K_{2}x}+B_{2}e^{-\frac{\kappa }{2}%
K_{2}x} \\ 
\phi _{-}(x)=A_{-}e^{\frac{\kappa }{2}K_{-}x}+B_{-}e^{-\frac{\kappa }{2}%
K_{-}x}%
\end{array}%
\right. \;\;.  \label{e-phisol}
\end{equation}%
where 
\begin{eqnarray}
&&K_{+}\equiv X+\frac{\epsilon }{4}(p_{1}+p_{2}+p_{3}),\qquad K_{1}\equiv X-%
\frac{\epsilon }{4}(p_{1}-p_{2}-p_{3}),\qquad K_{2}\equiv X-\frac{\epsilon }{%
4}(p_{1}+p_{2}-p_{3})  \label{K-const} \\
&&K_{-}\equiv X-\frac{\epsilon }{4}(p_{1}+p_{2}+p_{3})\;\;.  \nonumber
\end{eqnarray}%
For these solutions to be the actual solutions to (\ref{e-phi}) with delta
function source terms, they must satisfy the following matching conditions
at the locations of the particles $x=z_{1},z_{2},z_{3}$: {\ %
\setcounter{enumi}{\value{equation}} \addtocounter{enumi}{1} %
\setcounter{equation}{0} \renewcommand{\theequation}{\theenumi%
\alph{equation}} 
\begin{eqnarray}
&&\phi _{+}(z_{1})=\phi _{1}(z_{1})=\phi (z_{1})  \label{match-a} \\
&&\phi _{1}(z_{2})=\phi _{2}(z_{2})=\phi (z_{2})  \label{match-b} \\
&&\phi _{-}(z_{3})=\phi _{2}(z_{3})=\phi (z_{3})  \label{match-c} \\
&&\phi _{+}^{\prime }(z_{1})-\phi _{1}^{\prime }(z_{1})=\frac{\kappa }{4}%
\sqrt{p_{1}^{2}+m_{1}^{2}}\phi (z_{1})  \label{match-d} \\
&&\phi _{1}^{\prime }(z_{2})-\phi _{2}^{\prime }(z_{2})=\frac{\kappa }{4}%
\sqrt{p_{2}^{2}+m_{2}^{2}}\phi (z_{2})  \label{match-e} \\
&&\phi _{2}^{\prime }(z_{3})-\phi _{-}^{\prime }(z_{3})=\frac{\kappa }{4}%
\sqrt{p_{3}^{2}+m_{3}^{2}}\phi (z_{3})\;\;.  \label{match-f}
\end{eqnarray}%
\setcounter{equation}{\value{enumi}} } The conditions (\ref{match-a}) and (%
\ref{match-d}) lead to 
\begin{equation}
e^{\frac{\kappa }{2}K_{+}z_{1}}A_{+}+e^{-\frac{\kappa }{2}%
K_{+}z_{1}}B_{+}=e^{\frac{\kappa }{2}K_{1}z_{1}}A_{1}+e^{-\frac{\kappa }{2}%
K_{1}z_{1}}B_{1}
\end{equation}%
and 
\begin{eqnarray}
\lefteqn{e^{\frac{\kappa }{2}K_{+}z_{1}}A_{+}-e^{-\frac{\kappa }{2}%
K_{+}z_{1}}B_{+}}  \nonumber \\
&=&\frac{\sqrt{p_{1}^{2}+m_{1}^{2}}+2K_{1}}{2K_{+}}e^{\frac{\kappa }{2}%
K_{1}z_{1}}A_{1}+\frac{\sqrt{p_{1}^{2}+m_{1}^{2}}-2K_{1}}{2K_{+}}e^{-\frac{%
\kappa }{2}K_{1}z_{1}}B_{1}\;\;.
\end{eqnarray}%
Then {\ \setcounter{enumi}{\value{equation}} \addtocounter{enumi}{1} %
\setcounter{equation}{0} \renewcommand{\theequation}{\theenumi%
\alph{equation}} 
\begin{eqnarray}
A_{+} &=&\frac{\sqrt{p_{1}^{2}+m_{1}^{2}}+2(K_{+}+K_{1})}{4K_{+}}e^{-\frac{%
\kappa }{2}(K_{+}-K_{1})z_{1}}A_{1}  \nonumber \\
&&\makebox[5em]{}+\frac{\sqrt{p_{1}^{2}+m_{1}^{2}}+2(K_{+}-K_{1})}{4K_{+}}%
e^{-\frac{\kappa }{2}(K_{+}+K_{1})z_{1}}B_{1}  \label{A+} \\
B_{+} &=&-\frac{\sqrt{p_{1}^{2}+m_{1}^{2}}-2(K_{+}-K_{1})}{4K_{+}}e^{\frac{%
\kappa }{2}(K_{+}+K_{1})z_{1}}A_{1}  \nonumber \\
&&\makebox[5em]{}-\frac{\sqrt{p_{1}^{2}+m_{1}^{2}}-2(K_{+}+K_{1})}{4K_{+}}e^{%
\frac{\kappa }{2}(K_{+}-K_{1})z_{1}}B_{1}\;\;.  \label{B+}
\end{eqnarray}%
\setcounter{equation}{\value{enumi}} } Similarly from (\ref{match-b}) and (%
\ref{match-e}) we obtain%
\begin{equation}
e^{\frac{\kappa }{2}K_{1}z_{2}}A_{1}+e^{-\frac{\kappa }{2}%
K_{1}z_{2}}B_{1}=e^{\frac{\kappa }{2}K_{2}z_{2}}A_{2}+e^{-\frac{\kappa }{2}%
K_{2}z_{2}}B_{2}
\end{equation}%
and 
\begin{eqnarray}
\lefteqn{e^{\frac{\kappa }{2}K_{1}z_{2}}A_{1}-e^{-\frac{\kappa }{2}%
K_{1}z_{2}}B_{1}}  \nonumber \\
&=&\frac{\sqrt{p_{2}^{2}+m_{2}^{2}}+2K_{2}}{2K_{1}}e^{\frac{\kappa }{2}%
K_{2}z_{2}}A_{2}+\frac{\sqrt{p_{2}^{2}+m_{2}^{2}}-2K_{2}}{2K_{1}}e^{-\frac{%
\kappa }{2}K_{2}z_{2}}B_{2}\;\;.
\end{eqnarray}%
Then {\ \setcounter{enumi}{\value{equation}} \addtocounter{enumi}{1} %
\setcounter{equation}{0} \renewcommand{\theequation}{\theenumi%
\alph{equation}} 
\begin{eqnarray}
A_{1} &=&\frac{\sqrt{p_{2}^{2}+m_{2}^{2}}+2(K_{1}+K_{2})}{4K_{1}}e^{-\frac{%
\kappa }{2}(K_{1}-K_{2})z_{2}}A_{2}  \nonumber \\
&&\makebox[5em]{}+\frac{\sqrt{p_{2}^{2}+m_{2}^{2}}+2(K_{1}-K_{2})}{4K_{1}}%
e^{-\frac{\kappa }{2}(K_{1}+K_{2})z_{2}}B_{2}  \label{A1} \\
B_{1} &=&-\frac{\sqrt{p_{2}^{2}+m_{2}^{2}}-2(K_{1}-K_{2})}{4K_{1}}e^{\frac{%
\kappa }{2}(K_{1}+K_{2})z_{2}}A_{2}  \nonumber \\
&&\makebox[5em]{}-\frac{\sqrt{p_{2}^{2}+m_{2}^{2}}-2(K_{1}+K_{2})}{4K_{1}}e^{%
\frac{\kappa }{2}(K_{1}-K_{2})z_{2}}B_{2}\;\;.  \label{B1}
\end{eqnarray}%
\setcounter{equation}{\value{enumi}} } Finally, from (\ref{match-c}) and (%
\ref{match-f}) we get 
\begin{equation}
e^{\frac{\kappa }{2}K_{-}z_{3}}A_{-}+e^{-\frac{\kappa }{2}%
K_{-}z_{3}}B_{-}=e^{\frac{\kappa }{2}K_{2}z_{3}}A_{2}+e^{-\frac{\kappa }{2}%
K_{2}z_{3}}B_{2}
\end{equation}%
and 
\begin{eqnarray}
\lefteqn{e^{\frac{\kappa }{2}K_{-}z_{3}}A_{-}-e^{-\frac{\kappa }{2}%
K_{-}z_{3}}B_{-}}  \nonumber \\
&=&-\frac{\sqrt{p_{3}^{2}+m_{3}^{2}}-2K_{2}}{2K_{-}}e^{\frac{\kappa }{2}%
K_{2}z_{3}}A_{2}-\frac{\sqrt{p_{3}^{2}+m_{3}^{2}}+2K_{2}}{2K_{-}}e^{-\frac{%
\kappa }{2}K_{2}z_{3}}B_{2}\;\;.
\end{eqnarray}%
Then {\ \setcounter{enumi}{\value{equation}} \addtocounter{enumi}{1} %
\setcounter{equation}{0} \renewcommand{\theequation}{\theenumi%
\alph{equation}} 
\begin{eqnarray}
A_{-} &=&-\frac{\sqrt{p_{3}^{2}+m_{3}^{2}}-2(K_{2}+K_{-})}{4K_{-}}e^{\frac{%
\kappa }{2}(K_{2}-K_{-})z_{3}}A_{2}  \nonumber \\
&&\makebox[5em]{}-\frac{\sqrt{p_{3}^{2}+m_{3}^{2}}+2(K_{2}-K_{-})}{4K_{-}}%
e^{-\frac{\kappa }{2}(K_{2}+K_{-})z_{3}}B_{2}  \label{A-} \\
B_{-} &=&\frac{\sqrt{p_{3}^{2}+m_{3}^{2}}-2(K_{2}-K_{-})}{4K_{-}}e^{\frac{%
\kappa }{2}(K_{2}+K_{-})z_{3}}A_{2}  \nonumber \\
&&\makebox[5em]{}+\frac{\sqrt{p_{3}^{2}+m_{3}^{2}}+2(K_{2}+K_{-})}{4K_{-}}%
e^{-\frac{\kappa }{2}(K_{2}-K_{-})z_{3}}B_{2}\;\;.  \label{B-}
\end{eqnarray}%
\setcounter{equation}{\value{enumi}} }

Since the magnitudes of both $\phi $ and $\chi $ increase with increasing $%
|x|$, we must impose a boundary condition that ensures that the surface
terms which arise in transforming the action vanish and simultaneously
preserves the finiteness of the Hamiltonian. This condition can be shown to
be \cite{OR,2bd} 
\begin{equation}
\Psi ^{2}-4\kappa ^{2}\chi ^{2}=0\makebox[2em]{}\mbox{in $(\pm)$ regions}%
\;\;.  \label{bndcnd}
\end{equation}%
Since 
\begin{equation}
\chi =\left\{ 
\begin{array}{ll}
-\left\{ \epsilon X+\frac{1}{4}(p_{1}+p_{2}+p_{3})\right\} x+\epsilon
C_{\chi }+\frac{1}{4}(p_{1}z_{1}+p_{2}z_{2}+p_{3}z_{3}) & \makebox[3em]{}(+)%
\mbox{region} \\ 
-\left\{ \epsilon X-\frac{1}{4}(p_{1}+p_{2}+p_{3})\right\} x+\epsilon
C_{\chi }-\frac{1}{4}(p_{1}z_{1}+p_{2}z_{2}+p_{3}z_{3}) & \makebox[3em]{}(-)%
\mbox{region}%
\end{array}%
\right.  \label{chiregions}
\end{equation}%
the boundary condition implies 
\begin{eqnarray}
&&A_{-}=B_{+}=0  \label{bound1} \\
&&-\mbox{log}A_{+}-\frac{\kappa \epsilon }{8}%
(p_{1}z_{1}+p_{2}z_{2}+p_{3}z_{3})=\mbox{log}B_{-}+\frac{\kappa \epsilon }{8}%
(p_{1}z_{1}+p_{2}z_{2}+p_{3}z_{3})=\frac{\kappa }{2}C_{\chi }\;\;.
\label{bound2}
\end{eqnarray}%
The condition (\ref{bound1}) leads to 
\begin{equation}
\frac{A_{1}}{B_{1}}=-\frac{\sqrt{p_{1}^{2}+m_{1}^{2}}-2(K_{+}+K_{1})}{\sqrt{%
p_{1}^{2}+m_{1}^{2}}-2(K_{+}-K_{1})}e^{-\kappa K_{1}z_{1}}\;\;.  \label{AB1}
\end{equation}%
and 
\begin{equation}
\frac{A_{2}}{B_{2}}=-\frac{\sqrt{p_{3}^{2}+m_{3}^{2}}+2(K_{2}-K_{-})}{\sqrt{%
p_{3}^{2}+m_{3}^{2}}-2(K_{2}+K_{-})}e^{-\kappa K_{2}z_{3}}\;\;.  \label{AB2}
\end{equation}%
Continuing, from (\ref{A1}), (\ref{B1}), (\ref{AB1}) and (\ref{AB2}), we get 
\begin{equation}
\frac{\sqrt{p_{1}^{2}+m_{1}^{2}}-2(K_{+}+K_{1})}{\sqrt{p_{1}^{2}+m_{1}^{2}}%
-2(K_{+}-K_{1})}e^{-\kappa K_{1}z_{1}}=\frac{N}{D}  \label{N/D}
\end{equation}%
where 
\begin{eqnarray}
N &=&\left[ \sqrt{p_{2}^{2}+m_{2}^{2}}+2(K_{1}+K_{2})\right] \left[ \sqrt{%
p_{3}^{2}+m_{3}^{2}}+2(K_{2}-K_{-})\right] e^{-\frac{\kappa }{2}%
(K_{1}-K_{2})z_{2}-\kappa K_{2}z_{3}}  \nonumber \\
&&-\left[ \sqrt{p_{2}^{2}+m_{2}^{2}}+2(K_{1}-K_{2})\right] \left[ \sqrt{%
p_{3}^{2}+m_{3}^{2}}-2(K_{2}+K_{-})\right] e^{-\frac{\kappa }{2}%
(K_{1}+K_{2})z_{2}}  \nonumber \\
D &=&\left[ \sqrt{p_{2}^{2}+m_{2}^{2}}-2(K_{1}-K_{2})\right] \left[ \sqrt{%
p_{3}^{2}+m_{3}^{2}}+2(K_{2}-K_{-})\right] e^{\frac{\kappa }{2}%
(K_{1}+K_{2})z_{2}-\kappa K_{2}z_{3}}  \nonumber \\
&&-\left[ \sqrt{p_{2}^{2}+m_{2}^{2}}-2(K_{1}+K_{2})\right] \left[ \sqrt{%
p_{3}^{2}+m_{3}^{2}}-2(K_{2}+K_{-})\right] e^{\frac{\kappa }{2}%
(K_{1}-K_{2})z_{2}}\;\;.  \nonumber
\end{eqnarray}%
and so equation (\ref{N/D}) leads to 
\begin{eqnarray}
&&\left[ \sqrt{p_{1}^{2}+m_{1}^{2}}-2(K_{+}+K_{1})\right] \left[ \sqrt{%
p_{2}^{2}+m_{2}^{2}}-2(K_{1}+K_{2})\right] \left[ \sqrt{p_{3}^{2}+m_{3}^{2}}%
-2(K_{2}+K_{-})\right]  \nonumber \\
&&-\left[ \sqrt{p_{1}^{2}+m_{1}^{2}}-2(K_{+}-K_{1})\right] \left[ \sqrt{%
p_{2}^{2}+m_{2}^{2}}+2(K_{1}-K_{2})\right]  \nonumber \\
&&\makebox[10em]{}\times \left[ \sqrt{p_{3}^{2}+m_{3}^{2}}-2(K_{2}+K_{-})%
\right] e^{\kappa K_{1}z_{12}}  \nonumber \\
&&+\left[ \sqrt{p_{1}^{2}+m_{1}^{2}}-2(K_{+}-K_{1})\right] \left[ \sqrt{%
p_{2}^{2}+m_{2}^{2}}+2(K_{1}+K_{2})\right]  \nonumber \\
&&\makebox[10em]{}\times \left[ \sqrt{p_{3}^{2}+m_{3}^{2}}+2(K_{2}-K_{-})%
\right] e^{\kappa (K_{1}z_{12}+K_{2}z_{23}}  \nonumber \\
&&-\left[ \sqrt{p_{1}^{2}+m_{1}^{2}}-2(K_{+}+K_{1})\right] \left[ \sqrt{%
p_{2}^{2}+m_{2}^{2}}-2(K_{1}-K_{2})\right]  \nonumber \\
&&\makebox[10em]{}\times \left[ \sqrt{p_{3}^{2}+m_{3}^{2}}+2(K_{2}-K_{-})%
\right] e^{\kappa K_{2}z_{23}}  \nonumber \\
&=&0\;\;.  \label{e-Keq}
\end{eqnarray}

Insertion of (\ref{e-phisol}) into (\ref{ham1}) implies that $H=4X$. We can
rewrite (\ref{e-Keq}) in terms of $H$ as 
\begin{eqnarray}
&&\left[ H-\sqrt{p_{1}^{2}+m_{1}^{2}}+\epsilon (p_{2}+p_{3})\right] \left[ H-%
\sqrt{p_{2}^{2}+m_{2}^{2}}-\epsilon (p_{1}-p_{3})\right] \left[ H-\sqrt{%
p_{3}^{2}+m_{3}^{2}}-\epsilon (p_{1}+p_{2})\right]  \nonumber \\
&&-\left[ \sqrt{p_{1}^{2}+m_{1}^{2}}-\epsilon p_{1}\right] \left[ \sqrt{%
p_{2}^{2}+m_{2}^{2}}+\epsilon p_{2}\right] \left[ H-\sqrt{p_{3}^{2}+m_{3}^{2}%
}-\epsilon (p_{1}+p_{2})\right] e^{\frac{\kappa }{4}\left\{ H-\epsilon
(p_{1}-p_{2}-p_{3})\right\} z_{12}}  \nonumber \\
&&-\left[ \sqrt{p_{1}^{2}+m_{1}^{2}}-\epsilon p_{1}\right] \left[ H+\sqrt{%
p_{2}^{2}+m_{2}^{2}}-\epsilon (p_{1}-p_{3})\right] \left[ \sqrt{%
p_{3}^{2}+m_{3}^{2}}+\epsilon p_{3}\right]  \nonumber \\
&&\makebox[8em]{}\times e^{\frac{\kappa }{4}\left\{ H-\epsilon
(p_{1}-p_{2}-p_{3})\right\} z_{12}+\frac{\kappa }{4}\left\{ H-\epsilon
(p_{1}+p_{2}-p_{3})\right\} z_{23}}  \nonumber \\
&&-\left[ H-\sqrt{p_{1}^{2}+m_{1}^{2}}+\epsilon (p_{2}+p_{3})\right] \left[ 
\sqrt{p_{2}^{2}+m_{2}^{2}}-\epsilon p_{2}\right] \left[ \sqrt{%
p_{3}^{2}+m_{3}^{2}}+\epsilon p_{3}\right] e^{\frac{\kappa }{4}\left\{
H-\epsilon (p_{1}+p_{2}-p_{3})\right\} z_{23}}  \nonumber \\
&=&0  \label{detH}
\end{eqnarray}%
which is the determining equation of the Hamiltonian for the system of three
particles in the case of $z_{3}<z_{2}<z_{1}$.

The full determining equation is obtained with the permutation of suffices
1, 2 and 3. To find it, we begin be rewriting the somewhat cumbersome
expression (\ref{detH}) as 
\begin{eqnarray}
L_{1}L_{2}L_{3} &=&[M_{1}-\epsilon p_{1}][M_{2}+\epsilon p_{2}]L_{3}^{\ast
}e^{\kappa /4[(L_{1}+M_{1}-\epsilon p_{1})z_{13}-(L_{2}+M_{2}+\epsilon
p_{2})z_{23}]}  \label{detHL} \\
&&+[M_{1}-\epsilon p_{1}][M_{3}+\epsilon p_{3}][L_{2}^{\ast }]e^{\kappa
/4[(L_{1}+M_{1}-\epsilon p_{1})z_{12}+(L_{3}+M_{3}+\epsilon p_{3})z_{23}]} 
\nonumber \\
&&+L_{1}^{\ast }[M_{2}-\epsilon p_{2}][M_{3}+\epsilon p_{3}]e^{\kappa
/4[(L_{3}+M_{3}+\epsilon p_{3})z_{13}-(L_{2}+M_{2}-\epsilon p_{2})z_{12}]} 
\nonumber
\end{eqnarray}%
where 
\begin{eqnarray}
L_{1} &=&H-\sqrt{p_{1}^{2}+m_{1}^{2}}+\epsilon (p_{2}+p_{3})\textrm{ \ \ \ \ }%
L_{2}=H-\sqrt{p_{2}^{2}+m_{2}^{2}}-\epsilon (p_{1}-p_{3})  \label{Ldef1} \\
L_{3} &=&H-\sqrt{p_{3}^{2}+m_{3}^{2}}-\epsilon (p_{1}+p_{2})\textrm{ \ \ \ \ }%
L_{1}^{\ast }=L_{1};L_{3}^{\ast }=L_{3}  \label{Ldef2} \\
L_{2}^{\ast } &=&H+\sqrt{p_{2}^{2}+m_{2}^{2}}-\epsilon (p_{1}-p_{3})\textrm{ \
\ \ \ }M_{i}=\sqrt{p_{i}^{2}+m_{i}^{2}}  \label{Ldef3}
\end{eqnarray}

We obtain solutions when the particles are in a different order by permuting
the indices in the solution shown above. This leaves the solution
essentially the same, except for a number of sign interchanges. First,
consider what happens to the $L_{i}$ terms after the particles cross. Their
general form is: 
\begin{equation}
L_{i}=H-\sqrt{p_{i}^{2}+m_{i}^{2}}-\epsilon (\pm p_{k}\pm p_{j})
\label{Lintdef}
\end{equation}%
with $j\neq k$. To determine the signs in the third term, note that the $L$%
's above obey the following pattern: for $L_{i}$, we have {$-\epsilon p_{j}$
if $j<i$ (that is, if $z_{j}-z_{i}>0$) and $-\epsilon (-p_{j})$ if $j>i$
(that is, if $z_{j}-z_{i}<0$). }Careful inspection then shows that $%
L_{i}=H-M_{i}-\epsilon (\sum_{j}p_{j}s_{ji})$, where $s_{ij}=$sgn$%
(z_{i}-z_{j})$. \ 

The $L^{\ast }$'s are the same, except for the middle particle, for which
the $M_{i}$ term flips sign. \ This means that we can write $L_{i}^{\ast
}=(1-\prod_{j<k\neq i}s_{ij}s_{ik})M_{i}+L_{i}$, where the first term
vanishes unless the $i$th particle is in the middle.

Finally, consider what happens to terms of the form $M_{i}\pm \epsilon p_{i}$
when the particles are permuted. Notice that the sign of the second term is
always negative for the particle on the right, and always positive for the
particle on the left. For the particle in the middle, the sign is positive
when it is added to or multiplied by terms relating to the particle on its
right (in which case it plays the role of the leftmost particle), and
negative when it is added to or multiplied by terms relating to the particle
on its left (when it plays the role of the rightmost particle). We can
summarize this information by writing ${\frak M}_{ij}=M_{i}-\epsilon
p_{i}s_{ij}$.

Putting this information together, we obtain%
\begin{eqnarray}
L_{1}L_{2}L_{3} &=&{\frak M}_{12}{\frak M}_{21}L_{3}^{\ast }e^{\frac{\kappa 
}{4}s_{12}[(L_{1}+{\frak M}_{12})z_{13}-(L_{2}+{\frak M}_{21})z_{23}]} 
\nonumber \\
&&+{\frak M}_{23}{\frak M}_{32}L_{1}^{\ast }e^{\frac{\kappa }{4}%
s_{23}[(L_{2}+{\frak M}_{23})z_{21}-(L_{3}+{\frak M}_{32})z_{31}]}  \nonumber
\\
&&+{\frak M}_{31}{\frak M}_{13}L_{2}^{\ast }e^{\frac{\kappa }{4}%
s_{31}[(L_{3}+{\frak M}_{31})z_{32}-(L_{1}+{\frak M}_{13})z_{12}]}
\label{Htrans}
\end{eqnarray}%
or more compactly 
\begin{equation}
L_{1}L_{2}L_{3}=\sum_{ijk}\left| \epsilon ^{ijk}\right| {\frak M}_{ij}{\frak %
M}_{ji}L_{k}^{\ast }e^{\frac{\kappa }{4}s_{ij}[(L_{i}+{\frak M}%
_{ij})z_{ik}-(L_{j}+{\frak M}_{ji})z_{jk}]}  \label{Htranscomp}
\end{equation}%
for the full determining equation, where 
\begin{eqnarray}
{\frak M}_{ij} &=&M_{i}-\epsilon p_{i}s_{ij},\textrm{ \ \ \ \ \ \ \ \ \ \ \ \
\ \ \ \ \ \ \ \ \ \ \ }M_{i}=\sqrt{p_{i}^{2}+m_{i}^{2}}  \label{massM} \\
L_{i} &=&H-M_{i}-\epsilon (\sum_{j}p_{j}s_{ji})\textrm{ \ \ \ \ \ \ \ \ }%
L_{i}^{\ast }=(1-\prod_{j<k\neq i}s_{ij}s_{ik})M_{i}+L_{i}  \label{Ldef}
\end{eqnarray}%
with $z_{ij}=(z_{i}-z_{j})$ , $s_{ij}=$sgn$(z_{ij})$, and $\epsilon ^{ijk}$
is the 3-dimensional Levi-Civita tensor. \ It is straightforward (but
somewhat tedious) to check that eq. (\ref{Htranscomp}) indeed reproduces the
correct determining equation for any permutation of the particles.

The next task is to obtain the equations of motion from the Hamiltonian. For
the $N$-body case we can use the canonical equations \cite{OR}%
\begin{eqnarray}
\dot{z}_{a} &=&\frac{\partial H}{\partial p{_{a}}}  \label{zdot} \\
\dot{p}_{a} &=&-\frac{\partial H}{\partial z{_{a}}}  \label{pdot}
\end{eqnarray}%
where as previously mentioned the overdot denotes a derivative with respect
to $t$. Although we do not have a closed-form expression for $H$, we can
nevertheless obtain explicit expressions for $\left( \dot{z}_{a},\dot{p}%
_{a}\right) $ by implicit differentiation of both sides of (\ref{Htranscomp}%
).

For example, after differentiation of (\ref{Htrans}) with respect to $p_{1}$
we find after some algebra

\begin{eqnarray}
&&\dot{z}_{1}\left\{ L_{2}L_{3}+L_{1}L_{3}+L_{1}L_{2}\right. 
\nonumber \\
&&-[M_{2}-\epsilon p_{2}s_{21}][M_{1}-\epsilon p_{1}s_{12}][1+\frac{\kappa }{%
4}L_{3}^{\ast }|z_{12}|]e^{\frac{\kappa }{4}s_{12}[(L_{1}+{\frak M}%
_{12})z_{13}-(L_{2}+{\frak M}_{21})z_{23}]}  \nonumber \\
&&-[M_{3}-\epsilon p_{3}s_{31}][M_{1}-\epsilon p_{1}s_{13}][1+\frac{\kappa }{%
4}L_{2}^{\ast }|z_{13}|]e^{\frac{\kappa }{4}s_{13}[(L_{1}+{\frak M}%
_{23})z_{12}+(L_{3}+{\frak M}_{32})z_{23}]}  \nonumber \\
&&\left. -[M_{2}-\epsilon p_{2}s_{23}][M_{3}-\epsilon p_{3}s_{32}][1+\frac{%
\kappa }{4}L_{1}^{\ast }|z_{23}|]e^{\frac{\kappa }{4}s_{23}[(L_{3}+{\frak M}%
_{31})z_{13}-(L_{2}+{\frak M}_{13})z_{12}]}\right\}  \nonumber \\
&=&[M_{2}-\epsilon p_{2}s_{21}][(\frac{\partial M_{1}}{\partial p_{1}}%
-\epsilon s_{12})L_{3}^{\ast }-(M_{1}-\epsilon p_{1}s_{12})(\epsilon s_{13}+%
\frac{\kappa }{4}L_{3}^{\ast }(\epsilon z_{12}))]e^{\frac{\kappa }{4}%
s_{12}[(L_{1}+{\frak M}_{12})z_{13}-(L_{2}+{\frak M}_{21})z_{23}]}  \nonumber
\\
&&+[M_{3}-\epsilon p_{3}s_{31}][(\frac{\partial M_{1}}{\partial p_{1}}%
-\epsilon s_{13})L_{2}^{\ast }-(M_{1}-\epsilon p_{1}s_{13})\{\epsilon s_{12}+%
\frac{\kappa }{4}L_{2}^{\ast }(\epsilon z_{13})\}]e^{\frac{\kappa }{4}%
s_{13}[(L_{1}+{\frak M}_{23})z_{12}+(L_{3}+{\frak M}_{32})z_{23}]}  \nonumber
\\
&&+[M_{2}-\epsilon p_{2}s_{23}][M_{3}-\epsilon p_{3}s_{32}][-s_{12}s_{13}%
\frac{\partial M_{1}}{\partial p_{1}}+\frac{\kappa }{4}s_{23}L_{1}^{\ast
}[\epsilon |z_{12}|-\epsilon |z_{13}|]]e^{\frac{\kappa }{4}s_{23}[(L_{3}+%
{\frak M}_{31})z_{13}-(L_{2}+{\frak M}_{13})z_{12}]}  \nonumber \\
&&+\frac{\partial M_{1}}{\partial p_{1}}L_{2}L_{3}+\epsilon
(s_{12}L_{1}L_{3}+s_{13}L_{2}L_{1})  \label{z1dot}
\end{eqnarray}

The expressions for $\dot{z}_{2}$ and $\dot{z}_{3}$ are extremely similar
and we shall not reproduce them here. Similarly, differentiating (\ref%
{Htrans}) with respect to $z_{1}$ gives after some algebra

\begin{eqnarray}
&&\dot{p}_{1}\left\{ L_{2}L_{3}+L_{1}L_{3}+L_{1}L_{2}\right. 
\nonumber \\
&&-[M_{2}-\epsilon p_{2}s_{21}][M_{1}-\epsilon p_{1}s_{12}][1+\frac{\kappa }{%
4}L_{3}^{\ast }|z_{12}|]e^{\frac{\kappa }{4}s_{12}[(L_{1}+{\frak M}%
_{12})z_{13}-(L_{2}+{\frak M}_{21})z_{23}]}  \nonumber \\
&&-[M_{3}-\epsilon p_{3}s_{31}][M_{1}-\epsilon p_{1}s_{13}][1+\frac{\kappa }{%
4}L_{2}^{\ast }|z_{13}|]e^{\frac{\kappa }{4}s_{13}[(L_{1}+{\frak M}%
_{23})z_{12}+(L_{3}+{\frak M}_{32})z_{23}]}  \nonumber \\
&&\left. -[M_{2}-\epsilon p_{2}s_{23}][M_{3}-\epsilon p_{3}s_{32}][1+\frac{%
\kappa }{4}L_{1}^{\ast }|z_{23}|]e^{\kappa /4s_{23}[(L_{3}+M_{3}-\epsilon
p_{3}s_{32})z_{13}-(L_{2}+M_{2}-\epsilon p_{2}s_{23})z_{12}]}\right\} 
\nonumber \\
&=&[M_{2}-\epsilon p_{2}s_{21}][M_{1}-\epsilon p_{1}s_{12}][\frac{\kappa }{4}%
s_{12}L_{3}^{\ast }[H+\epsilon (p_{2}-p_{1})s_{12}+\epsilon p_{3}s_{13}]]e^{%
\frac{\kappa }{4}s_{12}[(L_{1}+{\frak M}_{12})z_{13}-(L_{2}+{\frak M}%
_{21})z_{23}]}  \nonumber \\
&&+[M_{3}-\epsilon p_{3}s_{31}][M_{1}-\epsilon p_{1}s_{13}][\frac{\kappa }{4}%
s_{13}L_{2}^{\ast }[H+\epsilon p_{2}s_{12}+\epsilon (p_{3}-p_{1})s_{13}]]e^{%
\frac{\kappa }{4}s_{13}[(L_{1}+{\frak M}_{23})z_{12}+(L_{3}+{\frak M}%
_{32})z_{23}]}  \nonumber \\
&&+[M_{2}-\epsilon p_{2}s_{23}][M_{3}-\epsilon p_{3}s_{32}][\frac{\kappa }{4}%
s_{23}L_{1}^{\ast }p_{1}(s_{12}-s_{13})]e^{\frac{\kappa }{4}s_{23}[(L_{3}+%
{\frak M}_{31})z_{13}-(L_{2}+{\frak M}_{13})z_{12}]}  \label{pdot1}
\end{eqnarray}%
The other expressions are similar and we shall omit them here.

The components of the metric are determined from the equations (\ref{feqpi}%
),(\ref{feq2}),(\ref{feq5}) and (\ref{feq6}) under the coordinate conditions
(\ref{cc}). It is straightforward to verify that insertion of the solutions
of the metric and dilaton fields also solve the particle equations (\ref%
{feq7}) and (\ref{feq8}), as in the two-body case \cite{OR}. We shall not
present a derivation of the explicit solutions of these fields here, but
shall instead defer this calculation to a more complete study of the unequal
mass case \cite{justin}.

\section{General Properties of the Equations of Motion}

In this section we undertake a general analysis of the determining equation
for the Hamiltonian (\ref{Htrans}) and the equations of motion (\ref{zdot},%
\ref{pdot}) before proceeding to (numerically) solve them.

Consider first the determining equation (\ref{Htrans}). Its solution yields\
the Hamiltonian, which is a function of only four independent variables: the
two separations between the particles and their conjugate momenta. \ Hence a
simpler description can be given by employing the following change of
coordinates%
\begin{eqnarray}
z_{1}-z_{2}=\sqrt{2}\rho &&  \label{hexrho} \\
z_{1}+z_{2}-2z_{3}=\sqrt{6}\lambda &&  \label{hamlam}
\end{eqnarray}%
which in turn implies%
\begin{equation}
z_{12}=\sqrt{2}\rho \textrm{ \ \ \ \ }z_{13}=\frac{1}{\sqrt{2}}(\sqrt{3}%
\lambda +\rho )\textrm{ \ \ \ \ \ }z_{23}=\frac{1}{\sqrt{2}}(\sqrt{3}\lambda
-\rho )  \label{zijrholam}
\end{equation}%
The coordinates $\rho $ and $\lambda $ describe the motions of the three
particles about their center of mass. Their conjugate momenta can be
straightforwardly obtained by imposing the requirement%
\begin{equation}
\{q_{\textrm{{\sc A}}},p_{\textrm{{\sc b}}}\}=\delta _{\textrm{{\sc AB}}}
\label{Poissonbrack}
\end{equation}%
where{\bf \ }{\sc A}{\bf ,}{\sc B}$=\rho ,\lambda ,Z${\bf .} \ This yields 
\begin{eqnarray}
p_{\rho } &=&\frac{1}{\sqrt{2}}(p_{1}-p_{2})  \label{prho} \\
p_{\lambda } &=&\frac{1}{\sqrt{6}}(p_{1}+p_{2}-2p_{3})  \label{plam} \\
p_{Z} &=&\frac{1}{3}(p_{1}+p_{2}+p_{3})  \label{pX}
\end{eqnarray}%
where $Z=z_{1}+z_{2}+z_{3}$ is the remaining irrelevant coordinate degree of
freedom: the Hamiltonian is independent of $Z$ and $p_{Z}$. \ In the
non-relativistic limit $\left( Z,p_{Z}\right) $ corresponds to the centre of
mass and its conjugate momenta. While it is not possible to fix the value of 
$Z$ relativistically, it is possible to fix the center of inertia; in other
words we can set $p_{Z}=0$ without loss of generality. In this case we obtain

\begin{eqnarray}
p_{1} &=&\frac{1}{\sqrt{6}}p_{\lambda }+\frac{1}{\sqrt{2}}p_{\rho }
\label{p1rholam} \\
p_{2} &=&\frac{1}{\sqrt{6}}p_{\lambda }-\frac{1}{\sqrt{2}}p_{\rho }
\label{p2rholam} \\
p_{3} &=&-\sqrt{\frac{2}{3}}p_{\lambda }  \label{p3rholam}
\end{eqnarray}%
upon inversion of the preceding relations.

The Hamiltonian can then be regarded as a function $H=H\left( \rho ,\lambda
,p_{\rho },p_{\lambda }\right) $, determined by replacing the variables $%
\left( z_{1},z_{2},z_{3},p_{1},p_{2},p_{3}\right) $ with $\left( \rho
,\lambda ,p_{\rho },p_{\lambda }\right) $ from eqs. (\ref{zijrholam}) and (%
\ref{p1rholam}-\ref{p3rholam}). \ The resultant expression is rather
cumbersome, but can be written compactly (eq. (\ref{Htransrholam})) using a
judicious choice of notation, as shown in appendix A. \ 

A post-Newtonian expansion \cite{OR} of (\ref{Htrans}) (or really, eq. (\ref%
{Htransrholam})) in these variables in the equal-mass case yields%
\begin{eqnarray}
H &=&3mc^{2}+\frac{p_{\rho }^{2}+p_{\lambda }^{2}}{2m}+\frac{\kappa
m^{2}c^{4}}{\sqrt{8}}\left[ \left| \rho \right| +\frac{\sqrt{3}}{2}\left(
\left| \lambda +\frac{\rho }{\sqrt{3}}\right| +\left| \lambda -\frac{\rho }{%
\sqrt{3}}\right| \right) \right] -\frac{(p_{\rho }^{2}+p_{\lambda }^{2})^{2}%
}{16m^{3}c^{2}}+\frac{\kappa c^{2}}{\sqrt{8}}|\rho |p_{\rho }^{2}  \nonumber
\\
&&+\frac{\kappa c^{2}}{16\sqrt{2}}\left[ 3\left( \left| \lambda +\frac{\rho 
}{\sqrt{3}}\right| +\left| \lambda -\frac{\rho }{\sqrt{3}}\right| \right)
\left( \sqrt{3}p_{\lambda }^{2}+p_{\rho }^{2}\right) +6\left( \left| \lambda
+\frac{\rho }{\sqrt{3}}\right| -\left| \lambda -\frac{\rho }{\sqrt{3}}%
\right| \right) p_{\rho }p_{\lambda }\right]  \nonumber \\
&&+\frac{\kappa ^{2}m^{3}c^{6}}{16}\left[ \frac{\left| \rho \right| \sqrt{3}%
}{2}\left( \left| \lambda +\frac{\rho }{\sqrt{3}}\right| +\left| \lambda -%
\frac{\rho }{\sqrt{3}}\right| \right) +\frac{3}{4}\left| \lambda +\frac{\rho 
}{\sqrt{3}}\right| \left| \lambda -\frac{\rho }{\sqrt{3}}\right| -\frac{3}{4}%
\left( \lambda ^{2}+\rho ^{2}\right) \right]  \label{HpN}
\end{eqnarray}%
where factors of the speed of light $c$ have been restored (recall $\kappa =%
\frac{8\pi G}{c^{4}}$). \ The first three terms on the right-hand-side of
eq. (\ref{HpN}) are 
\begin{equation}
H=3mc^{2}+\frac{p_{\rho }^{2}+p_{\lambda }^{2}}{2m}+\frac{\kappa m^{2}c^{4}}{%
\sqrt{8}}\left[ \left| \rho \right| +\frac{\sqrt{3}}{2}\left( \left| \lambda
+\frac{\rho }{\sqrt{3}}\right| +\left| \lambda -\frac{\rho }{\sqrt{3}}%
\right| \right) \right]  \label{HN}
\end{equation}%
and are equivalent to the hexagonal-well Hamiltonian of a single particle
studied in{\it \ }\cite{Butka}, the first term being the total rest mass of
the system. \ This quantity is irrelevant non-relativistically, but we shall
retain it so that we can straightforwardly compare the motions and energies
of the relativistic and non-relativistic systems. The Hamiltonian (\ref{HN})
describes the motion of a single particle of mass $m$ (which we shall refer
to as the hex-particle) in a linearly-increasing potential well in the $%
\left( \rho ,\lambda \right) $ plane whose cross-sectional shape is that of
a regular hexagon.

If we regard the potential as being defined by the relation $V\left( \rho
,\lambda \right) =H\left( p_{\rho }=0,p_{\lambda }=0\right) $, we can make a
comparison of the non-relativistic, post-Newtonian, and exact relativistic
cases at any given value of the conserved Hamiltonian. The pN potential
takes the form \ 
\begin{eqnarray}
&&V_{\textrm{pN}}=3mc^{2}+\kappa m^{2}c^{4}\frac{R\sqrt{2}}{4}\left( \left|
\sin \theta \right| +\left| \sin \left( \theta +\frac{\pi }{3}\right)
\right| +\left| \sin \left( \theta -\frac{\pi }{3}\right) \right| \right)  
\nonumber \\
&&+\frac{\kappa ^{2}m^{3}R^{2}c^{6}}{16}\left( \left| \sin \theta \sin
\left( \theta +\frac{\pi }{3}\right) \right| -\sin \theta \sin \left( \theta
+\frac{\pi }{3}\right) +\left| \sin \theta \sin \left( \theta -\frac{\pi }{3}%
\right) \right| +\sin \theta \sin \left( \theta -\frac{\pi }{3}\right)
\right.   \nonumber \\
&&\left. +\left| \sin \left( \theta -\frac{\pi }{3}\right) \sin \left(
\theta +\frac{\pi }{3}\right) \right| -\sin \left( \theta -\frac{\pi }{3}%
\right) \sin \left( \theta +\frac{\pi }{3}\right) \right) 
\label{pNRthetpot}
\end{eqnarray}%
where we have made the hexagonal symmetry manifest by writing 
\begin{equation}
\rho =R\sin \theta \textrm{ \ \ \ \ \ \ \ }\lambda =R\cos \theta 
\label{polartrans}
\end{equation}%
As $c\rightarrow \infty $, $\kappa \rightarrow 0$ and the potential of the
hexagonal well in the N-system is recovered. \ The\ R version of this is
straightforwardly calculated from (\ref{Htrans})%
\begin{eqnarray}
&&\left( V_{\textrm{R}}-m_{1}c^{2}\right) \left( V_{\textrm{R}%
}-m_{2}c^{2}\right) \left( V_{\textrm{R}}-m_{3}c^{2}\right) =\left( V_{\textrm{R}%
}-s_{31}s_{32}m_{3}c^{2}\right) m_{1}m_{2}c^{4}\exp \left[ \frac{\sqrt{2}%
\kappa R}{4}V_{\textrm{R}}\left| \sin \theta \right| \right] 
\label{Rthetaplot} \\
&&+\left( V_{\textrm{R}}-s_{12}s_{13}m_{1}c^{2}\right) m_{2}m_{3}c^{4}\exp %
\left[ \frac{\sqrt{2}\kappa R}{4}V_{\textrm{R}}\left| \sin \left( \theta -%
\frac{\pi }{3}\right) \right| \right]   \nonumber \\
&&+\left( V_{\textrm{R}}-s_{21}s_{23}m_{2}c^{2}\right) m_{3}m_{1}c^{4}\exp %
\left[ \frac{\sqrt{2}\kappa R}{4}V_{\textrm{R}}\left| \sin \left( \theta +%
\frac{\pi }{3}\right) \right| \right]   \nonumber
\end{eqnarray}%
and also retains the hexagonal symmetry of the N system, as well as the
appropriate $c\rightarrow \infty $ limit.

\begin{figure}[tbp]
\begin{center}
\epsfig{file=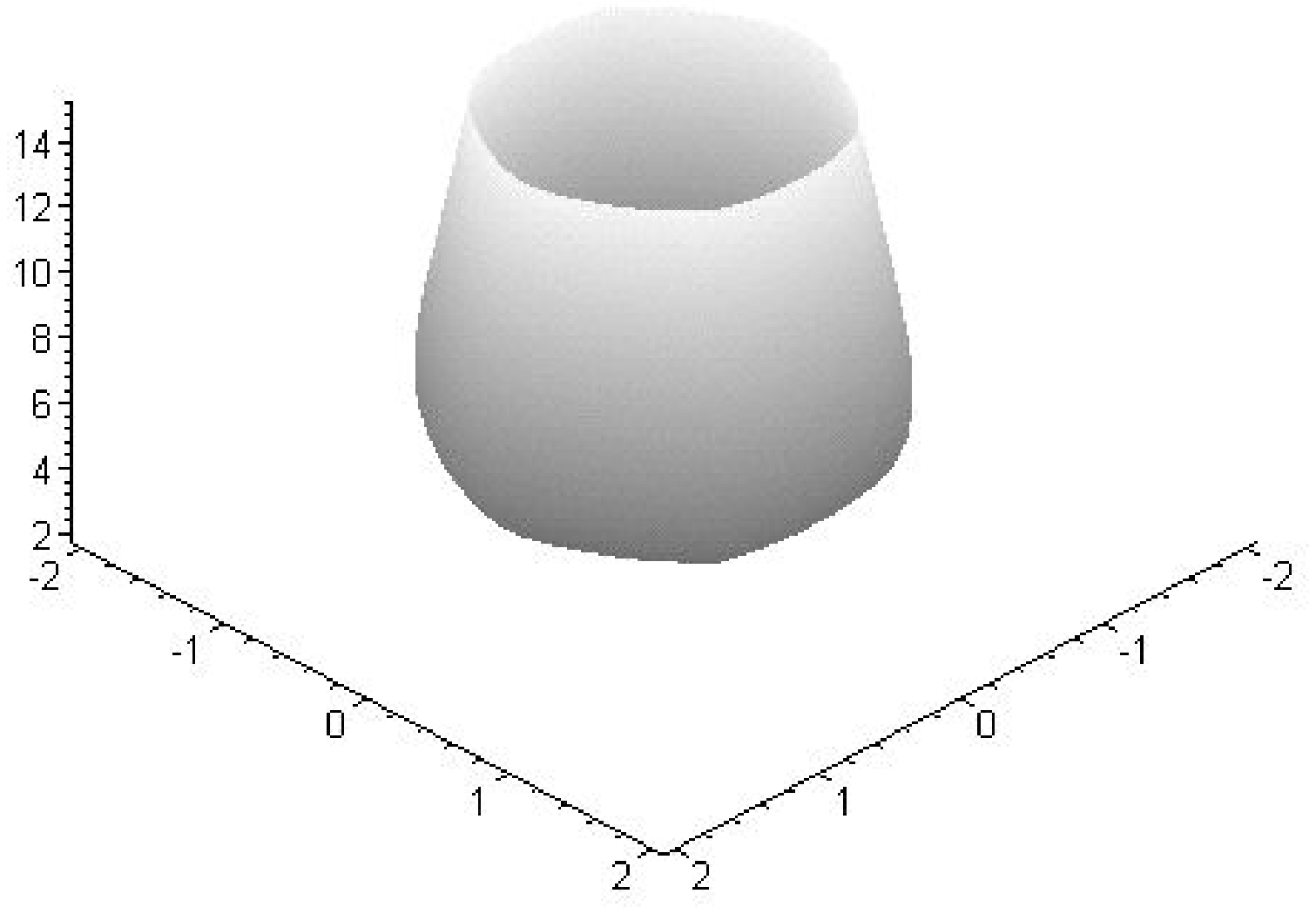,width=0.8\linewidth}
\end{center}
\caption{{}The shape of the relativistic potential in the equal mass case. }
\label{potentialgraphs_1}
\end{figure}
\begin{figure}[tbp]
\begin{center}
\epsfig{file=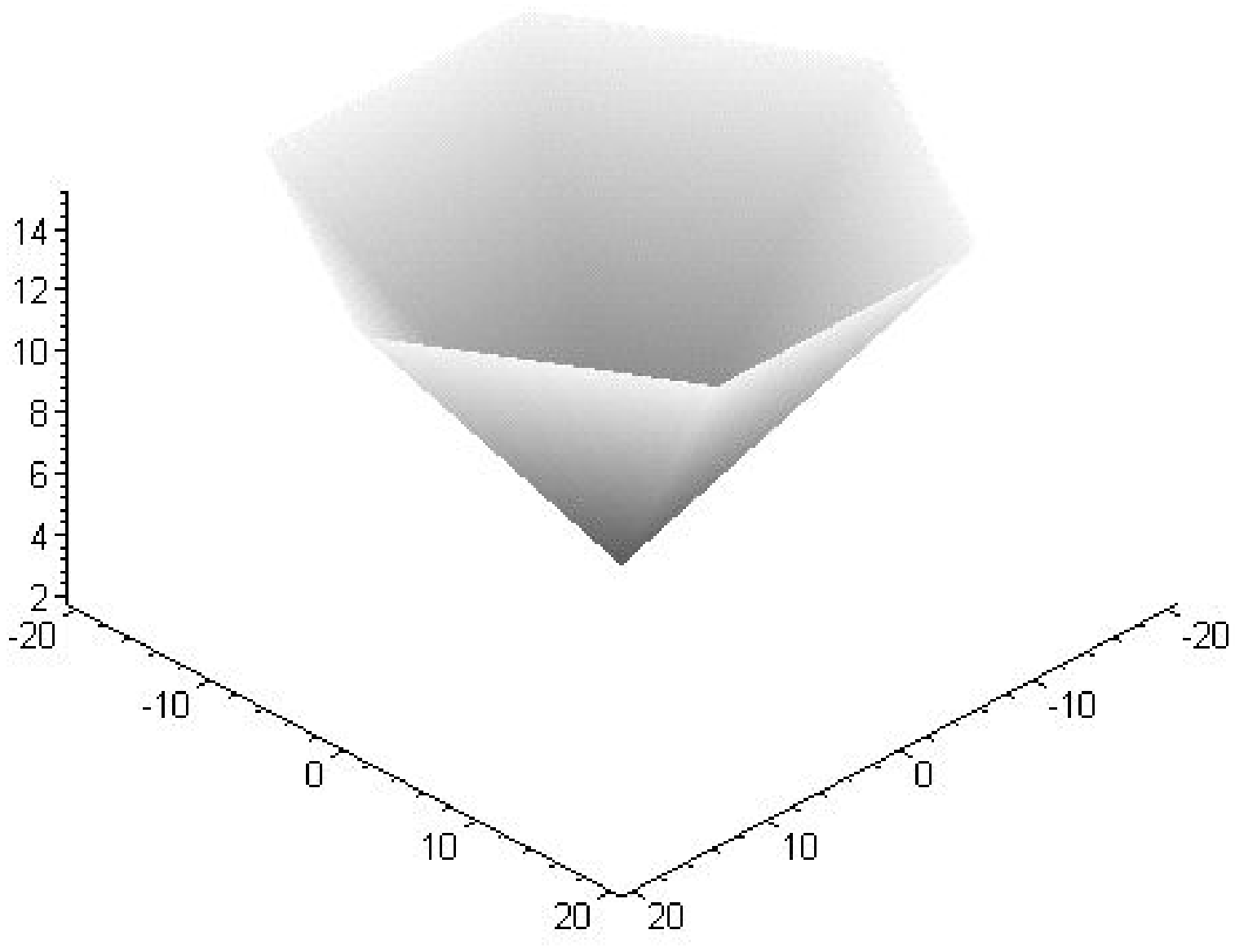,width=0.8\linewidth}
\end{center}
\caption{{}The shape of the non-relativistic potential in the equal mass
case. }
\label{potentialgraphs_2}
\end{figure}

At very low energies these cases are indistinguishable. However striking
differences between them develop quite rapidly with increasing energy, as
figures \ref{potentialgraphs_1} and \ref{potentialgraphs_2} illustrate for
the R and N systems. For all energies the Newtonian potential takes the form
of the hexagonal-well potential noted earlier: equipotential lines form the
shape of a regular hexagon in the $\left( \rho ,\lambda \right) $ plane,
with the sides rising linearly in all directions. The post-Newtonian
potential retains this basic hexagonal symmetry, but distorts the sides to
be parabolically concave. The growth of the potential is more rapid, with
the sides of the potential growing quadratically with $\left( \rho ,\lambda
\right) $. \ 

The exact potential differs substantively from both of these cases. \ It
retains the hexagonal symmetry, but the sides of the hexagon become convex,
even at energies only slightly larger than the rest mass. The growth of the
potential with increasing $V_{\textrm{R}}$ is extremely rapid compared to the
other two cases, and so the overall size of the hexagon at a given value of $%
V_{\textrm{R}}$ is considerably smaller. \ The size of the cross-sectional
hexagon reaches a maximum at $V_{\textrm{R}}=V_{\textrm{R}c}=6.711968022mc^{2}$,
after which it decreases in diameter like $\ln \left( V_{\textrm{R}}\right)
/V_{\textrm{R}}$ with increasing $V_{\textrm{R}}$. \ \ 

The part of the potential on the branch with $V_{\textrm{R}}>V_{\textrm{R}c}$ is
in an intrinsically non-perturbative relativistic regime: the motion for
values of $V_{\textrm{R}}$ larger than this cannot be understood as a
perturbation from some classical limit of the motion. The non-relativistic
hexagonal cone becomes a hexagonal carafe in the relativistic case, with a
neck that narrows as $V_{\textrm{R}}$ increases. 
\begin{figure}[tbp]
\begin{center}
\epsfig{file=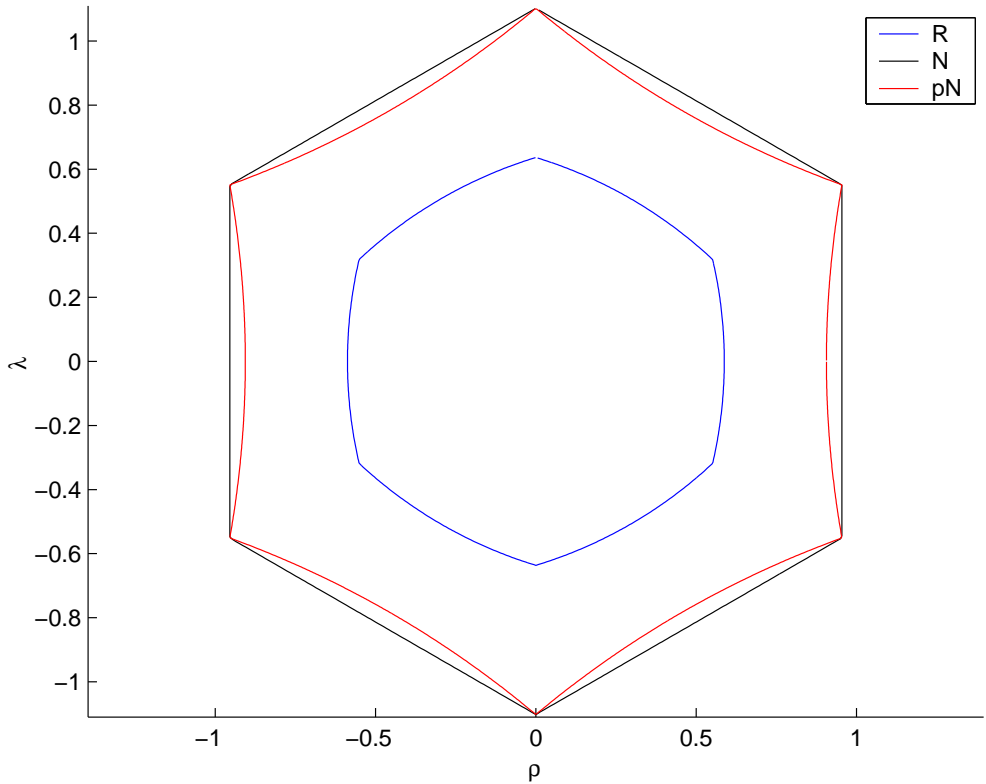,width=0.8\linewidth}
\end{center}
\caption{Equipotential lines at $V\simeq 4mc^{2}$ for each of the N, pN and
R systems in the equal mass case.}
\label{potentialgraphs_3}
\end{figure}

Of course in both the pN and R systems the potential does not fully govern
the motion since there are couplings between the momentum and position of
the hex-particle. \ In the post-Newtonian case we see that to leading order
in $c^{-2}$ there is a momentum-dependent steepening of the walls of the
hexagon. \ 

For unequal masses the hexagon becomes squashed, with two opposite corners
moving inward, changing both the slopes of the straight edges and their
relative lengths; relativistic corrections maintain this basic distortion,
but with the straight edges becoming parabolic. We shall not discuss the
unequal mass case any further.

\section{{\bf M}ethods for Solving the Equations of Motion}

We begin our analysis of the $3$-body system by analyzing (studying) the
motion of the hex-particle in the $\left( \rho ,\lambda \right) $\ plane. We
shall consider this motion in the non-relativistic (N), post-Newtonian (pN)
and exact relativistic (R) cases described in the previous section. \ In all
three cases the bisectors joining opposite vertices of the hexagon
correspond to particle crossings in the full 3-particle system, and denote a
discontinuous change in the hex-particle's acceleration. Thus the
hex-particle's motion, in the Hamiltonian formalism, is described by a pair
of differential equations that are continuous everywhere except across the
three hexagonal bisectors $\rho =0$, $\rho -\sqrt{3}\lambda =0$, and $\rho +%
\sqrt{3}\lambda =0$. These bisectors divide the hexagon into sextants and
correspond to the crossings of particles 1 and 2, 2 and 3, or 1 and 3
respectively.

An analogous system has been studied in the N system by Miller and Lehtihet,
who considered the motions of a ball under a constant gravitational force
elastically colliding with a wedge \cite{LMiller}. They established that
such motions can be analyzed using a discrete mapping that describes the
particle's angular and radial velocities each time it collides with the
edges of the wedge, which corresponds in our case to crossing one of the
three hexagonal bisectors. The two systems differ in that in the wedge
system the hex-particle collides elastically with the wedge (equivalent to
an elastic collision between a pair of particles in the $3$-body system),
whereas in our system the particle crosses the hexagon's bisectors,
equivalent to a pair of particles passing through each other in the $3$%
-particle system. In the equal mass case the systems are nearly identical,
since an elastic collision between two equal mass particles cannot be
distinguished from a crossing between two equal mass particles. We do,
however, observe a distinction between the two systems in a certain class of
orbits that we shall discuss later.

It is the non-smoothness of the potential along these bisectors that in all
three cases yields interesting dynamics for the system. In our subsequent
analysis we shall distinguish between two distinct types of motion \cite%
{LMiller}: $A$-motion, corresponding to the same pair of particles crossing
twice in a row (the hex-particle crossing a single bisector twice in
succession), and $B$-motion, in which one particle crosses each of its
compatriots in succession (the hex-particle crossing two successive sextant
boundaries). We can characterize a given motion by a sequence of letters $A$
and $B$ (called a symbol sequence), with a finite exponent $n$ denoting $n$%
-repeats and an overbar denoting an infinite repeated sequence. For example
the expression $A^{4}B^{3}$ denotes four $A$-motions (two adjacent particles
cross twice in a row four times in succession) followed by three $B$-motions
(one particle crosses the other two in succession, which then cross each
other). In the $\left( \rho ,\lambda \right) $ plane this will correspond to
a curve that \ crosses (for example) the $\rho =0$ axis $4$ times before
crossing one of the other sextant boundaries, after which it crosses two
more sextant boundaries in succession, ending up in a sextant $180^{o}$
opposite to the one in which it began. \ The expression $\left(
A^{n}B^{m}\right) ^{p}$ denotes $p$-repeats in succession of the motion $%
A^{n}B^{m}$, and $\overline{\left( A^{n}B^{m}\right) }$ denotes infinitely
many repeats of this motion. Note that the classification of a crossing
motion as $A$\ or $B$\ is contingent upon the previous crossing, and so
there is an ambiguity in classification of either the final or the initial
crossing. \ We shall resolve this ambiguity by taking the initial crossing
of any sequence of motions as being unlabeled -- as we are considered
arbitrarily large sequences of motions, this ambiguity in practice causes no
difficulties. \ 

\bigskip

We have carried out three methods of analysis to study the motions of this
system. First, we plot trajectories of the hex-particle in the $\rho $- $%
\lambda $ plane, comparing the motions of the N, pN, and R systems for a
variety of initial conditions. Second, we plot the motions of the three
particles as a function of time for each case. This provides an alternate
means of visualizing the difference between the various types of motion that
can arise in the system. Third, we construct Poincare sections by recording
the radial momentum ($p_{R}$, labelled as $x$) and the square of the angular
momentum ($p_{\theta }^{2}$, labelled as $z$) of the hex-particle each time
it crosses one of the bisectors. When all three particles have the same
mass, all bisectors are equivalent, so that all the crossings may be plotted
on the same surface of section. \ \ This allows us to find regions of \
periodicity, quasi-periodicity and chaos, and we shall discuss these
features in turn. \ 

One issue that arises upon comparison between the three systems is that the
same initial conditions do not yield the same conserved energy. \ There is
therefore some ambiguity in comparing trajectories between each of these
three cases: one can either compare at fixed values of the energy, modifying
the initial conditions as appropriate (as required by the conservation laws
for each system), or else fix the initial conditions, comparing trajectories
at differing values of $H$. \ We shall consider both methods of comparison.
In the former case we fix the initial values of $H,\rho ,\lambda $ and $%
p_{\rho }$, adjusting $p_{\lambda }$ so that the Hamiltonian constraint (\ref%
{Htrans}) is satisfied. We shall refer to these conditions as fixed-energy
(FE) conditions. In the latter situation we set the initial values of all
four phase-space coordinates $\left( \rho ,\lambda ,p_{\rho },p_{\lambda
}\right) $ in each system, allowing the energy to differ for each of the N,
pN, and R systems according to the respective constraints (\ref{HN}),\ (\ref%
{HpN}) and (\ref{Htrans}) for each. We shall refer to these conditions as
the fixed-momenta (FM) conditions.

\bigskip

There is no closed-form solution to either the determining equation (\ref%
{Htrans}) or to the equations of motion (\ref{zdot},\ref{pdot}), and so to
analyze the motion we must solve these equations numerically. We did this by
numerically integrating the equations of motion using a Matlab ODE routine
(ode45, or for the exact solutions, ode113). To generate Poincare sections,
we stopped the integration each time the hex-particle crossed one of the
bisectors by using an `events' function, saving the values of the radial and
angular momentum for plotting. Ideally, for each chaotic trajectory the
Poincare section should be allowed to run for a very long time in order to
determine as accurately as possible which regions of the plane it may visit,
and which are off-limits.

As a control over errors, we imposed absolute and relative error tolerances
of $10^{-8}$ for the numerical ODE solvers. For the values of $\eta $ we
studied, this yielded numerically stable solutions. We tested this by
checking that the energy remained a constant of the motion for all three
systems to within a value no larger than $10^{-6}$; for nearly all of our
runs it was comparable to the error tolerances ($10^{-8}$ ) that we imposed.
\ 

However we found that for $\eta \succapprox 1$ that the ODE solver was
unable to carry out the integration for more than a few time steps for the R
system before exceeding the allowed error tolerances (this problem remains
even if the error tolerances are lowered significantly). This value of $\eta 
$ is approximately the value $V_{\textrm{R}c}$ at which the equipotential
hexagon reaches its maximal size. We were unable to find an algorithm
capable of handling the numerical instabilities, which the ODE solvers we
employed could not deal with, for these larger values of $\eta $. \ A full
numerical solution in this larger $\eta $ regime remains an open problem.

We also found that the pN system had diverging trajectories for values of $%
\eta $ larger than $0.3$. We believe that this is due to an intrinsic
instability in the pN system, but we have not confirmed this.

In addition to plots, a few other routines were used to record information.
While running Poincare sections, the name of the edge crossed at each
collision (corresponding to the pair of particles that pass through each
other along that edge) can be recorded in addition to the velocity. From
this information, the symbol sequence of the trajectory can be extracted. In
addition, we recorded a frequency of return: this function measures the time
interval separating the hex-particle's successive returns to within some
small distance of its original location, and takes its inverse to find a
`frequency' at each time. The value of frequency depends strongly on how
small the specified area is; nonetheless, these `frequencies' give us an
approximate idea of how long the hex-particle takes to complete one full
cycle in its `orbit'.

In performing our numerical analysis we rescale the variables

\begin{equation}
p_{i}=M_{tot}c\hat{p_{i}}\textrm{ \ \ \ \ \ \ \ \ \ \ }z_{i}=\frac{4}{\kappa
M_{tot}c^{2}}\hat{z_{i}}  \label{rescale}
\end{equation}%
where $M_{tot}=3m$ is the total mass of the system and {$\hat{p}_{i}$ and $%
\hat{z}_{i}$ are the dimensionless momenta and positions respectively}.
Writing {$\eta +1=H/M_{tot}c^{2}$, $\hat{m_{i}}=(\frac{m_{i}}{M_{tot}})$, \
we have } 
\begin{eqnarray}
M_{i} &=&M_{tot}c^{2}\left( \sqrt{\hat{p_{i}}^{2}+\hat{m_{i}}^{2}}+\hat{p_{i}%
}\right) =M_{tot}c^{2}\hat{M}_{i}  \label{Miscale} \\
L_{i} &=&M_{tot}c^{2}\left( \eta +1-\sqrt{\hat{p_{i}}^{2}+\hat{m_{i}}^{2}}%
-\epsilon (\sum_{j}\hat{p}_{j}s_{ji})\right) \textrm{\ \ }=M_{tot}c^{2}\hat{L}%
_{i}\textrm{\ \ \ \ }  \label{Liscale}
\end{eqnarray}%
which in turn yields 
\begin{eqnarray}
\hat{L}_{1}\hat{L}_{2}\hat{L}_{3} &=&{\frak \hat{M}}_{12}{\frak \hat{M}}_{21}%
\hat{L}_{3}^{\ast }e^{s_{12}[(\hat{L}_{1}+{\frak \hat{M}}_{12})\hat{z}_{13}-(%
\hat{L}_{2}+{\frak \hat{M}}_{21})\hat{z}_{23}]}  \nonumber \\
&&+{\frak \hat{M}}_{23}{\frak \hat{M}}_{32}\hat{L}_{1}^{\ast }e^{s_{23}[(%
\hat{L}_{2}+{\frak \hat{M}}_{23})\hat{z}_{21}-(\hat{L}_{3}+{\frak \hat{M}}%
_{32})\hat{z}_{31}]}  \nonumber \\
&&+{\frak \hat{M}}_{31}{\frak \hat{M}}_{13}\hat{L}_{2}^{\ast }e^{s_{31}[(%
\hat{L}_{3}+{\frak \hat{M}}_{31})\hat{z}_{32}-(\hat{L}_{1}+{\frak \hat{M}}%
_{13})\hat{z}_{12}]}  \label{Htransscale}
\end{eqnarray}%
for the rescaled determining equation. Similarly the equations of motion
become 
\begin{eqnarray}
\frac{\partial \eta }{\partial \hat{p_{i}}} &=&\frac{1}{c}\frac{\partial H}{%
\partial p_{i}}=\frac{4}{\kappa M_{tot}c^{3}}\frac{d\hat{z}_{i}}{dt}=\frac{d%
\hat{z}_{i}}{d\hat{t}}  \label{pdotscale} \\
\frac{\partial \eta }{\partial \hat{z_{i}}} &=&(\frac{4}{\kappa
M_{tot}^{2}c^{4}})\frac{\partial H}{\partial z_{i}}=-\frac{4}{\kappa
M_{tot}c^{3}}\frac{d\hat{p_{i}}}{dt}=-\frac{d\hat{p_{i}}}{d\hat{t}}
\label{zdotscale}
\end{eqnarray}%
where $t=\frac{4}{\kappa M_{tot}c^{3}}\hat{t}$. \ A time step in the
numerical code has a value $\hat{t}=1$. \ All diagrams will be shown using
the rescaled coordinates (\ref{rescale}) unless otherwise stated.

We close this section with some final comments regarding the time variable $%
t $. This parameter is a coordinate time and it would be desirable to
describe the trajectories of the particles in terms of some invariant
parameter. The natural candidate is the proper time $\tau _{a}$\ of each
particle. From equations (\ref{feq8}), the proper time is 
\begin{eqnarray}
d\tau _{a}^{2} &=&dt^{2}\left\{ N_{0}(z_{a})^{2}-(N_{1}(z_{a})+\dot{z}%
_{a})^{2}\right\} \;,  \nonumber \\
&=&dt^{2}N_{0}(z_{a})^{2}\frac{m_{a}^{2}}{p_{a}^{2}+m_{a}^{2}}\qquad \qquad
(a=1,2,3)\;\;.  \label{dtprop}
\end{eqnarray}%
for the $a$th particle. \ Unfortunately this is in general different for
each particle, even in the equal mass case. This is quite unlike the
two-body situation, in which the symmetry of the system yields the same
proper time for each particle in the equal mass case (though not for the
unequal mass case) \cite{2bdcoslo}.

There are several different choices available at this stage. One could
choose to work with the proper time of a single particle in the system, in
which case invariance is recovered, but the manifest permutation symmetry
between particles is lost. Another possibility is to construct a
`fictitious' fourth particle that does not couple to the other three, but
moves along a geodesic of the system, and make use of its proper time. \
Rather than consider these or other possible options, we shall postpone
their consideration for future research and work with $t$, keeping in mind
that it is a coordinate time.

\bigskip

\section{Equal Mass Trajectories}

The three systems (N, pN, R) we consider here constitute a family of related
systems whose dynamics one would expect to be similar (but not identical) at
small energies, with increasingly different behaviour emerging as the energy
increases. However, our examination of the symbolic sequence of various
equal mass trajectories reveals a strong set of similarities, in that
certain types of sequences are common\ to all three systems over the energy
range we considered. We find that the types of motion exhibited by this
family of systems may thus be divided into three principal classes, which we
denote by the names annuli, pretzel and chaotic.

Before discussing each of these classes in detail, we make a few general
remarks. First, within each class a further distinction must be made between
those orbits which eventually densely cover the portion of $\left( \rho
,\lambda \right) $ space they occupy, and those which do not. (By
definition, all of the chaotic orbits densely cover their allowed portion of
phase space). The latter situation corresponds to regular orbits in which
the symbol sequence consists of a finite sequence repeated infinitely many
times. The former situation corresponds to orbits that are quasi-regular:
the symbol sequence consists of repeats of the same finite sequence, but
with an $A$-motion added or removed at irregular intervals.

These two types of orbits are separated in phase space by separatrixes
(trajectories joining a pair of hyperbolic fixed points). Regular orbits lie
inside the `elliptical' region surrounding an elliptical fixed point; the
quasi-regular orbits lie outside such a region.

It is useful to further distinguish between the quasi-periodic and periodic
regular orbits. Quasi-periodic trajectories closely resemble the related
periodic trajectories, except that the orbit fails to exactly repeat itself
and hence eventually densely covers some region of phase space. Thus a
quasi-periodic trajectory displays a high degree of regularity. In the
system we study, this regularity is manifest by its periodic symbol
sequence. The classic example of this is a particle moving on a torus $%
S^{1}\times S^{1}$. The motion is characterized by its angular velocity
around each copy of $S^{1}$: if the ratio of these is rational, the motion
will be periodic; if it is irrational, the motion will be quasi-periodic.
For the system considered here, non-periodic orbits with fixed symbol
sequences are quasi-periodic. They appear as a collection of closed circles,
ovals, or crescents in the Poincare section. Orbits with symbol sequences
that are not fixed, however, are distinctly not regular; we shall refer to
them as quasi-regular as noted above.

We also note that, although the orbits of the non-relativistic and
relativistic systems realize the same symbol sequences, important
qualitative differences exist between these orbits for both the trajectories
and the Poincare sections. Consider a comparison between each system at
identical values of the total energy $H=E_{T}$ and the initial values of $%
\left( \rho ,\lambda ,p_{\rho }\right) $, with the remaining initial value
of $p_{\lambda }$ chosen to satisfy (\ref{Htrans},\ref{HpN},\ref{HN}) in
each respective case. \ As $\eta =H/\left( 3mc^{2}\right) -1$\ increases,
distinctions between the exact relativistic and non-relativistic cases
become substantive, both qualitatively and quantitatively. The relativistic
trajectories have higher frequencies and extend over a smaller region of the 
$\left( \rho ,\lambda \right) $ plane than their non-relativistic
counterparts. \ Their trajectory patterns in the pretzel class also develop
a slight ``hourglass'' shape (narrowing with increasing $\eta $\ in the
small-$\lambda $ region) in comparison to the cylindrical shapes of their
non-relativistic counterparts.\ 

At FM initial conditions, we find in general that the R system has greater
energy than its corresponding N and pN counterparts. Hence in these cases we
find that the R orbits cover a correspondingly larger region of the $\left(
\rho ,\lambda \right) $ plane and have a higher frequency.

\subsection{Annulus Orbits}

The annuli are orbits in which the hex-particle never re-crosses the same
bisector twice. All such orbits have the symbol sequence $\overline{B}$, and
describe an annulus encircling the origin in the $\rho -\lambda $ plane.

\begin{figure}[tbp]
\begin{center}
\epsfig{file=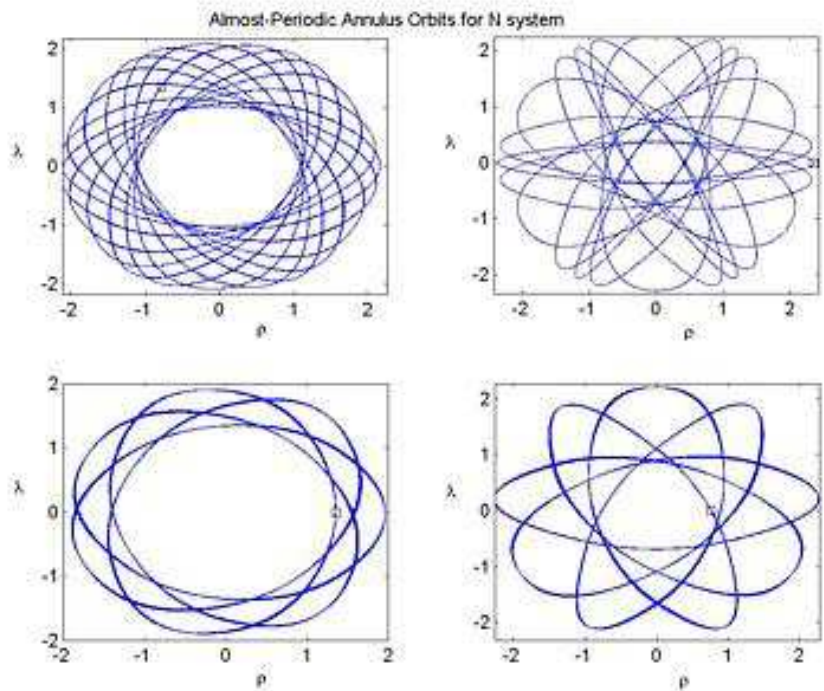,width=0.8\linewidth}
\end{center}
\caption{{}Examples of almost-periodic annuli for the N system: trajectories
do not cover the entire annulus band (run for 200 time units). \ A wide
variety of complex patterns can be found. The square indicates the intial
values of $\left( \protect\rho ,\protect\lambda \right) $; FM initial
conditions were employed.}
\label{NRanuper}
\end{figure}

\begin{figure}[tbp]
\begin{center}
\epsfig{file=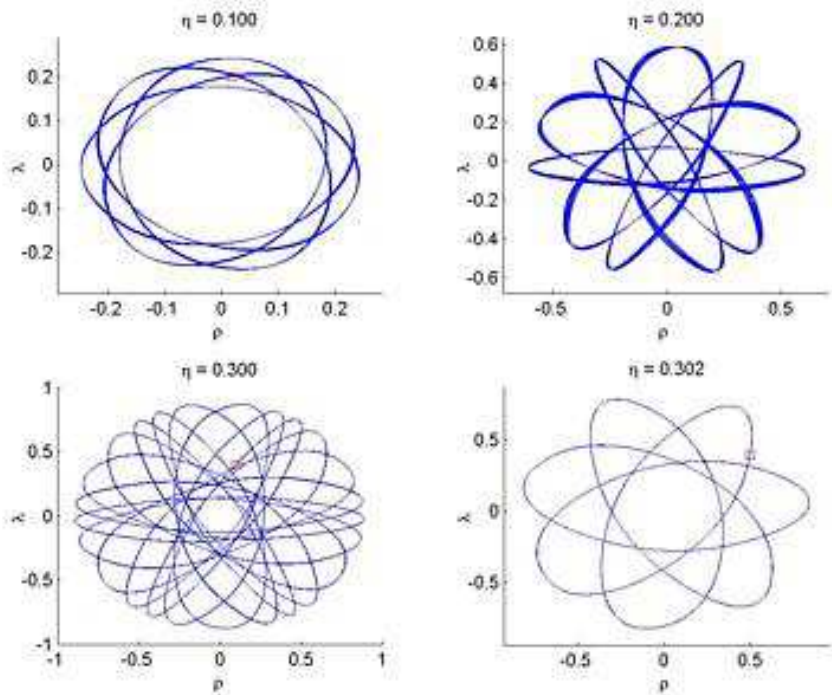,width=0.8\linewidth}
\end{center}
\caption{{}Examples of almost-periodic annuli for the pN case (trajectories
run for 200 time units). \ Up to the values of $\protect\eta $ attainable
numerically, no differences were observed between PN and N annulus-type
orbits. The square indicates the intial values of $\left( \protect\rho ,%
\protect\lambda \right) $; FM initial conditions were employed.}
\label{PNanuper}
\end{figure}

As noted above, we see that some annulus orbits are quasi-periodic and fill
in the entire ring (generating one of the closed triangle-like shapes in the
middle of the Poincare section) while a choice few apparently repeat
themselves after some number of rotations about the origin. This latter
situation is illustrated in figs. \ref{NRanuper}, \ref{PNanuper}\ and \ref%
{Rannuli} for the N, pN and R cases respectively. \ In each of these cases a
wide variety of patterns emerge contingent upon the initial conditions but
independent of the system in question. Both these orbits and those that fill
in the ring (not illustrated) are `close to' an elliptic fixed point; the
difference between them is that in some cases the normally quasi-periodic
orbits have commensurate winding numbers, producing an eventually periodic
orbit. As periodic orbits are difficult to find numerically, the orbits in
the figures are actually orbits that are very close to periodic orbits, so
that the pattern of the periodic orbit is still visible. In fact they are
quasi-periodic orbits about these higher-period fixed points, which means
that they will not cover the entire annulus, only bands of phase space.

\begin{figure}[tbp]
\begin{center}
\epsfig{file=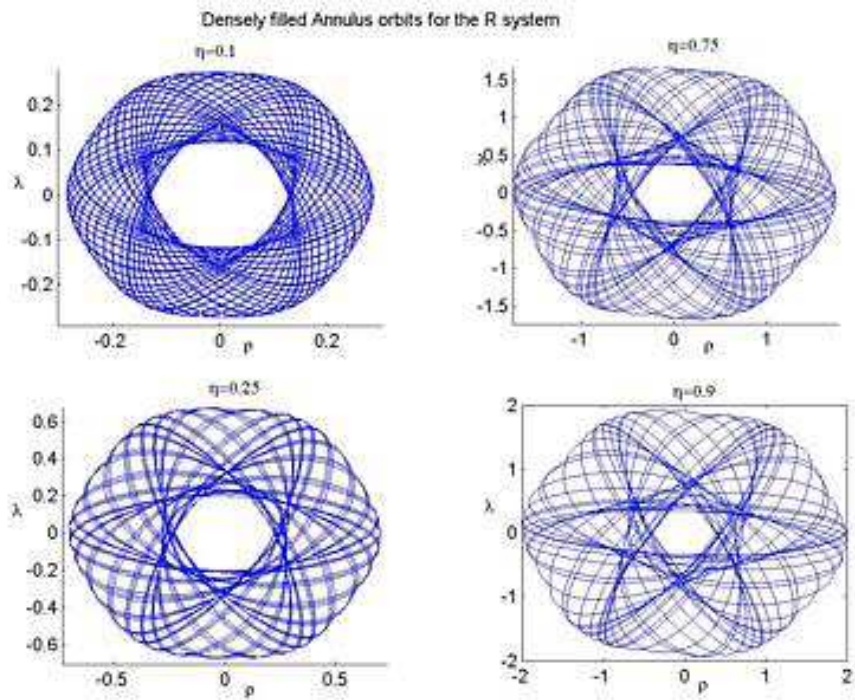,width=0.8\linewidth}
\end{center}
\caption{A sampling of densely-filled annuli for the R solution. All orbits
were run for 200 time steps using FE initial conditions. Densely filled
annuli can be found for all values of $\protect\eta $. \ As $\protect\eta $\
increases, the R annuli rotate with increasing angle relative to their N
counterparts; the orientation (clockwise or anticlockwise) depends upon the
initial conditions. }
\label{Rannuli}
\end{figure}

We find no qualitative distinctions between the N and pN annuli up to the
values of $\eta $ that we can attain numerically. However we do find that
the R cases appear to undergo a slight rotation relative to their N
counterparts as $\eta $ increases. This is noticeable in the right-hand
diagrams of fig.\ref{Rannuli}, for $\eta =0.75$ and $\eta =0.9$.


\begin{figure}[tbp]
\begin{center}
\epsfig{file=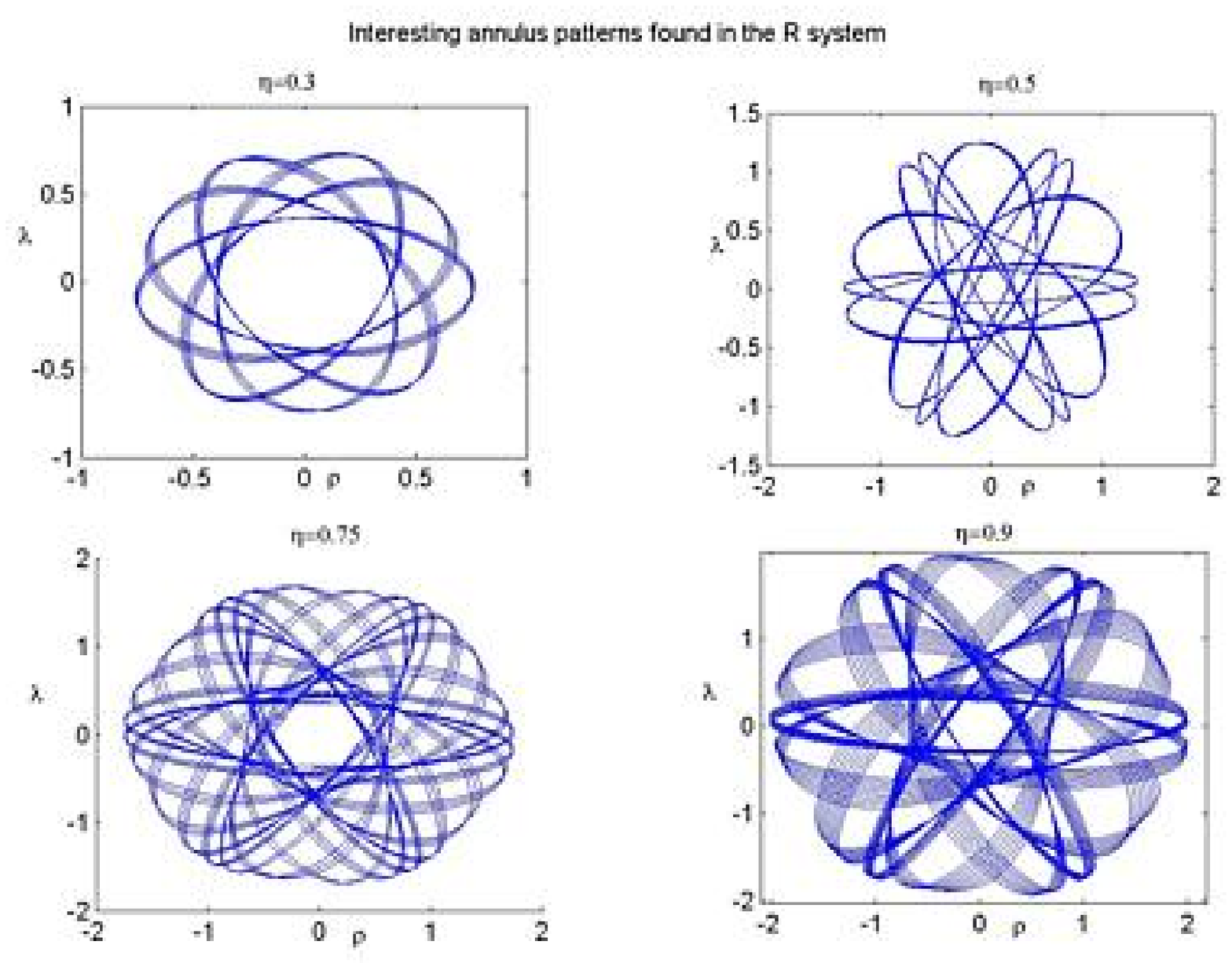,width=0.8\linewidth}
\end{center}
\caption{A sampling of quasi-periodic annuli for the R system, run for 200
time steps and FE initial conditions. \ As in the N and pN systems, various
periodic, quasi-periodic, and densely filled annuli can be found for all $%
\protect\eta $. \ }
\label{Rannu2}
\end{figure}

\begin{figure}[tbp]
\begin{center}
\epsfig{file=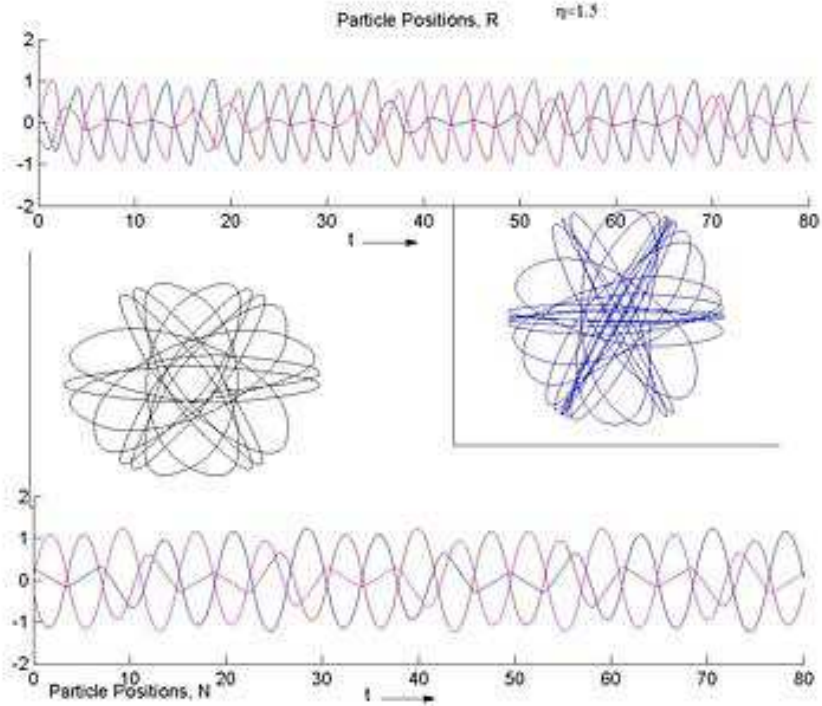,width=0.8\linewidth}
\end{center}
\caption{{}Near-chaotic annulus orbits (N-black, R-blue) shown in
conjunction with their corresponding 3-particle trajectories. These
near-chaotic orbits have been run for 200 time steps using FE initial
conditions. We have truncated the 3-particle trajectory plot after 80 time
steps. The R trajectory is closer to chaos than the N trajectory.}
\label{1anut}
\end{figure}

\begin{figure}[tbp]
\begin{center}
\epsfig{file=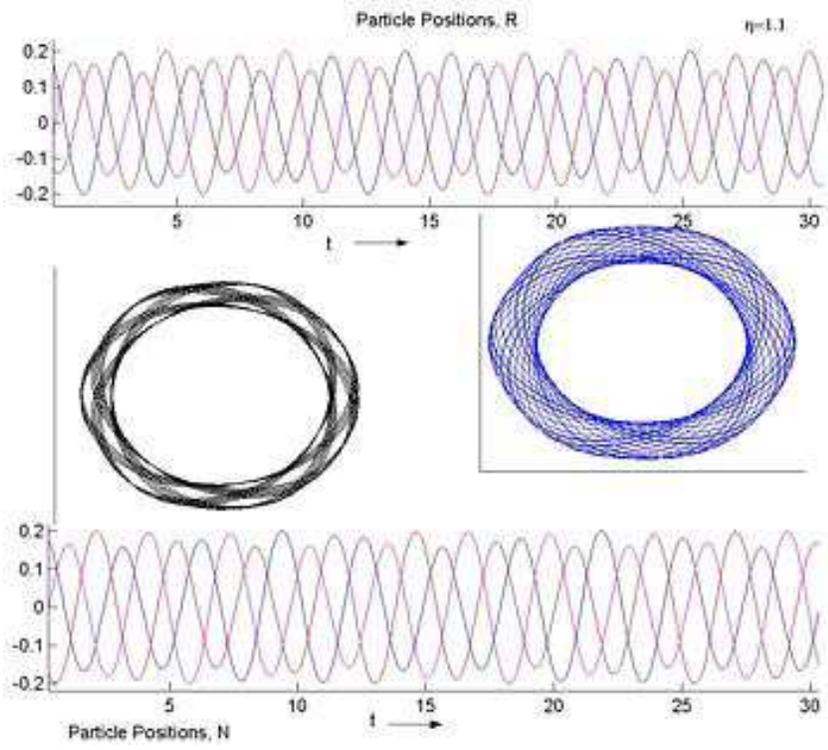,width=0.8\linewidth}
\end{center}
\caption{{}Annulus orbits (N-black, R-blue) shown in conjunction with their
corresponding 3-particle trajectories. These quasi-regular orbits have been
run for 200 time steps using FE initial conditions and are far from being
chaotic. We have truncated the 3-particle trajectory plot after 30 time
steps. \ The R motion is further from periodicity, leaving far fewer open
regions in the $\left( \protect\rho ,\protect\lambda \right) $ plane.}
\label{2anut}
\end{figure}

In figures \ref{1anut},\ref{2anut} we plot the positions of each of the
three bodies as a function of time in conjunction with their corresponding
trajectories in the $\left( \rho ,\lambda \right) $ plane for FE conditions
in both the N and R cases. \ \ \ We see that at similar energies a $3$-body
system experiencing relativistic gravitation covers the $\left( \rho
,\lambda \right) $ plane in the hex-particle representation more densely
than its non-relativistic counterpart, and induces a higher frequency of
oscillation. This higher frequency is also characteristic of the $2$-body
system \cite{2bd,2bdchglo}, and\ we have observed it to be a general
phenomenon for all FE conditions we have studied. \ The increased trajectory
density for FE conditions is also a phenomenon we observe to generally occur
for the $3$-body R system, presumably because both systems were run for the
same number of time steps, and the relativistic one has a higher frequency.


\begin{figure}[tbp]
\begin{center}
\epsfig{file=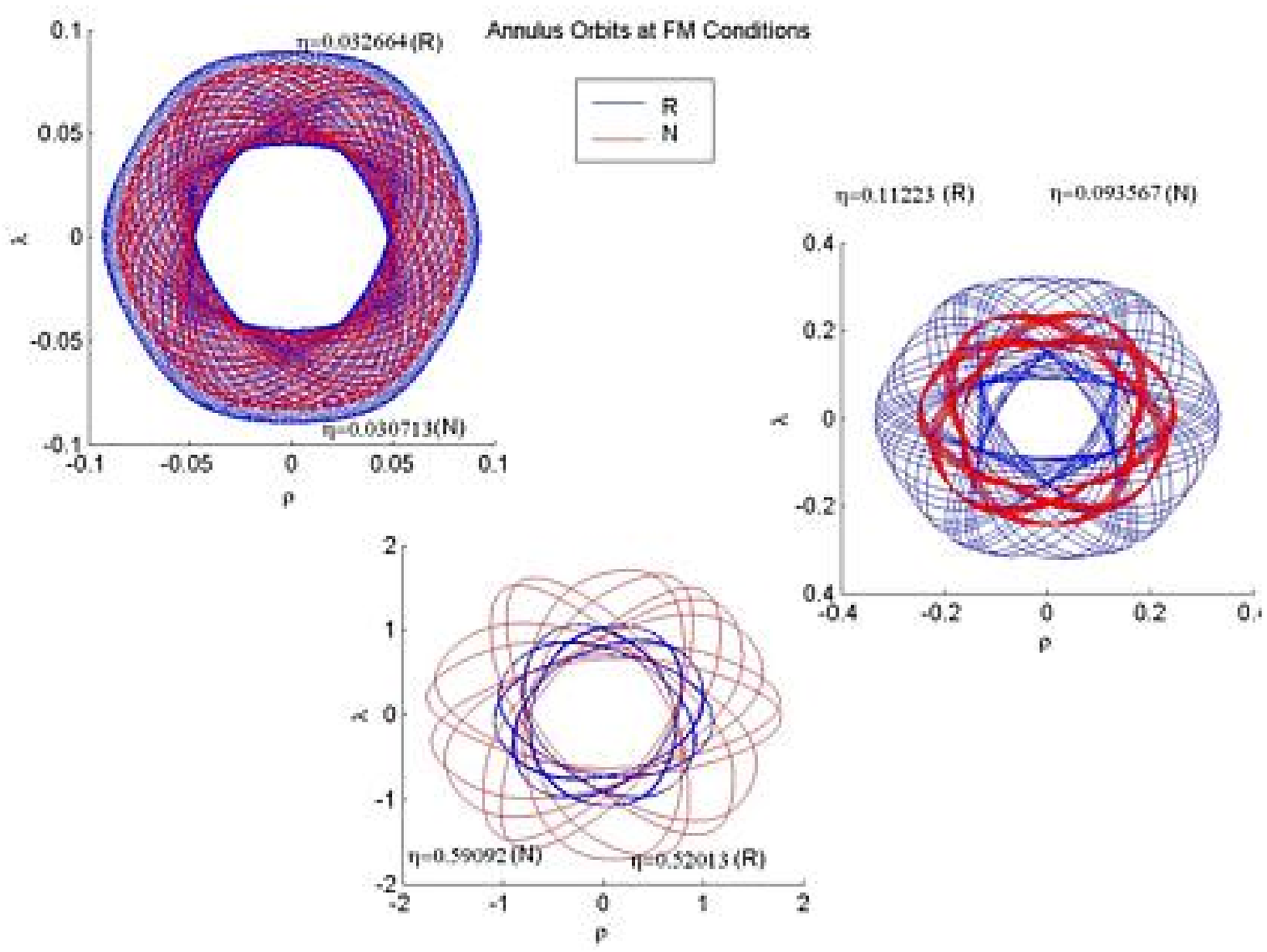,width=0.8\linewidth}
\end{center}
\caption{{}A comparsion of annulus orbits at identical FM conditions, for
three similar values of $\protect\eta $. \ With identical initial
conditions, the relativistic trajectories typically have greater energy and
so cover a larger region of the $\left( \protect\rho ,\protect\lambda %
\right) $ plane. However for some initial conditions the N system has a
larger energy and so covers a correspondingly larger region. All orbits were
run for 200 time steps.}
\label{annucompmany}
\end{figure}

A comparison of orbits using FM initial conditions is also instructive; fig. %
\ref{annucompmany} provides an example. At FM conditions the R system
typically has slightly higher energy, and so covers a considerably larger
region of the $\left( \rho ,\lambda \right) $ plane more densely than its
non-relativistic counterpart, venturing slightly closer to the origin. This
effect increases with increasing $\eta $, provided that the R energy remains
larger than its N counterpart. \ However as $\eta $\ gets larger it becomes
increasingly more difficult to find initial conditions such that both the N
and R annuli are close to periodic orbits. \ The bottom diagram in fig. \ref%
{annucompmany} is an example at $\eta \simeq 0.5$; the N system has about
14\% more energy than its R counterpart, and so it now covers a larger
region. 

\begin{figure}[tbp]
\begin{center}
\epsfig{file=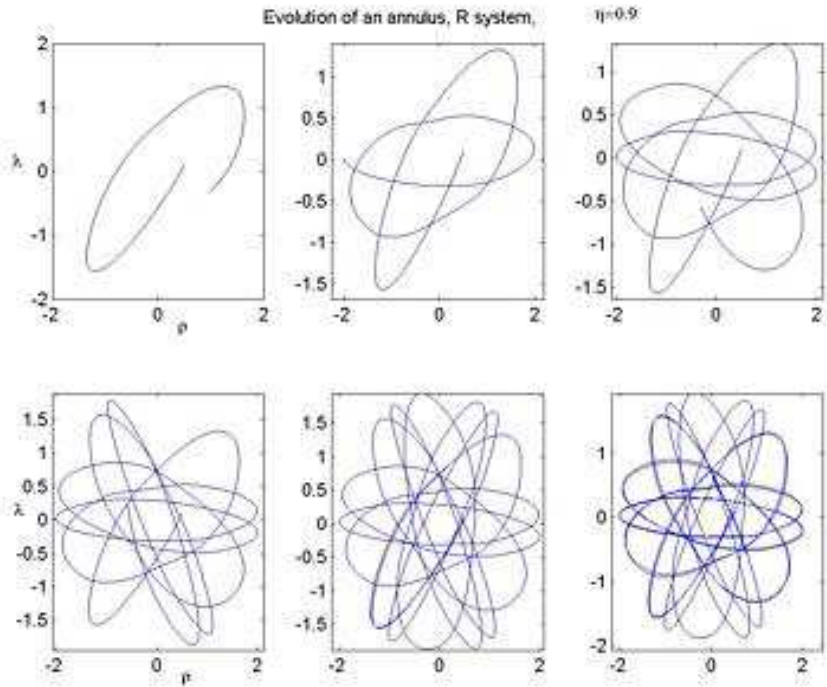,width=0.8\linewidth}
\end{center}
\caption{{}Early time development of a densely filled annulus in the R
system with $\protect\eta =0.9$, shown at $t=5,12,20,30,45$ and $60$ time
steps. The evolution of annuli is qualitatively the same in the N, pN and R
systems at all values of $\protect\eta $ investigated, although the
orientation and scale of the anulus itself changes in the R case. \ }
\label{TimeAnuR}
\end{figure}

Figure \ref{TimeAnuR} illustrates the temporal development of an annulus
pattern for $\eta =0.9$ in the R system. At all values of $\eta $
investigated, annuli evolution is qualitatively the same in the N, pN and R
systems, although the shape of the annulus itself changes.

\subsection{Pretzel Orbits}

Pretzel orbits are those in which the hex-particle essentially oscillates
back and forth about one of the three bisectors, corresponding to a stable
or quasi-stable bound subsystem of two particles. Symbolically such orbits
can be written as $\prod_{ijk}\left( A^{n_{i}}B^{3m_{j}}\right) ^{l_{k}}$\ ,
where $n_{i},m_{j},l_{k}\in Z^{+}$, with some $l_{k}$\ possibly infinite.
The resulting collection of trajectories is extremely diverse. Many families
of regular orbits exist. Such families contain one base element (for example 
$AB^{3}$) and a sequence of elements formed by appending an $A$ to each
existing sequence of $A$'s (for example, $\{AB^{3},A^{2}B^{3},A^{3}B^{3},...%
\}$. The result is that the phase space has an extremely complex structure
that we shall discuss further in section 7. It is differences in this
structure and the shapes of the corresponding orbits that show the most
remarkable distinctions between the R, PN and N systems.

In the above sequences, the $B^{3}$\ sequence corresponds to a 180-degree
swing of the hex-particle around the origin, and the resultant figures in
the $\left( \rho ,\lambda \right) $ plane comprise a broad variety of
twisted, pretzel-like figures, from whence their name. This situation is a
key distinction between the systems we study and the wedge system \cite%
{LMiller}\ discussed earlier. In the wedge system $B$ and $B^{2}$ sequences
are observed in addition to $B^{3}$\ sequences; we observe only the latter
in all pretzel orbits.

Before proceeding to a detailed description of this class, we summarize the
main results of our investigation. Again we have both regular orbits (with
the symbol sequence above repeating ad infinitum) and non-regular orbits
that densely fill a cylindrical tube in the $\left( \rho ,\lambda \right) $\
plane. The periodic and quasi-periodic orbits we find in the N system appear
for the most part to have counterparts with the same symbol sequence in the
R system (though not in the pN system). In general orbits in the\ R system
have kinks about the $\lambda =0$\ line relative to their N and pN
counterparts; for example a cylindrical-shaped trajectory in the N system
looks like an hourglass in the R system. \ The pN system exhibits chaotic
behaviour not seen in the N and R systems, a point we shall discuss in a
subsequent section.

\begin{figure}[tbp]
\begin{center}
\epsfig{file=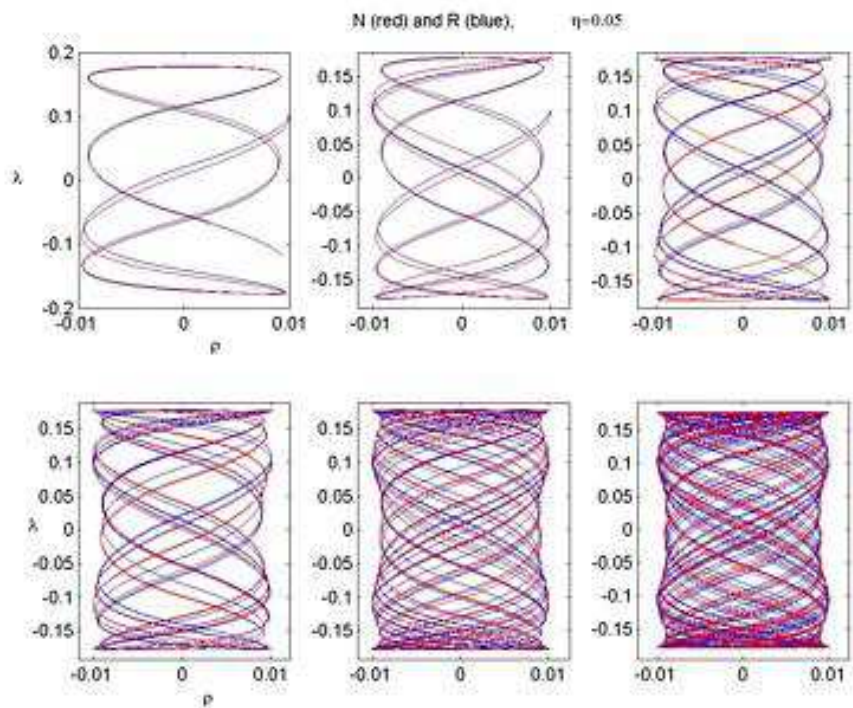,width=0.8\linewidth}
\end{center}
\caption{A time series evolution for a pretzel orbit shown simultaneously in
the N and R systems at t=3,6,11,16,25,and 35 units for $\protect\eta =0.05$
at FE conditions. \ The trajectories in the two systems are very similar at
such low energies. \ }
\label{Timeserloweta}
\end{figure}

\begin{figure}[tbp]
\begin{center}
\epsfig{file=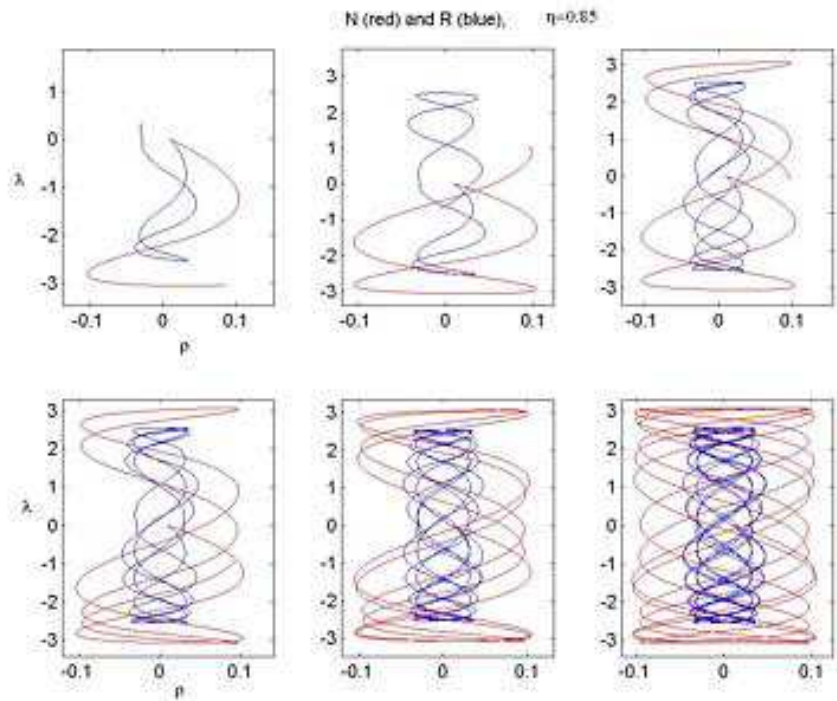,width=0.8\linewidth}
\end{center}
\caption{N and R pretzel orbits evolving simultaneously at t=3,6,11,16, 26,
and 36 units for $\protect\eta =0.85$ with FE conditions. The N trajectory
extends considerably further from the origin, while the R orbit evolves with
a higher collision frequency. The R orbit has stabilized into a
quasi-periodic pattern, whereas the N orbit will eventually form a densely
filled cylinder.}
\label{Timeserhigheta}
\end{figure}

Figures \ref{Timeserloweta} and \ref{Timeserhigheta} illustrate the
development of a trajectory in the $\left( \rho ,\lambda \right) $ plane for
FE conditions at small and large values of $\eta $. As expected we see that
for small $\eta $ ($\eta =0.05$) there is very little distinction between
the N and R motions, consistent with the smooth non-relativistic limit of (%
\ref{Htrans}). \ However at larger $\eta $ ($\eta =0.85$) the hex-particle
traces out significantly different trajectories in the N and R systems. The
oscillation frequency is higher and the trajectory is more tightly confined,
features commensurate with $2$-body motion in the R system \cite%
{2bd,2bdchglo,2bdcossh}. \ We also see a considerably different weave
pattern in fig \ref{Timeserhigheta} for each case, with the N pattern
exhibiting a near-cylindrical shape in contrast to its R counterpart with
oscillating sides.

\begin{figure}[tbp]
\begin{center}
\epsfig{file=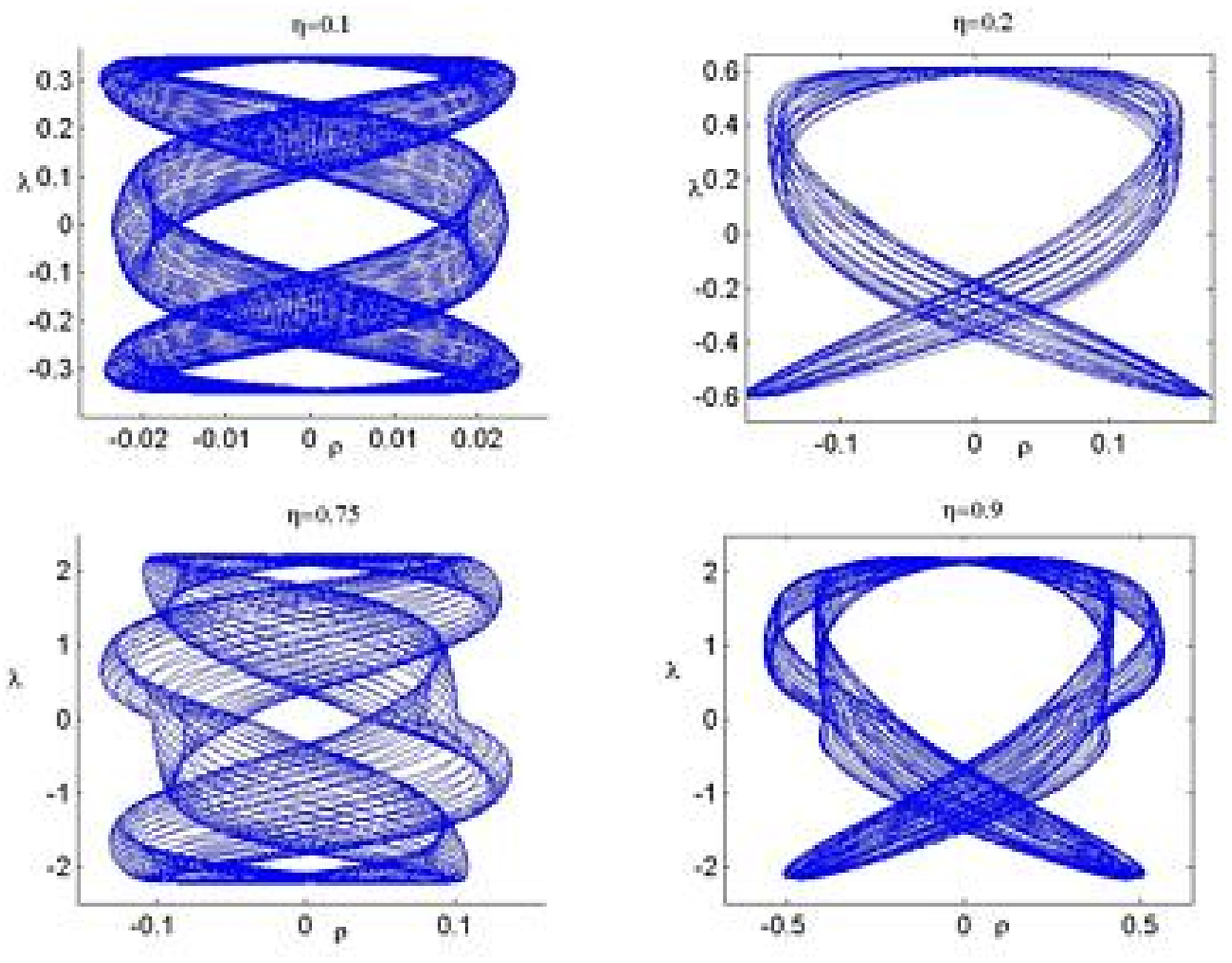,width=0.8\linewidth}
\end{center}
\caption{A comparison of similar quasi-regular pretzel-type orbits for the R
case at different values of $\protect\eta $. \ The symbol sequence is $%
\overline{\left( A^{2}B^{3}\right) }$ for the trajectories on the left, and $%
\overline{\left( AB^{6}\right) }$ for the trajectories on the right.\ As $%
\protect\eta $ increases the asymmetry due to relativity becomes
increasingly apparent. \ (Note that as $\protect\eta $ changes, the initial
conditions required to find a given orbit also alter.)}
\label{pretRcomp.fig}
\end{figure}

\bigskip

\begin{figure}[tbp]
\begin{center}
\epsfig{file=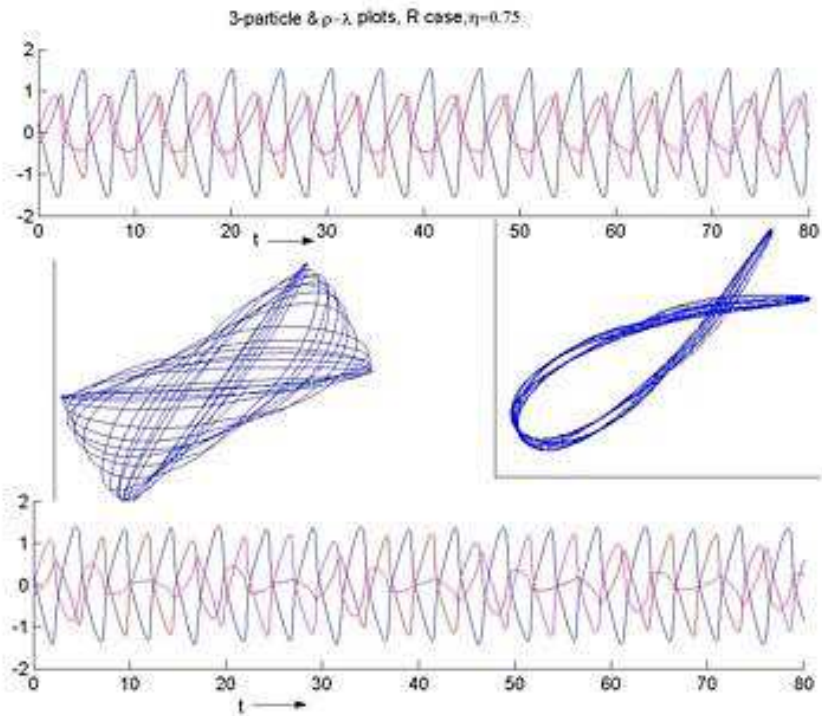,width=0.8\linewidth}
\end{center}
\caption{Pretzel orbits of the relativistic system with the corresponding $3$%
-particle time evolution: a regular $AB^{6}$ orbit pattern (top), and
similar irregular orbit pattern, at slightly different initial conditions
(bottom). Both cases employ FE conditions, and were run for 200 time steps,
with the $3$-particle trajectory plot truncated after 80 time steps. }
\label{3Seq1}
\end{figure}

In fig. \ref{3Seq1} we compare the positions of each of the three bodies as
a function of time in conjunction with their corresponding trajectories in
the $\left( \rho ,\lambda \right) $ plane for two slightly different FE
conditions in the R system. The fish-like diagram corresponds to an $AB^{6}$
symbol sequence: we see that two of the particles oscillate quasi-regularly
about each other, this pair undergoing larger-amplitude and lower-frequency
oscillations with the third. \ A slight change of initial conditions yields
the strudel-like figure; here we see that one particle alternates its
oscillations with the other two, maintaining a near-constant amplitude
throughout.

\begin{figure}[tbp]
\begin{center}
\epsfig{file=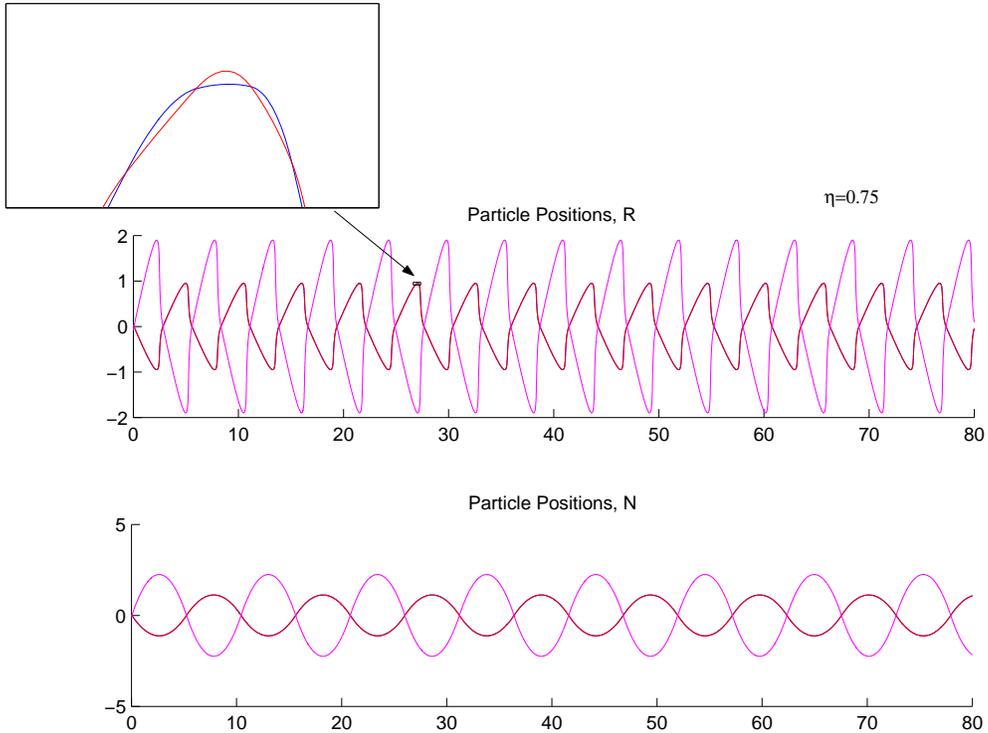,width=0.8\linewidth}
\end{center}
\caption{A pretzel-orbit with a large number of $A$ collisions before the $%
B^{3}$ collision sequence, run with FE intial conditions. \ Particles 1 and
2 (red and blue lines respectively) remain very close together, colliding
frequently. \ They act much like a single body from the viewpoint of
particle 3; the above trajectory bears a strong resemblance to the
trajectories found in the 2-body case. The inset shows detail near the one
of the peaks in the R system.}
\label{3_2bd}
\end{figure}

We compare in figure \ref{3_2bd} pretzel orbits with the same FE conditions
in the R and N cases plotted as trajectories in the $3$-body system. These
orbits are distinguished by having a very high-frequency low-amplitude
oscillation between two of the particles; this pair in turn undergoes a
low-frequency high-amplitude oscillation with the third. \ The red and blue
lines are nearly indistinguishable due to their close proximity; the inset
in the figure provides a close-up of the oscillations in this two-body
subsystem near one of its extrema. The N oscillations are parabolic in shape
whereas the R oscillations have the shoulder-like distortion seen previously
in the $2$-body system \cite{2bd}. These diagrams illustrate that under
appropriate initial conditions two bodies can tightly and stably bind
together in both the N and R systems (even at substantively large $\eta $),
behaving like a single body relative to the third.

\begin{figure}[tbp]
\begin{center}
\epsfig{file=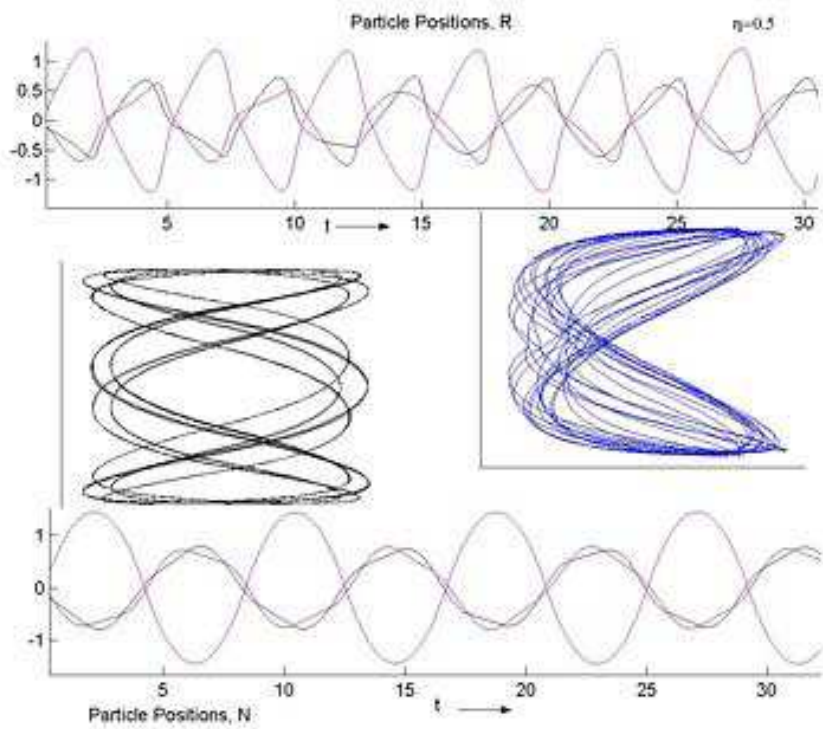,width=0.8\linewidth}
\end{center}
\caption{Regular pretzel orbits (run for 120 time steps) of the R and N
systems with the corresponding $3$-particle time evolution (truncated at 35
time steps). These orbits correspond to the collision sequences $AB^{3}$ (R)
and $A^{2}B^{3}$ (N). \ Since we employ FE initial conditions the collision
sequences differ.}
\label{3Seq2}
\end{figure}

A similar situation is shown in fig. \ref{3Seq2}, where FE initial
conditions were employed. The oscillations for the $2$-body subsystem are
now of lower frequency and larger amplitude than in fig. \ref{3Seq1}, and
the symbol sequences differ between the N and R cases. The respective
parabolic regularity and shoulder-like distortions are evident, and the $2$%
-body subsystem oscillations in the R case are\ of slightly larger amplitude
and higher frequency than in its N counterpart.

\begin{figure}[tbp]
\begin{center}
\epsfig{file=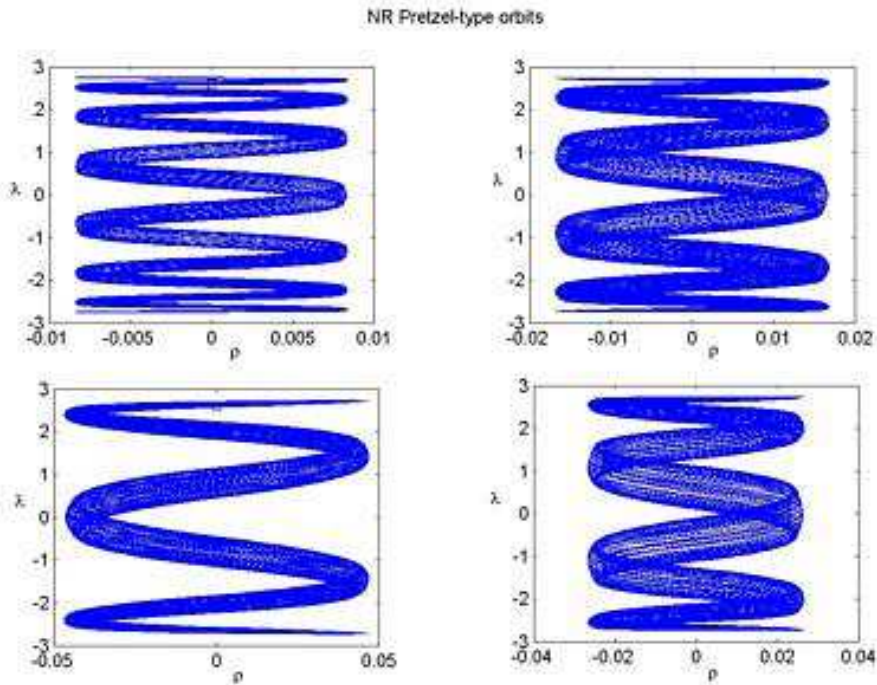,width=0.8\linewidth}
\end{center}
\caption{A family of quasi-regular snake-like orbits for the N system, each
with two sharp turning points separated by some number $n$ of bumps. All
were run for 200 time steps with FM initial conditions; the square indicates
the starting point in the $\left( \protect\rho ,\protect\lambda \right) $
plane. These orbits have the symbol sequence $A^{m}B^{3}$ for $m$ odd, and
correspond to sequences of even numbers ($2(n+2)$) of circles in the lower
portion of the Poincare section. \ The value of $n$ increases with
decreasing {\it initial} angular momentum. It appears that such orbits exist
for arbitrary $n$; we have found them up to $n=15$.}
\label{Nbump}
\end{figure}

\begin{figure}[tbp]
\begin{center}
\epsfig{file=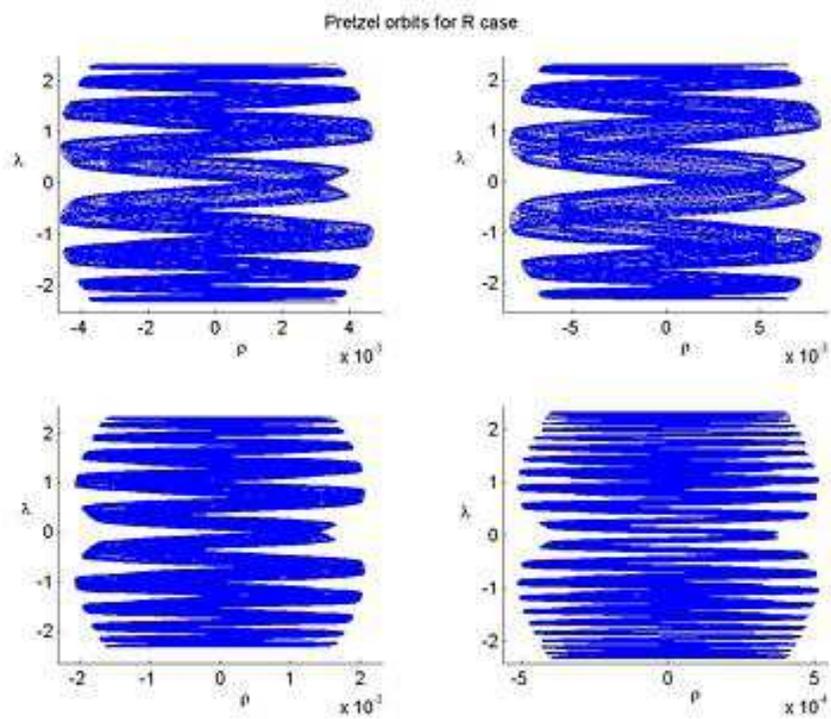,width=0.8\linewidth}
\end{center}
\caption{Orbits from the same family for the R system with $\protect\eta %
=0.75$.{\bf \ }(Here FE initial conditions were used, so only qualitative
features may be compared.) \ The N and R orbits in this family have
collision sequences of the form $A^{m}B^{3}$ ( $m$ odd); however, the R
orbits have an hourglass shape, while the N orbits in fig. \ref{Nbump}\ lie
in a cylinder. \ }
\label{Rbump}
\end{figure}

We can obtain interesting sequences of orbits of the hex-particle by
controlling the FM initial conditions. Consider for example fig. \ref{Nbump}%
, which consists of members of a family of quasi-regular orbits given by $%
\left\{ \overline{A^{i}B^{3}}\right\} $\ for the N case. \ These snake-like
orbits have two sharp turning points separated by some number $n\ $ of
bumps, and correspond to sequences of $2(n+2)$ circles in the lower portion
of the Poincare section. Such orbits have been shown to exist for arbitrary $%
n$\ in the N system \cite{LMiller}. \ We have found such orbits up to $n=15$%
\ and conjecture that they also exist for arbitrary $n$\ in the R and pN
systems below the threshold of chaos. In the pN system, orbits of higher $n$%
\ are gradually destroyed by chaos as $\eta $\ increases, with more and more
of the pretzel region becoming chaotic. We have found some evidence (see the
next section) that this may also occur in the R system; if so, the onset of
chaos will be much less dramatic than in the pN case. The corresponding
situation for the R system is shown in fig. \ref{Rbump}, with $\eta =0.75$.
\ The collision sequences are the same, as is their correspondence with the
circles in the lower portion of the Poincare section. However the figures in
the R system develop an hourglass shape, narrowing about $\lambda =0$,
whereas the N orbits are circumscribed by a cylinder. \ 


\begin{figure}[tbp]
\begin{center}
\epsfig{file=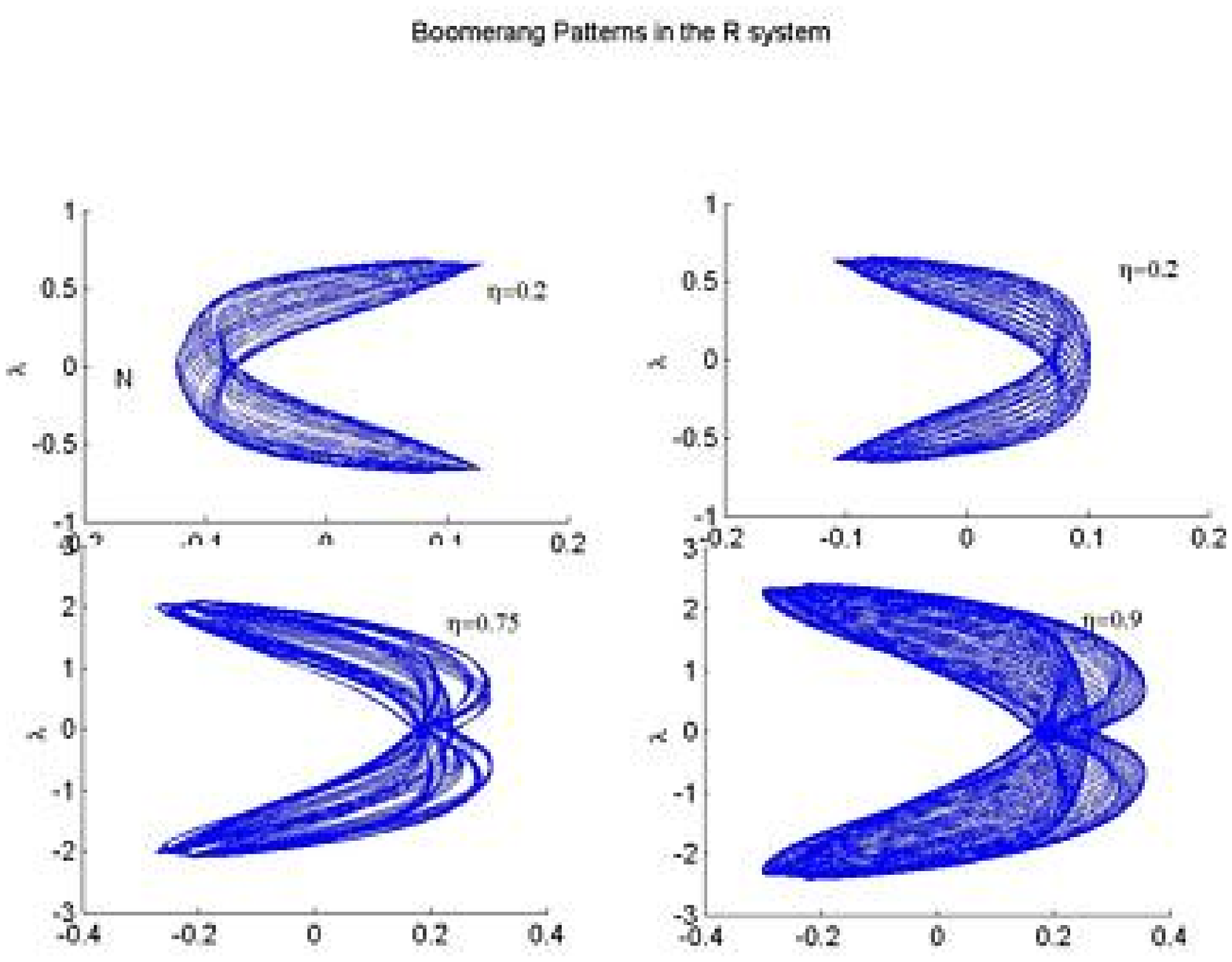,width=0.8\linewidth}
\end{center}
\caption{An orbit with the symbol sequence $AB^{3}$, shown for the N system
(plot 1), and the R system at different $\protect\eta $ values (plots 2-4).
\ All figures were run for 200 time steps with FM initial conditions. \ Note
that as $\protect\eta $ increases, the R trajectories develop a kink along
the $\protect\lambda =0$ axis, and begin to display a double-banding pattern
with two turning points at two distinct distances from the $\protect\rho $
axis about $\protect\lambda =0$.}
\label{pretR175}
\end{figure}

In fig. \ref{pretR175} we compare orbits with the symbol sequence $AB^{3}$.
\ Here we see another example of how relativistic effects induce
qualitatively different features not seen in the N system. \ As $\eta $
increases, orbits in the R system develop two distinct turning points at
different distances from the $\rho =0$ axis. This is particularly evident
for $\eta =0.75$. There is also the development of a kink at the
right-hand-side of the Boomerang figure that becomes increasingly more
pronounced with increasing $\eta $. The underlying reason behind the
development of this structure is not clear to us; however we do not see it
in the N system.

\begin{figure}[tbp]
\begin{center}
\epsfig{file=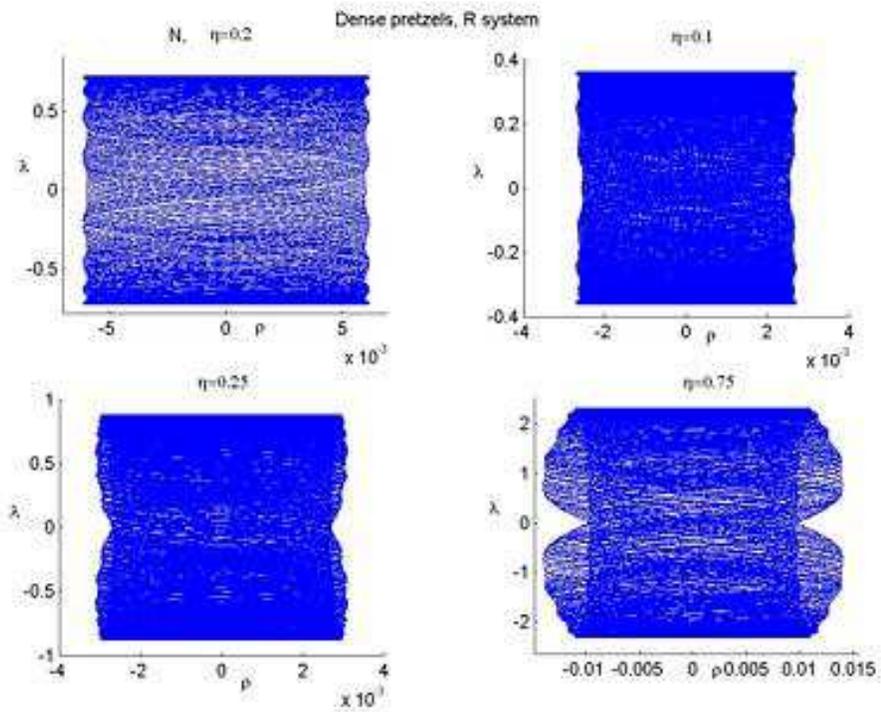,width=0.8\linewidth}
\end{center}
\caption{Densely filled orbits in the pretzel class for the R system at
varying values of $\protect\eta $, run for 200 steps with FE initial
conditions. A diagram for the N system at $\protect\eta =0.2$\ is included
in the upper left for comparative purposes. As $\protect\eta $ increases,
the R orbits take on an increasingly pronounced hourglass shape. \ In the R
system, we do not observe the breakdown to chaos seen in the pN case (fig.%
\ref{NtopNchaos} ). }
\label{pretcompR2}
\end{figure}

Overall the variety of orbits that appear in the R system appears to have a
richer and more detailed structure than that of the N system; for example
there are indentations in the bowtie patterns, the cylindrical shapes in the
N system become hourglass shapes in the R system and so on. Fig. \ref%
{Rprezteleta19a} \ illustrates this for $\eta =0.9$; we see variants of the
hourglass effect in two of these figures, along with different
manifestations of the development of distinct turning points.

\begin{figure}[tbp]
\begin{center}
\epsfig{file=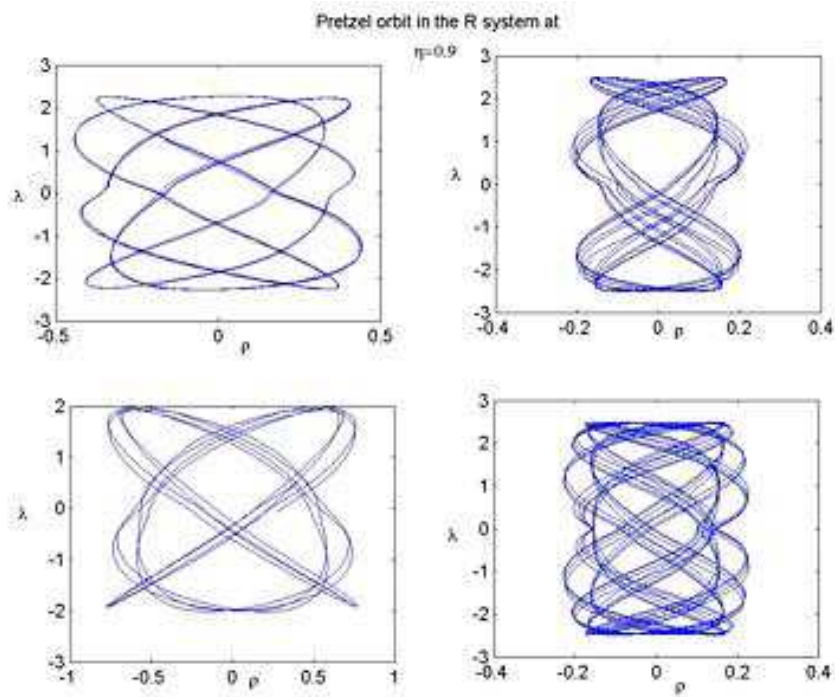,width=0.8\linewidth}
\end{center}
\caption{A survey of the broad variety of figures in the R system that arise
in the pretzel class for $\protect\eta =0.9$. \ \ All orbits were run for
200 time steps with FE initial conditions. }
\label{Rprezteleta19a}
\end{figure}

\subsection{Chaotic Orbits}

The chaotic orbits are those in which the hex-particle wanders between $A$%
-motions and $B$-motions in an apparently irregular fashion. Such orbits
eventually wander into all areas of the $\rho -\lambda $ plane-- a trait
neither the annuli nor the pretzels possess. The major area of chaos found
in all 3 systems occurs at the transition between annulus and pretzel
orbits, where the hex-particle passes very close to the origin.

\bigskip

The most striking feature of this class of motions is the distinction
between the pN system and its N and R counterparts. \ We find that the pN
system possesses an additional area of chaos in the pretzel region, a
phenomenon we shall discuss in more detail in the next section.

\begin{figure}[tbp]
\begin{center}
\epsfig{file=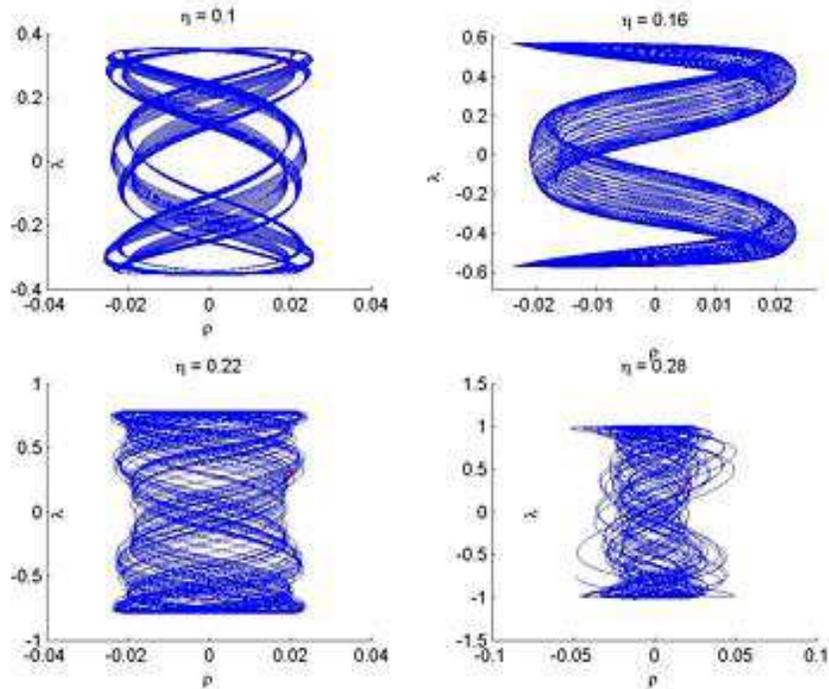,width=0.8\linewidth}
\end{center}
\caption{Transition to chaos for pretzel-type orbits in the pN case. All
four trajectories were run for 200 time steps with the same FE initial
conditions (with varying $\protect\eta $). As $\protect\eta $ increases, we
see the densely filled pretzel regions become less ordered. At $\protect\eta %
=0.28$ (lower right plot) we see the onset of chaos signified by an erratic
trajectory exploring a much larger area in the $\protect\rho $-$\protect%
\lambda $ plane.}
\label{NtopNchaos}
\end{figure}

We can observe the transition to chaos in the pretzel region of the pN
system by slowly adjusting the value of $\eta $ for FE initial conditions.
Figure \ref{NtopNchaos} illustrates an example. \ We begin with a pretzel
diagram at $\eta =0.1$. As $\eta $ increases, the trajectory changes shape
but remains regular until $\eta \simeq 0.22$ where the diagram appears
slightly less ordered. At $\eta \simeq 0.28$ the hex-particle begins to
irregularly traverse increasingly larger regions of the $\left( \rho
,\lambda \right) $ plane, signifying the onset of chaos. \ 

\begin{figure}[tbp]
\begin{center}
\epsfig{file=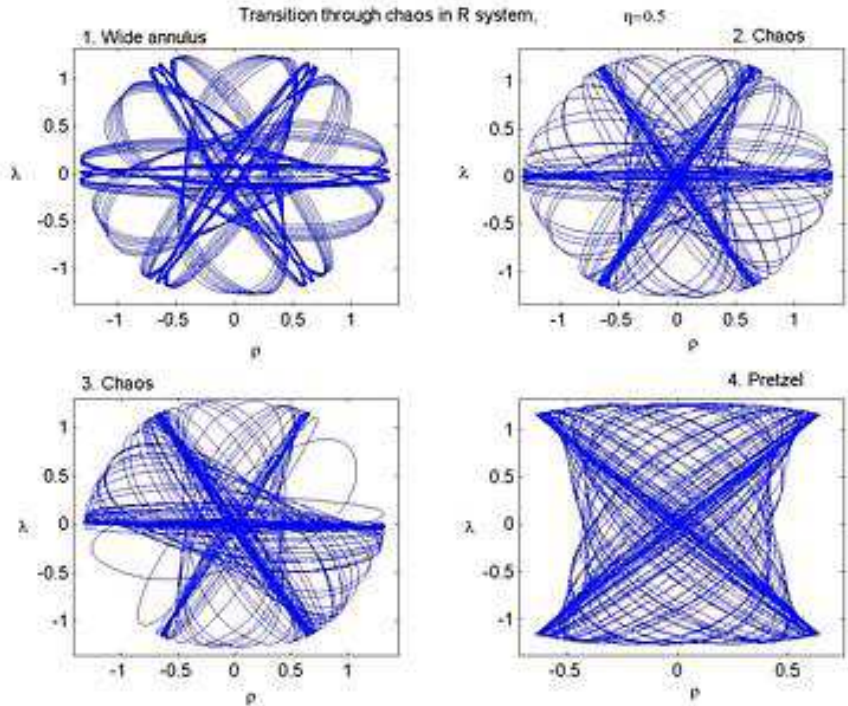,width=0.8\linewidth}
\end{center}
\caption{Transition from an annulus to a pretzel orbit through a chaotic
region in the R system for $\protect\eta =0.5$ as initial angular momentum
in the $\left( \protect\rho ,\protect\lambda \right) $ plane decreases. All
diagrams were run for 450 time steps with FE initial conditions. Note that
the chaotic trajectories pass very close to or through the origin. }
\label{Relchaostrans}
\end{figure}

We plot in fig. \ref{Relchaostrans} the transition in the R system with $%
\eta =0.5$ from an annulus to a pretzel orbit. \ The transition, which goes
from left-to-right and top-to-bottom with decreasing {\it initial} angular
momentum, passes through a chaotic set of orbits. The chaotic trajectories
pass very close to or through the origin, a characteristic feature for this
region of chaos in all three systems.

\begin{figure}[tbp]
\begin{center}
\epsfig{file=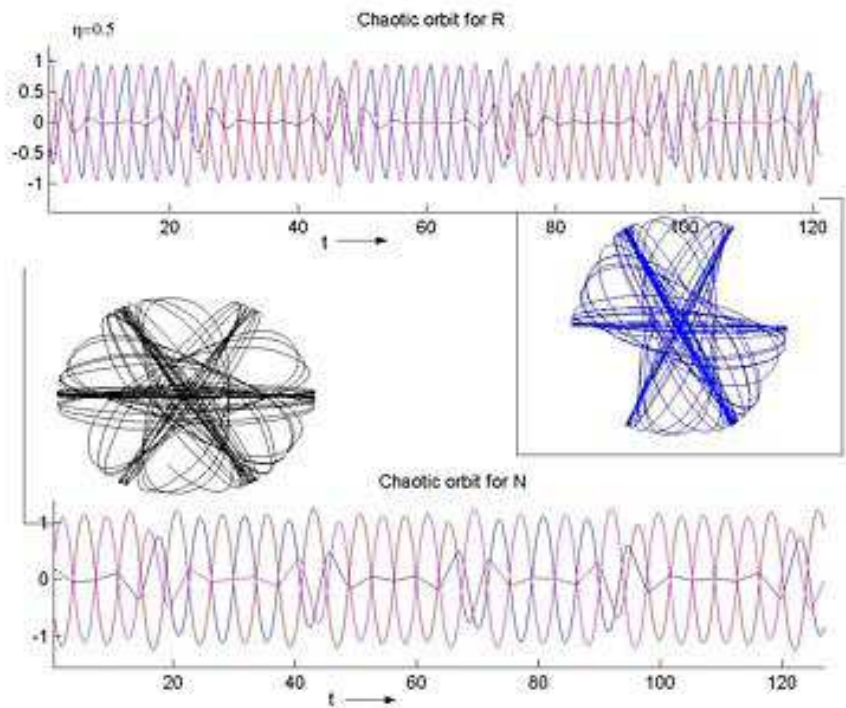,width=0.8\linewidth}
\end{center}
\caption{Chaotic orbits in the region of phase space separating annulus and
pretzel trajectories. \ Shown here are trajectories for R and N\ systems in
the $\left( \protect\rho ,\protect\lambda \right) $ plane (run for 300 time
steps) and for the 3-body system (truncated at 120 time steps). \ FE initial
conditions were employed, but the initial values of $\left( \protect\rho ,%
\protect\lambda ,p_{\protect\rho }\right) $ differ for the R and N
trajectories shown.\ In the 3-body system, the particles spend most of their
time in a configuration where one middle (`m') particle remains essentially
motionless. \ The motion appears very close to regular, its erraticity
apparent in slight irregularities between the number of crossings for which
one particle remains almost stationary. This causes the m-particle to
alternate in an irregular fashion.}
\label{3bdRnNchaos}
\end{figure}

\bigskip

\begin{figure}[tbp]
\begin{center}
\epsfig{file=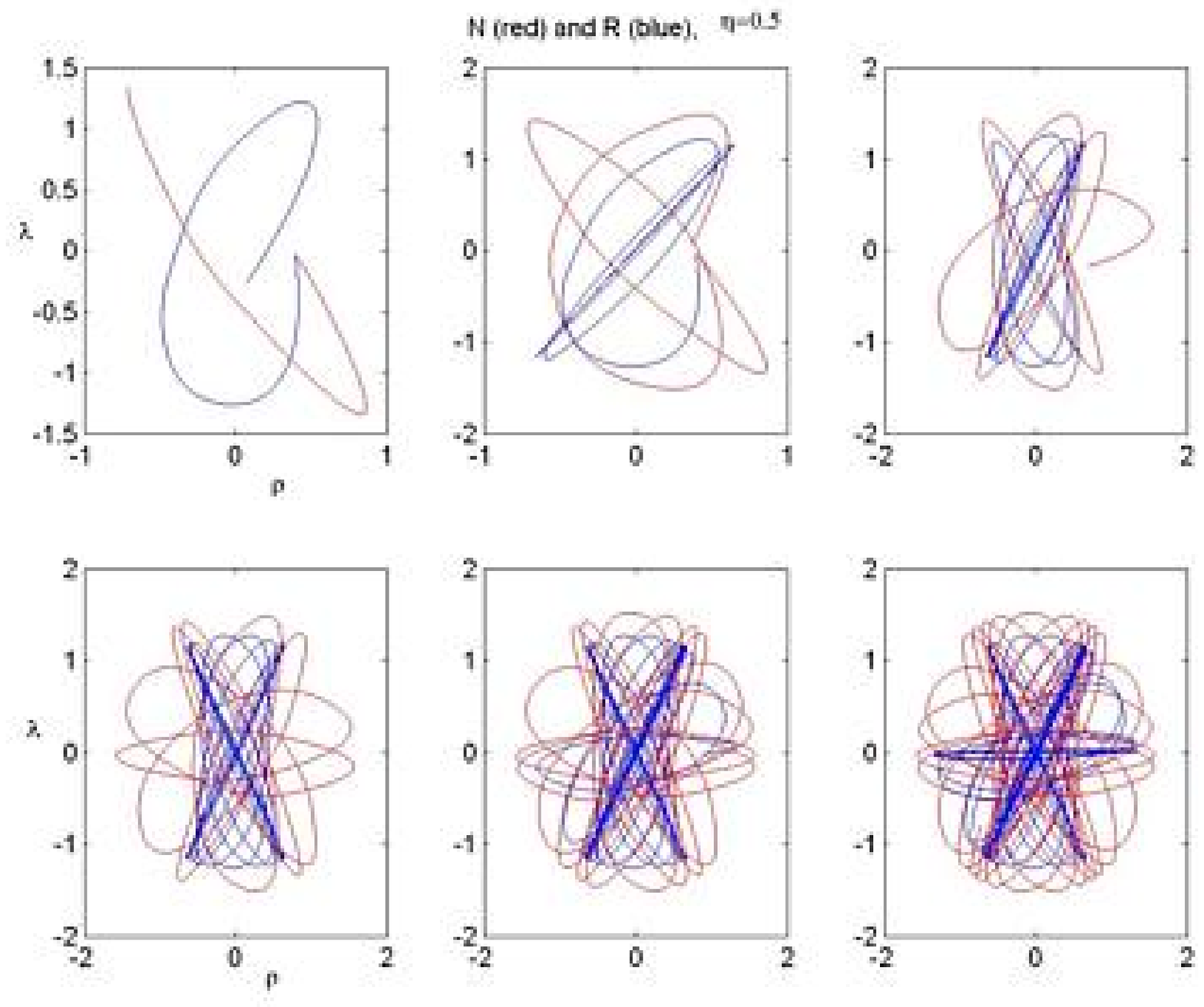,width=0.8\linewidth}
\end{center}
\caption{A comparison of the time-development of a chaotic R trajectory at $%
\protect\eta =0.5$ with its N counterpart at the same energy (FE intial
conditions), shown at $t=5,15,30,50,80$ and $110$ units.{\bf \ }The N
trajectory forms a densely filled annulus whilst its R counterpart is
chaotic.}
\label{timechaos2}
\end{figure}

\begin{figure}[tbp]
\begin{center}
\epsfig{file=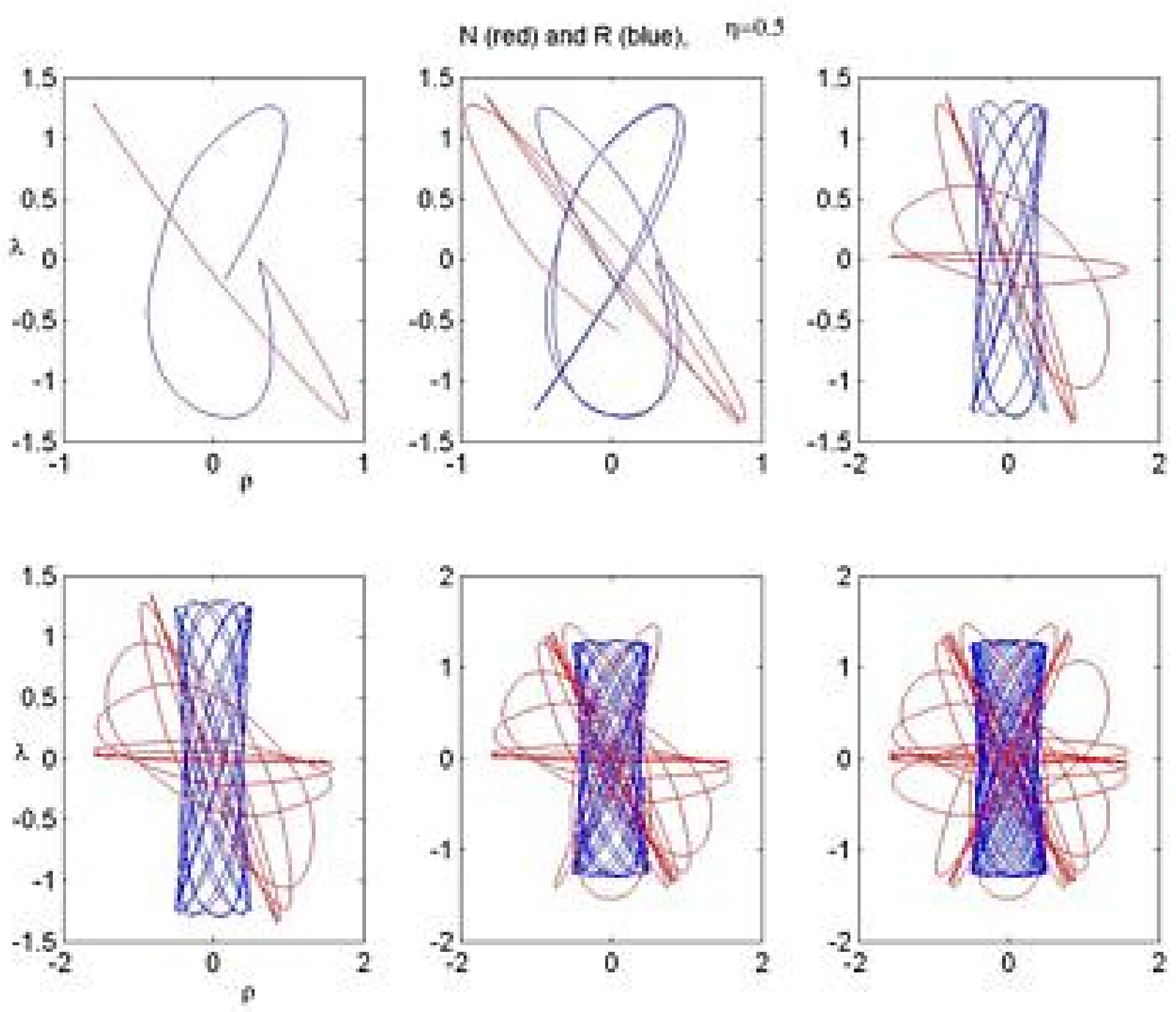,width=0.8\linewidth}
\end{center}
\caption{A comparison of the time-development of a chaotic N trajectory at $%
\protect\eta =0.5$ with its R counterpart at the same energy (FE intial
conditions), shown at $t=5,15,30,50,80$ and $110$ units. The R trajectory
forms a densely filled cylinder in the pretzel class whilst its N
counterpart is chaotic.}
\label{timechaos3}
\end{figure}

Figures \ref{timechaos2}\ and \ref{timechaos3} illustrate the
time-development of a corresponding pair of chaotic trajectories in the R
and N systems for different initial conditions. We see that different
initial conditions at the same energy can yield a chaotic R trajectory whose
N counterpart is an annulus, as well as a chaotic N trajectory whose R
counterpart is a cylinder in the pretzel class. \ In both cases we see that
the R trajectory approaches its final pattern much more rapidly than its N
counterpart (again, probably due to the difference in frequencies),
regardless of whether or not the final state is chaotic. 

In all three systems there is a region of chaos (R1) between the pretzel and
annulus type orbits, though in the R system it appears to shrink as $\eta $\
increases. Secondly, in the pN system, chaotic pretzel orbits are also
observed, becoming wilder and more prevalent with increasing $\eta $. These
chaotic pretzel orbits, unlike their R counterparts, do not cover the entire 
$\left( \rho ,\lambda \right) $\ plane, as can be seen from the figure \ref%
{NtopNchaos}.

\section{Poincare plots}

We turn now to consider the Poincare sections for the three systems N, pN,
R. These are constructed by plotting the square of the angular momentum ($%
p_{\theta }^{2}$, labelled as $z$) of the hex-particle against its radial
momentum ($p_{R}$, labelled as $x$) each time it crosses one of the
bisectors. Our conventions for these quantities are the same as in ref.\cite%
{LMiller}, apart from an overall normalization for each section that we
plot. All bisectors are equivalent since all three particles have the same
mass, and we can plot all crossings on the same surface of section, allowing
us to find regions of periodicity, quasi-periodicity and chaos.

Each of these systems is governed by a time-independent\thinspace\
Hamiltonian with four degrees of freedom. Hence the total energy is a
constant of the motion, and the phase space for each system is a
3-dimensional hypersurface in 4 dimensions. If an additional constant of the
motion exists, the system is said to be integrable, and its trajectories are
restricted to two-dimensional surfaces in the available phase space. Since
trajectories may never intersect, such a constraint imposes severe
limitations on the types of motion that integrable systems can exhibit:
trajectories may be periodic, repeating themselves after a finite interval
of time, or quasi-periodic. The trajectories of an integrable system always
appear as lines or dots, for periodic (1-dimensional) orbits on the Poincare
section, as they comprise by definition the intersection of two
2-dimensional surfaces. This contrasts sharply with the case when a system
is completely non-integrable, so that all orbits move freely in three
dimensions. The extra degree of freedom permits orbits to visit all regions
of phase space, and the system typically displays strongly chaotic behavior.
Such trajectories appear as filled in areas on the Poincare map.

When an integrable system is given a sufficiently small perturbation, most
of its orbits remain confined to two-dimensional surfaces . However small
areas of chaos appear, sandwiched between the remaining two-dimensional
surfaces. As the magnitude of the perturbation is increased, the chaotic
regions grow, and eventually become connected areas on the Poincare section.
This phenomenon is called a Kolmogorov, Arnold and Moser (KAM) transition %
\cite{KAM}. Islands of regularity may remain for quite some time, and
generally have an intricate fractal structure. For sufficiently large
perturbations, however, systems typically become almost fully ergodic \cite%
{Rzheng}.

\bigskip

The structure of the Poincare section in the N system has already been
studied to a certain extent in the wedge problem. In the equal mass case
3-body motion in the N system corresponds to motion of a body falling toward
a wedge whose sides are each at angles $30^{o}$ relative to the vertical
axis \cite{LMiller}. \ \ The outer boundary of the plot is determined by the
energy conservation relation (\ref{HN}), which is 
\begin{equation}
x^{2}\leq 1-z  \label{Poinenergy1}
\end{equation}%
where the energy $H-3mc^{2}$ has been normalized to unity (more generally, $%
x^{2}\leq \frac{2}{3}\eta -z$ for the unconstrained normalizations we
employ). Equality in (\ref{Poinenergy1}) holds when the hex-particle is at
the origin, and yields the phase-space limit since any departure from the
origin will reduce the values of $\left( x,z\right) $ relative to this
bound. \ \ Another relevant boundary is that given by 
\begin{equation}
\left( x-2\sqrt{3z}\right) ^{2}\leq 1-z  \label{Poinenergy2}
\end{equation}%
which is the energy constraint after an $A$-collision has taken place.
Equality corresponds to the point at which all three particles are
coincident (the hex-particle is at the origin). \ All points in phase space
satisfying (\ref{Poinenergy2}) will undergo an $A$-collision (the $A$%
-region) whereas those violating this inequality will undergo a $B$%
-collision (the $B$-region). \ Inevitably a point in the $A$-region will
venture into the $B$-region since the interaction is gravitational and
collisions with the third particle cannot be avoided. \ Hence the $A$-region
has no fixed points. \ However the $B$-region has a subregion containing a
fixed point in which the $B$-collisions are infinitely repeated (the $%
\overline{B}$ motion),\ corresponding to the annulus orbits.

\begin{figure}[tbp]
\begin{center}
\epsfig{file=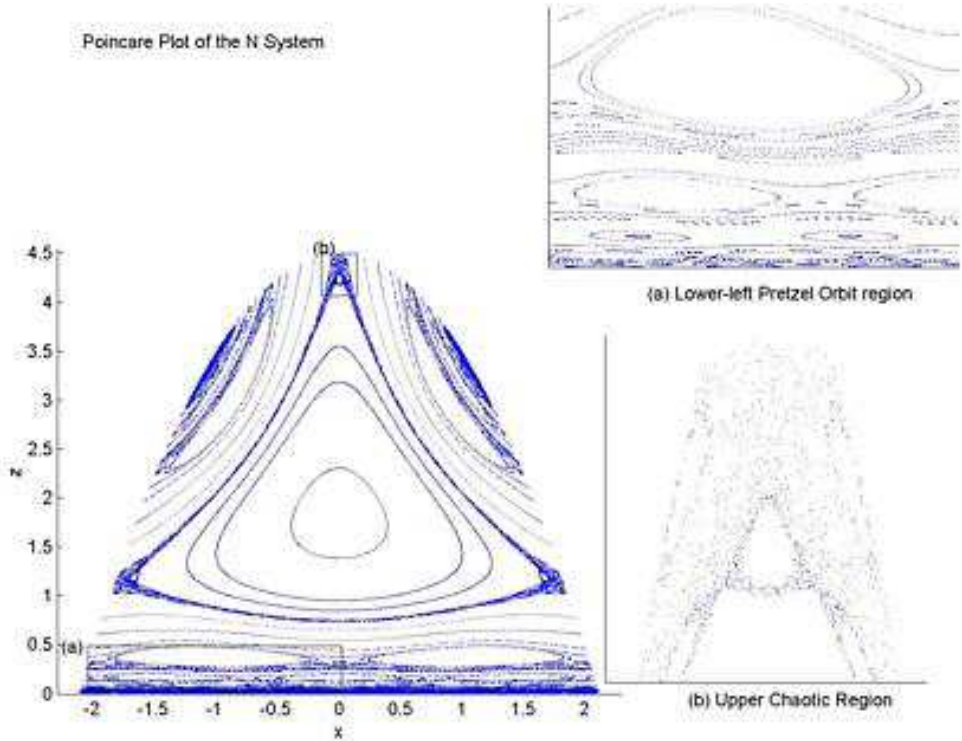,width=0.8\linewidth}
\end{center}
\caption{The Poincare plot of the N system. The squares denote the parts of
the plot magnified in the insets. }
\label{Poincare0b}
\end{figure}

The Poincare section for the N system is shown in fig. \ref{Poincare0b}.
There is a fixed point at the centre of the plot surrounded by a subregion
of near-integrable curves. All of the annuli are contained within the large
triangle surrounding this region; its boundary contains a thin region of
chaos, beyond which is the pretzel region. \ 
\begin{figure}[tbp]
\begin{center}
\epsfig{file=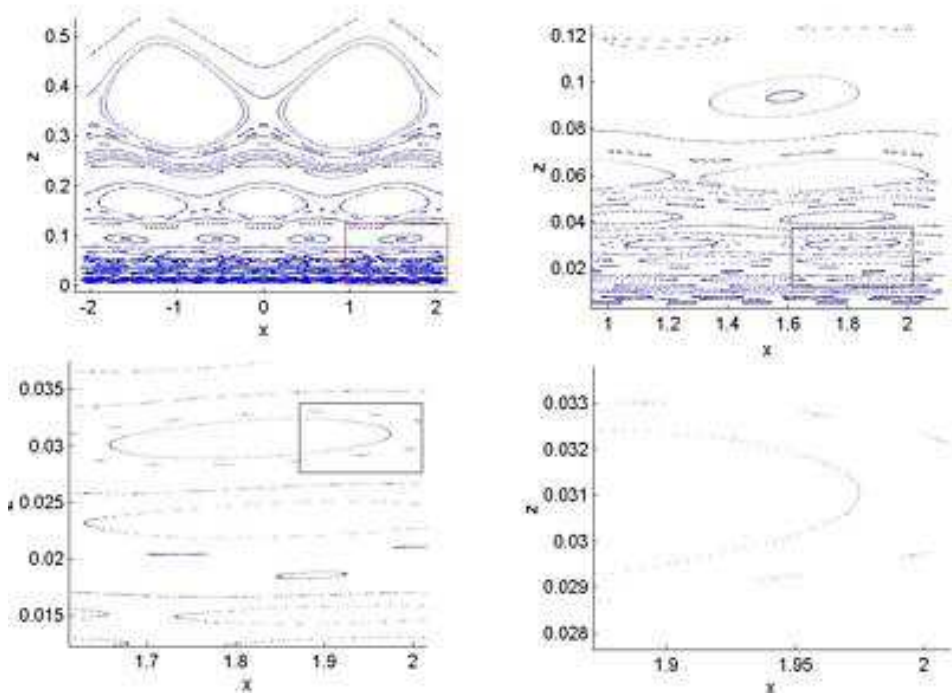,width=0.8\linewidth}
\end{center}
\caption{A series of successive close-ups of the lower section of the
Poincare plot of the N system. This illustrates the self-similar structure
in the pretzel region that repeats at increasingly small scales. \ The
limiting factor at very small scales is the number of trajectories that we
included in the plot.}
\label{Poincare0_z}
\end{figure}

The structure of the lower part and upper corners of fig. \ref{Poincare0b}
is extremely complicated and intricate, as illustrated by the insets. \ The
chaotic regions are confined and not simply connected. Though not
integrable, the N system shows a high degree of regularity. There is a
self-similarity within the pretzel region as illustrated in fig.\ref%
{Poincare0_z}, with the circles bounding the quasi-periodic near-integrable
regions repeating themselves on increasingly small scales. We find that
motions in the N system are completely regular, as evidenced by the absence
of dark areas in the Poincare section (except for the one region of chaos
mentioned previously).\ These results are all commensurate with those of the
wedge system \cite{LMiller}.

\bigskip

We pause here to comment on the symbol sequences corresponding to particular
patterns. For example, the two large circles observed just below the annulus
region correspond to the boomerang-shaped orbits ($\overline{AB^{3}}$). The
next set of circles will be $\overline{A^{2}B^{3}}$, and so on. The
collections of crescents between these sets of circles correspond to
sequences $\overline{AB^{3}A^{2}B^{3}}$, $\overline{AB^{3}AB^{3}A^{2}B^{3}}$%
, and so on. Of course, for each circle in the Poincare plot there is in
fact a continuum of possible circles, whose diameter depends on the initial
conditions. At the center of this family of circles is a dot corresponding
to the periodic orbit in question.

Another observable feature in the Poincare plot is a series of closed
circles that lie in a triangular pattern in the annulus region. These
correspond to quasi-periodic orbits about the periodic annuli with higher
period; for example figs. \ref{NRanuper}, \ref{PNanuper}\ and \ref{Rannuli}.

\bigskip 
\begin{figure}[tbp]
\begin{center}
\epsfig{file=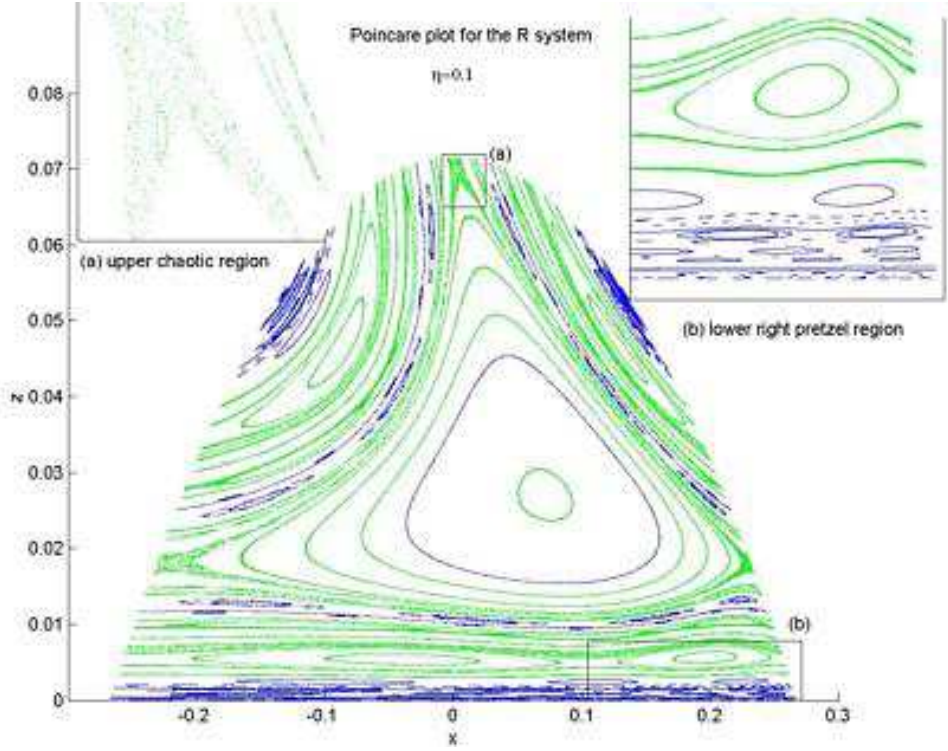,width=0.8\linewidth}
\end{center}
\caption{The Poincare plot of the R system at $\protect\eta =0.1$. The right
insets provides a close-up of the chaotic region at the top of the diagram;
note that it is similar to the N system, but distorted in shape. The left
inset is a close-up of the structure in a pretzel region in the lower right
of the diagram; it is similarly distorted relative to the N system.}
\label{Poincare110}
\end{figure}
\begin{figure}[tbp]
\begin{center}
\epsfig{file=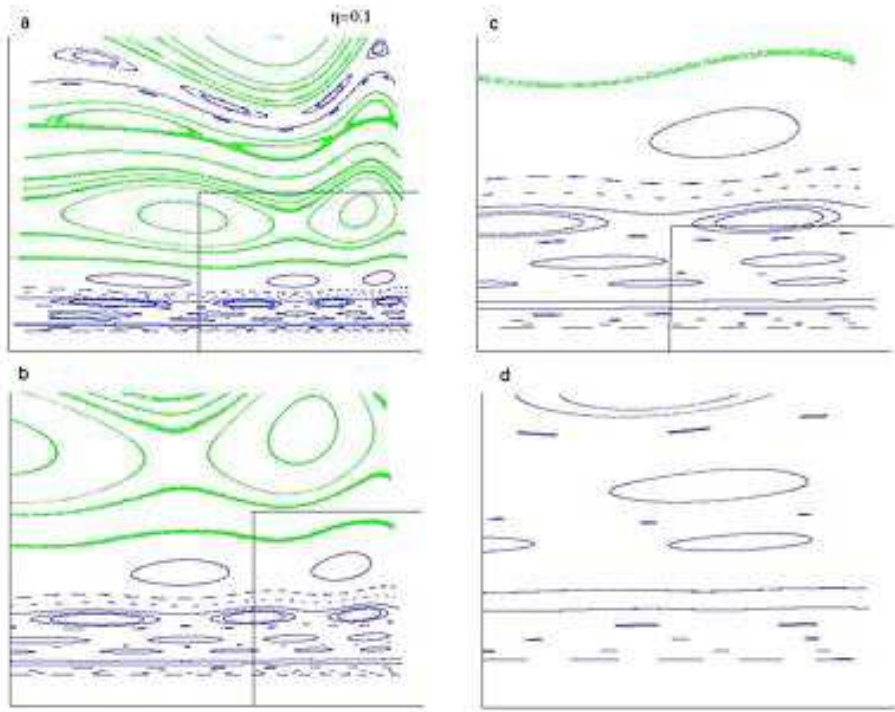,width=0.8\linewidth}
\end{center}
\caption{A series of successive close-ups of the lower section of the
Poincare plot of the R system at $\protect\eta =0.1$. We see no evidence of
a breakdown to chaotic behaviour. Rather the self-similar structure in the
pretzel region apparently repeats at increasingly small scales as in the N
system. \ }
\label{Poincare110_z}
\end{figure}

Turning next to the R system we find the result that all of the features of
its Poincare plot are qualitatively similar to the N system over the range
of $\eta $\ that we were able to investigate. This is remarkable considering
the high degree of non-linearity of the relativistic Hamiltonian given by (%
\ref{Htrans}). \ The annulus, pretzel and chaotic regions all retain their
same basic structure, as seen in fig. \ref{Poincare110}. \ 

\begin{figure}[tbp]
\begin{center}
\epsfig{file=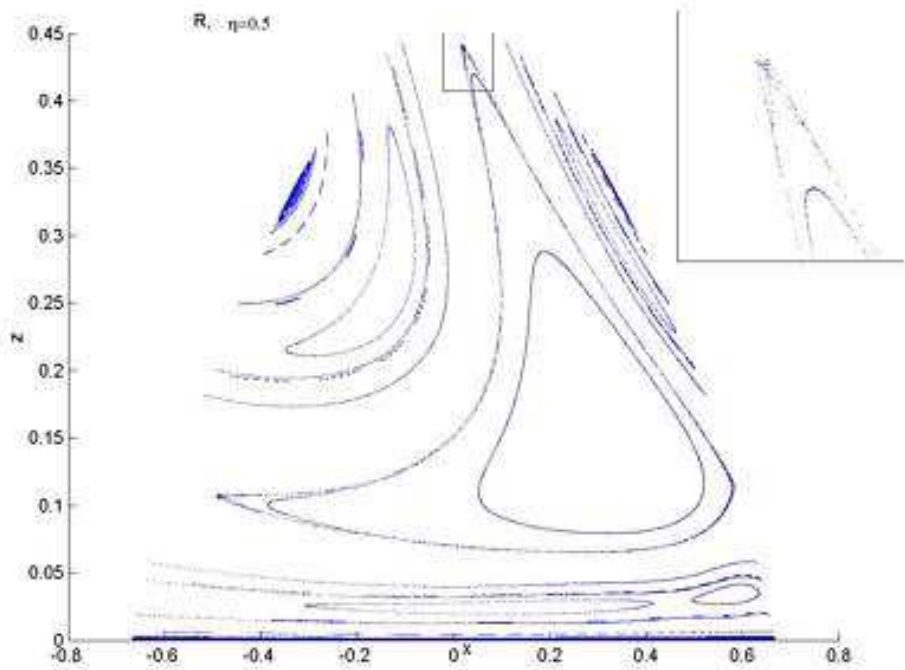,width=0.8\linewidth}
\end{center}
\caption{The Poincare plot of the R system at $\protect\eta =0.5$. The right
inset provides a close-up of the chaotic region at the top of the diagram.
The chaotic region is smaller and the distortion increased relative to the $%
\protect\eta =0.1$\ case. Fewer points appear on this figure because fewer
trajectories were run to generate it.}
\label{Poincare150}
\end{figure}

\begin{figure}[tbp]
\begin{center}
\epsfig{file=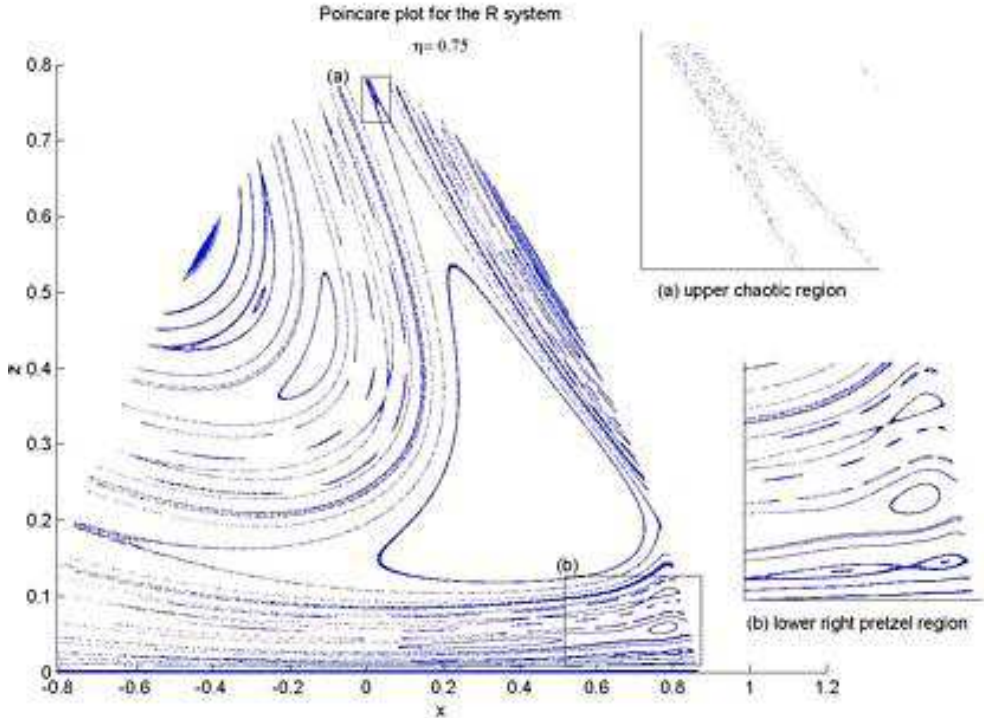,width=0.8\linewidth}
\end{center}
\caption{The Poincare plot of the R system at $\protect\eta =0.75$. The
upper right inset provides a close-up of the chaotic region at the top of
the diagram; it is now considerably narrower than for lower values of $%
\protect\eta $.\ The lower-right inset is a close-up of the structure in a
pretzel region in the lower right of the diagram. The lines between the
various ellipses have slightly thickened, possibly suggesting preliminary
stages of KAM breakdown.}
\label{Poincare175}
\end{figure}

However we find that the plot is no longer symmetric with respect to $%
p_{R}=0 $, and that the Poincare plot is asymmetrically deformed relative to
its counterpart in the N system, the deformation increasing with increasing $%
\eta $, as figs. \ref{Poincare110}, \ref{Poincare150} and \ref{Poincare175}
illustrate. Superficially this deformation is somewhat puzzling: the
trajectories of a subset of the annulus-type orbits always have positive
radial velocities when they intersect one of the hexagon's edges (and the
tendency of all annulus orbits is to have $p_{R}>0$ at the bisectors).
However it occurs because the Hamiltonian given by eq. (\ref{Htrans}) is not
invariant under the discrete symmetry $p_{i}\rightarrow -p_{i}$, but rather
is invariant only under the weaker discrete symmetry $\left( p_{i},\epsilon
\right) \rightarrow \left( -p_{i},-\epsilon \right) $. \ The parameter $%
\epsilon =\pm 1$ is a discrete constant of integration that is a measure of
the flow of time of the gravitational field relative to the particle
momenta. \ We have chosen $\epsilon =+1$ throughout, which has the effect of
making the principal features of the Poincare plot `squashed' towards the
lower right-hand side of the figure relative to its counterpart in the N
system. This deformation would be toward the lower-left had we chosen $%
\epsilon =-1$. It is reminiscent of the situation for two particles, in
which the gravitational coupling to the kinetic-energy of the particles
causes a distortion of the trajectory from an otherwise symmetric pattern %
\cite{2bd,2bdchglo}, becoming more pronounced as $\eta $ increases.

\begin{figure}[tbp]
\begin{center}
\epsfig{file=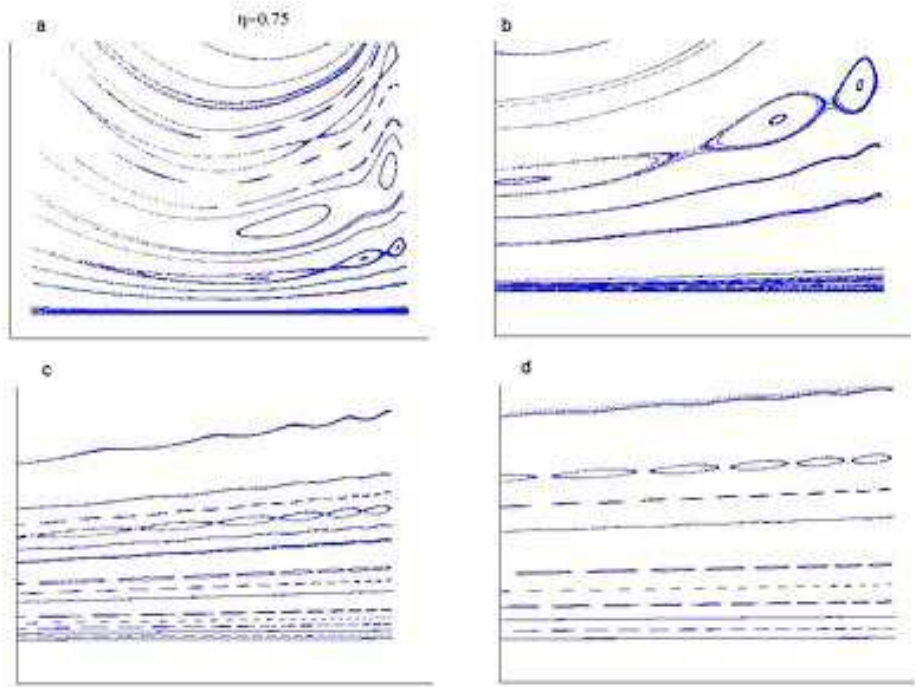,width=0.8\linewidth}
\end{center}
\caption{A series of successive close-ups of the lower right section
(pretzel region) of the Poincare plot of the R system at $\protect\eta =0.75$%
. While this region is still highly regular, the lines joining the ellipses
are suggestive of the early stages of KAM breakdown, as is the appearance of
waviness in the solid lines. \ }
\label{Poincare175_z}
\end{figure}

Remarkably we do not find a breakdown from regular to chaotic motion as $%
\eta $ increases in the R case. The lower regions of the Poincare map
clearly display the same patterns of series of circles as occurs in the
non-relativistic case and no sizeable connected areas of chaos are present.
This is clear from figs. \ref{Poincare110_z} and \ref{Poincare175_z}, which
respectively show a sequence of successive close-ups of the pretzel region
for the $\eta =0.1$ and $\eta =0.75$ cases. However we do find that at $\eta
=0.75$ the lines between the near-integrable elliptic regions increases,
suggesting either the onset of KAM breakdown or a relativistic
generalization of the fractal pattern seen in the N system.{\bf \ }
Unfortunately we have not been able to investigate whether or not KAM
breakdown occurs for higher $\eta $\ values in the R system due to a lack of
computer resources.

\begin{figure}[tbp]
\begin{center}
\epsfig{file=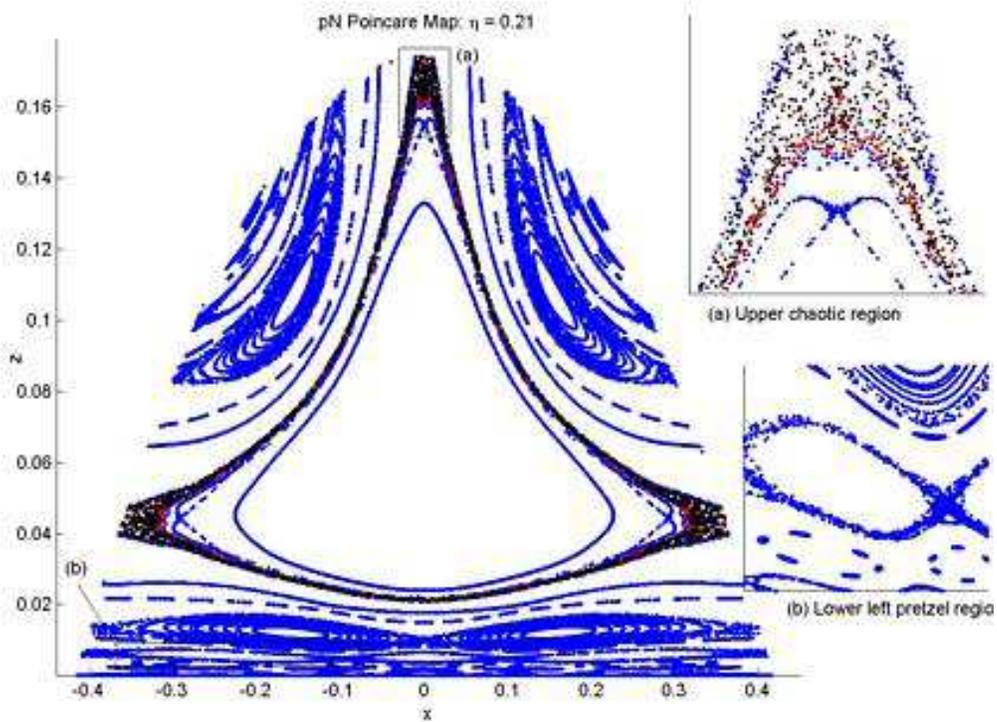,width=0.8\linewidth}
\end{center}
\caption{The Poincare plot for the pN system at $\protect\eta =0.21$. \
Unlike the R system, it is qualitatively similar to the N system in terms of
symmetry. However the chaotic regions have increased in size, with the
pretzel region being on the threshold of KAM breakdown.}
\label{PNPoinc2}
\end{figure}
\begin{figure}[tbp]
\begin{center}
\epsfig{file=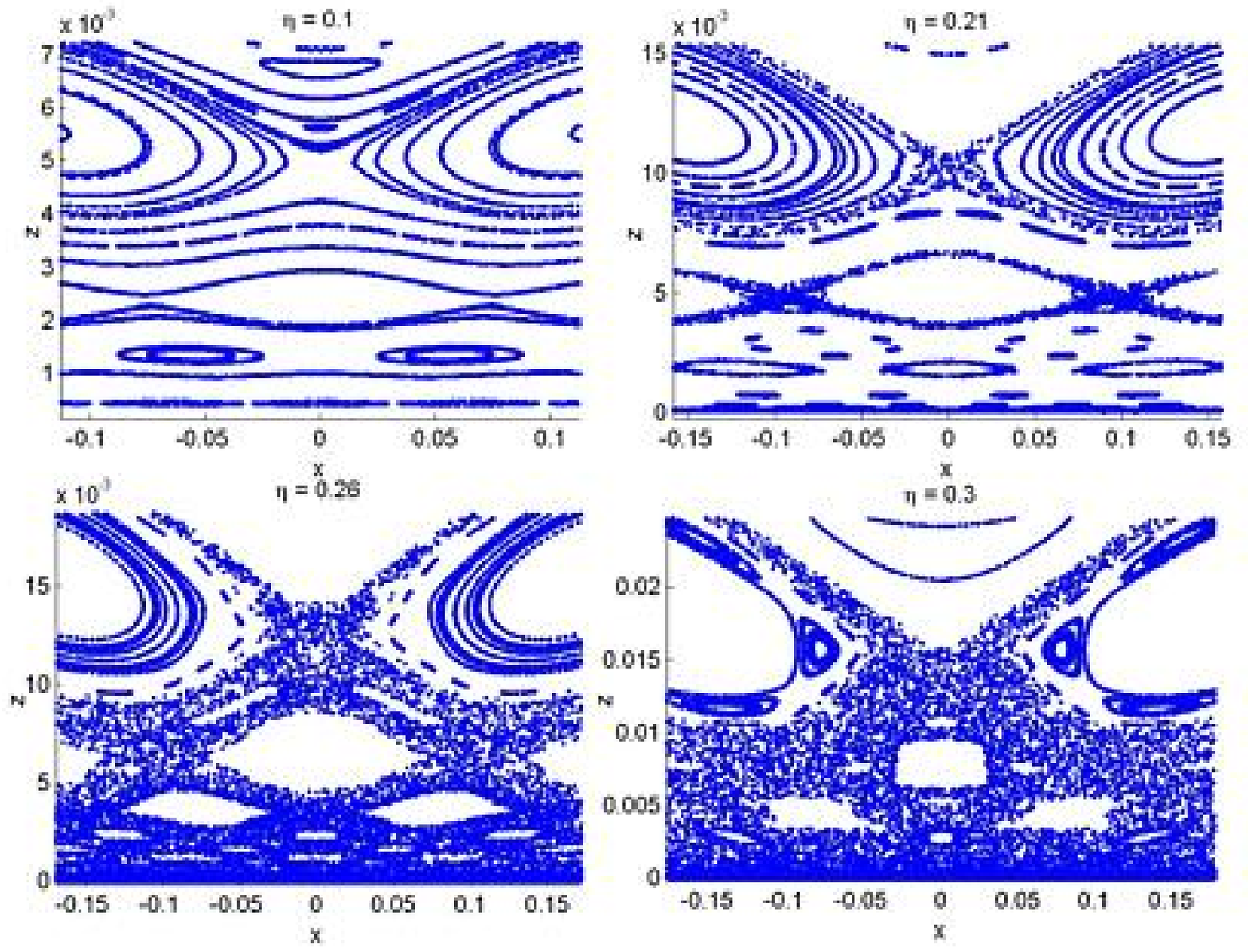,width=0.8\linewidth}
\end{center}
\caption{A close-up of the Poincare plots for the pN system for increasing
values of $\protect\eta $. \ The diagrams are all of the same lower part of
the section. \ We see clear evidence of KAM breakdown as $\protect\eta $
gets larger. }
\label{PNbreakseries}
\end{figure}

The pN system has a considerably different Poincare plot, shown in fig. \ref%
{PNPoinc2}. \ \ While it retains the $p_{i}\rightarrow -p_{i}$ symmetry of
its N-system predecessor, it appears to undergo a KAM transition from
relatively orderly behavior in the N system, to highly chaotic behavior at $%
\eta =.3$, as the series of Poincare sections in figure \ref{PNbreakseries}
demonstrates. At $\eta =.21$, the lines across the bottom of the figure have
widened slightly, though the overall behavior is still quite regular. As $%
\eta $ increases to $.26$, larger regions of chaos become evident around the
edges of the groups of ellipses that traverse the lower regions of the
figure. At $\eta =.3$, most of the lower section of the Poincare section has
been engulfed by a chaotic sea; only a few non-connected islands of regular
motion remain. This contrasts sharply with the behavior of the R system at
similar values of $\eta $.

The differences between the R and pN cases are not artifacts of the
difference in scalings; when trajectories with the same energy are compared,
the pN ones are visibly more chaotic than both the N ones and the R ones.
The apparent dearth of chaos in the R system is somewhat surprising, as it
indicates that most trajectories are effectively restricted to move on
two-dimensional surfaces in phase space, as in the N case. This occurs
despite the fact that the R system appears not to be integrable (chaotic
orbits separating the annulus and pretzel regions do seem to exist) for any $%
\eta $ within the range investigated. Nonetheless, clearly some underlying
feature enforces considerable structure on the phase space-- a feature that
is absent from the pN system.

\section{Discussion}

We consider in this section some general features of the motion of the three
systems we have studied.

We find for each system that the $\overline{B}$ symbol sequence always
occurs, for all values of $\eta $ that we were able to investigate. This
leads to a rich variety of annulus diagrams, symmetric about the $\rho =0$
axis for the N and pN systems, but with the axis of symmetry rotated
slightly for the R system, the rotation increasing with increasing $\eta $.
\ We conjecture that $\overline{B}$ motion takes place for arbitrarily large 
$\eta $ in each of the pN and R systems. It would be interesting to test
this conjecture -- were it not to hold it would mean that a highly
relativistic system must either experience a full KAM breakdown or else
repeatedly develop temporary quasi-bound 2-body subsystems. \ One thing
substantiating this conjecture is that there is no evidence that the annulus
region is shrinking with increasing $\eta $\ in either the pN or R systems.
\ However to prove this would require a relativistic equivalent of the
discrete mappings for the N case constructed in ref. \cite{LMiller}.

We found that the pretzel-type orbits display a remarkable richness of
dynamics for all three systems. As the angular momentum of the trajectory in
question decreases, the number of successive $A$\ collisions increases
before the hex-particle sweeps around the origin in the $B^{3}$-sequence. \
For example the trajectory $\overline{AB^{3}}$ (the simplest sequence after $%
\overline{B}$) corresponds to a boomerang-type orbit and appears as two
circles on the Poincare section. The next simplest sequence is $\overline{%
A^{2}B^{3}}$, which corresponds to a bow-tie like orbit, and generates three
slightly smaller circles on the Poincare section. \bigskip

Even\ in\ the\ small\ region\ of\ phase\ space\ between\ these\ two\ simple\
orbits\ a\ complex\ tangle\ of\ periodic\ and\ quasi-periodic\ orbits\
exists.\ For each of the patterns above, a family of orbits exists
corresponding to different widths of the `bands' of phase space that the
trajectory covers, and correspondingly different radii of circles in the
Poincare section. Between these regions, the orbits' sequences are mixtures
of $AB^{3}$\ and $A^{2}B^{3}$. This reasoning can be extended to more
general $A^{n}B^{3m}$-motions. We conjecture that the only allowed
non-chaotic orbits -- relativistic and non-relativistic -- are of the form $%
\prod_{i,j,k}\left( A^{n_{i}}B^{3m_{j}}\right) ^{l_{j}}$\ with $n_{i},m_{j}$%
\ finite, corresponding to increasingly complex weaving patterns. We expect
this conjecture to hold -- at least for the range of $\eta $\ that we could
access numerically -- for both the R and N systems (so that the pretzel
class can be divided into countably many distinct sub-classes: one for each
triple $(n_{i},m_{j},l_{k})$), but not for the pN system, as it experiences
KAM breakdown.

If the set of integers $l_{k}$\ is finite, then the sequence is regular,
leaving bands of phase space untraveled, and appearing as a series of closed
crescents or ellipsoids on the Poincare section. If, however, the sequence
of integers $l_{k}$\ never repeats itself, then the trajectory will fill the
available phase space densely, appearing as a wavy line on the surface of
section. We conjecture that there is a $1-1$ correspondence between rational
numbers and periodic orbits in this region of phase space, both for the N
and R systems. This would give the lower section of the Poincare plot a
fractal structure as the patterns of circles, ellipses and lines is repeated
on arbitrarily small scales as the hex-particle's angular momentum
approaches zero.

\section{Conclusions}

In (1+1) dimensions the degrees of freedom of the gravitational field are
frozen. One therefore expects the motion of a set of $N$ particles in curved
spacetime to be described by a conservative Hamiltonian. We find this to be
the case for the 3-body system we have studied. By canonically reducing the $%
N$-body action (\ref{act1}) to first-order form we derived an exact
determining equation of the Hamiltonian from the matching conditions. To our
knowledge this is the first such derivation for a relativistic
self-gravitating system. \ The canonical equations of motion given by the
Hamiltonian can be explicitly derived from this equation and then
numerically solved.

We recapitulate the main results of this paper:

\begin{enumerate}
\item We obtained the post-Newtonian expansion of the system we studied,
along with its non-relativistic limit. \ By comparing these two systems (pN
and N respectively) with their relativistic (R) counterpart we were able to
study quantitatively the distinctions between each of these systems. There
are two spatial degrees of freedom and two conjugate momenta in each, and so
the systems are most easily studied by making the transformations (\ref%
{hexrho}-\ref{zijrholam}). This yields the hex-particle representation of
the system: the 3-body N system is equivalent to that of a single particle
moving in a hexagonal linear well. \ The pN and R systems distort this well
by making the sides concave and convex respectively, with the latter system
inducing momentum-dependent changes to its shape.

\item We found that in the equal-mass case each system exhibited the same
three qualitative types of motion, that we classified in the hex-particle
representation as annulus, pretzel and chaotic. \ Annulus orbits correspond
to motions in which no two particles ever cross one another twice in
succession. \ Annuli can be either periodic, quasi-periodic, or densely
filled. \ Pretzel orbits correspond to motions in which a pair of particles
cross each other at least twice before either crosses the third. \ This
yields a very broad variety of increasingly intricate patterns for each
system, dependent upon the initial conditions. Stable bound subsystems of
two particles exist for each system. Chaotic orbits have no regular pattern,
and correspond to the case when the hex-particle crosses the origin. For
energies close to the total rest-energy we find that all of these types of
orbits are virtually indistinguishable for each of the N, pN and R systems.

\item We find that differences between each system become more pronounced as 
$\eta $\ increases. In general orbits in the R system are of higher
frequency and cover a smaller region of the $\left( \rho ,\lambda \right) $
plane than those of its N system counterparts at the same energy. \ If the
same initial conditions are posed for each system, the motions differ
considerably, with the R system having more energy and covering a larger
region of the $\left( \rho ,\lambda \right) $ plane. \ Annulus orbits in the
R system have a symmetry axis that is rotated slightly relative to their N
and pN counterparts. \ Pretzel orbits develop an hourglass shape in the R
system that is not seen in the N system, and additional turning points
appear for these orbits that are absent in the N system.\ 

\item We find that the qualitative features of the Poincare sections for the
R and N systems remain the same for all values of $\eta $ that we were able
to study. This is remarkable given the high degree of non-linearity in the
former. \ However the R system has a weaker symmetry than its N counterpart
and so its Poincare section develops an asymmetric distortion that increases
with increasing $\eta $. \ 

\item We find that the pN system experiences a KAM breakdown that is not
seen in the R and N systems. \ This takes place for $\eta \simeq 0.26$:
lines separating distinct near-integrable regions become increasingly wider
as $\eta $ increases, degenerating into chaos.
\end{enumerate}

\bigskip

A number of interesting questions arise from this work. \ First, it would be
of considerable interest to explore the R system in the large $\eta $
regime. \ This will require considerably more sophisticated numerical
algorithms than we have been using that avoid the numerical instabilities we
encountered, as well as perhaps employing a time parameter that is not the
coordinate time. \ Second, an investigation of the unequal mass case should
be carried out to see if the common features between the N, R and pN systems
are retained. In the N system, when masses are unequal, simply connected
regions of global chaos appear; the relationship of these regions to their
pN and R counterparts remains a subject for future consideration. Work on
the unequal mass case is in progress \cite{justin}.

\bigskip

\section*{APPENDIX A: The Determining equation in hexagonal coordinates}

The form of the determining equation is given by (\ref{Htrans}), and we wish
to rewrite it in terms of the four independent degrees of freedom $\left(
\rho ,\lambda ,p_{\rho },p_{\lambda }\right) $, using the relations (\ref%
{zijrholam}) and (\ref{p1rholam}-\ref{p3rholam}).

Consider first the expressions in the exponentials. Some algebra shows that 
\begin{eqnarray}
H_{0} &\equiv &(L_{1}+{\frak M}_{12})z_{13}-(L_{2}+{\frak M}_{21})z_{23} 
\nonumber \\
&=&\sqrt{2}H\rho -\epsilon \left( 2\left| \rho \right| p_{\rho }+\left[
\left| \lambda +\frac{\rho }{\sqrt{3}}\right| -\left| \lambda -\frac{\rho }{%
\sqrt{3}}\right| \right] p_{\lambda }\right)  \label{h0R} \\
H_{-} &\equiv &(L_{2}+{\frak M}_{23})z_{21}-(L_{3}+{\frak M}_{32})z_{31} 
\nonumber \\
&=&H\left( \lambda -\frac{\rho }{\sqrt{3}}\right) -\epsilon \left( \left[
\left| \rho \right| -\frac{\sqrt{3}}{2}\left| \lambda +\frac{\rho }{\sqrt{3}}%
\right| \right] \left( \frac{p_{\lambda }}{\sqrt{3}}+p_{\rho }\right) +\frac{%
3}{2}\left( p_{\lambda }-\frac{p_{\rho }}{\sqrt{3}}\right) \left| \lambda -%
\frac{\rho }{\sqrt{3}}\right| \right)  \label{hmR} \\
H_{+} &\equiv &(L_{3}+{\frak M}_{31})z_{32}-(L_{1}+{\frak M}_{13})z_{12} 
\nonumber \\
&=&-H\left( \lambda +\frac{\rho }{\sqrt{3}}\right) -\epsilon \left( \left[
\left| \rho \right| -\frac{\sqrt{3}}{2}\left| \lambda -\frac{\rho }{\sqrt{3}}%
\right| \right] \left( \frac{p_{\lambda }}{\sqrt{3}}-p_{\rho }\right) -\frac{%
3}{2}\left( p_{\lambda }+\frac{p_{\rho }}{\sqrt{3}}\right) \left| \lambda +%
\frac{\rho }{\sqrt{3}}\right| \right)  \label{hpR}
\end{eqnarray}

{\bf \bigskip }Writing 
\begin{equation}
m_{+}=m_{1}\textrm{, \ }m_{-}=m_{2}\textrm{, }m_{0}=m_{3}\textrm{ \ \ \ \ \ \ \ \
\ \ }s_{\pm }=\frac{\left| \lambda \pm \frac{\rho }{\sqrt{3}}\right| }{%
\lambda +\frac{\rho }{\sqrt{3}}}\textrm{ \ \ \ \ \ \ }s_{0}=\frac{\left| \rho
\right| }{\rho }  \label{msdefs}
\end{equation}%
we obtain

\begin{equation}
L_{+}L_{-}L_{0}={\frak M}_{+-}{\frak M}_{-+}L_{0}^{\ast }e^{\frac{\kappa }{4}%
s_{0}H_{0}}+{\frak M}_{-0}{\frak M}_{01}L_{+}^{\ast }e^{\frac{\kappa }{4}%
s_{-}H_{-}}+{\frak M}_{0+}{\frak M}_{+0}L_{-}^{\ast }e^{\frac{\kappa }{4}%
s_{+}H_{+}}  \label{Htransrholam}
\end{equation}%
where 
\begin{eqnarray}
\textrm{\ }M_{0} &=&\sqrt{\frac{2}{3}}\sqrt{p_{\lambda }^{2}+m_{0}^{2}}\textrm{\
\ \ \ \ \ \ \ \ \ \ }M_{\pm }=\frac{1}{\sqrt{2}}\sqrt{\left( \frac{%
p_{\lambda }}{\sqrt{3}}\pm p_{\rho }\right) ^{2}+2m_{\pm }^{2}} \\
L_{\pm } &=&H-M_{\pm }\pm \frac{\epsilon }{\sqrt{2}}\left[ \left( \frac{%
p_{\lambda }}{\sqrt{3}}\mp p_{\rho }\right) s_{0}\pm \frac{2}{\sqrt{3}}%
p_{\lambda }s_{\pm }\right] \textrm{ \ \ \ \ \ } \\
L_{0} &=&H-M_{0}-\frac{\epsilon }{\sqrt{2}}\left[ \left( \frac{p_{\lambda }}{%
\sqrt{3}}+p_{\rho }\right) s_{+}+\left( \frac{p_{\lambda }}{\sqrt{3}}%
-p_{\rho }\right) s_{-}\right] \\
{\frak M}_{\pm \mp } &=&M_{\pm }\mp \frac{\epsilon }{\sqrt{2}}\left( \frac{%
p_{\lambda }}{\sqrt{3}}\pm p_{\rho }\right) s_{0},\textrm{ \ \ }{\frak M}_{\pm
0}=M_{\pm }-\frac{\epsilon }{\sqrt{2}}\left( \frac{p_{\lambda }}{\sqrt{3}}%
\pm p_{\rho }\right) s_{\pm }\textrm{\ \ \ \ \ \ \ \ \ } \\
\textrm{\ \ \ \ \ }{\frak M}_{0\pm } &=&M_{o}+\frac{\epsilon \sqrt{2}}{\sqrt{3}%
}p_{\lambda }s_{\pm }\textrm{\ \ \ \ }L_{0}^{\ast }=(1-s_{+}s_{-})M_{0}+L_{0}
\\
L_{\pm }^{\ast } &=&(1\mp s_{0}s_{\pm })M_{\pm }+L_{\pm }
\end{eqnarray}

{\bf \bigskip }{\Large Acknowledgements}

This work was supported by the Natural Sciences and Engineering\ Research
Council of Canada.


\bigskip

\begin{thebibliography}{99}
\bibitem{Rybicki} G. Rybicki, Astrophys. Space.\ Sci {\bf 14} (1971) 56.

\bibitem{yawn} See B.N. Miller and P. Youngkins, Phys. Rev. Lett. {\bf 81}
4794 (1998); K.R. Yawn and B.N. Miller, Phys. Rev. Lett. {\bf 79} 3561
(1997) and references therein.

\bibitem{fractal} H. Koyama and T. Kinoshi, Phys Lett. {\bf A} (in press;
astro-ph/0008208).

\bibitem{OR} T. Ohta and R.B. Mann, Class. Quant. Grav. {\bf 13} (1996) 2585.

\bibitem{2bd} R.B. Mann and T. Ohta, Phys. Rev. {\bf D57} (1997) 4723;
Class. Quant. Grav. {\bf 14} (1997) 1259.

\bibitem{2bdcossh} R.B. Mann, D. Robbins and T. Ohta, Phys. Rev. Lett. {\bf %
82} (1999) 3738.

\bibitem{2bdcoslo} R.B. Mann, D. Robbins and T. Ohta, Phys. Rev. {\bf D60}
(1999) 104048.

\bibitem{2bdchglo} R.B. Mann, D. Robbins, T. Ohta and M.\ Trott, Nucl. Phys. 
{\bf B590} \ 367.

\bibitem{statbal} R.B. Mann and T. Ohta, Class.\ Quant. Grav. {\bf 17}
(2000) 4059.

\bibitem{pchak} R.B. Mann and P. Chak, Phys. Rev. {\bf E65} 026128 (2002).

\bibitem{circle} R.B. Mann, Class.Quant.Grav.{\bf 18} (2001) 3427.

\bibitem{ryan} R. Kerner and R.B. Mann, gr-qc/0206029.

\bibitem{LMiller} H.E. Lehtihet and B.N. Miller, Physica {\bf 21D}, 93
(1987).

\bibitem{Goodings} N. D. Whelan, D. A. Goodings, and J. K. Cannizzo, Phys.
Rev. {\bf A42}, 742 (1990)

\bibitem{Butka} D. Butka, G. Karl and B. Nickel, Can J. Phys {\bf 78}, 449
(2000).

\bibitem{optbill} V.Millner, J.L.Hanssen, W.C. Campbell, and M.G. Raizen,
Phys. Rev. Lett. {\bf 86}, 1514 (2001).

\bibitem{r3} R.B. Mann, Found. Phys. Lett. {\bf 4} (1991) 425; R.B. Mann,
Gen. Rel. Grav. {\bf 24} (1992) 433.

\bibitem{jchan} S.F.J. Chan and R.B. Mann, Class. Quant. Grav. {\bf 12}
(1995) 351.

\bibitem{JT} R. Jackiw, Nucl. Phys. B {\bf 252}, 343 (1985); C. Teitelboim,
Phys. Lett. B {\bf 126}, 41, (1983).

\bibitem{BanksMann} T. Banks and M. O' Loughlin, Nucl. Phys. {\bf B362 }
(1991) 649; R.B. Mann, Phys. Rev.{\bf D47} (1993) 4438.

\bibitem{OK} T. Ohta and T. Kimura, Phys. Letters {\bf 63A} (1977) 193; {\bf %
90A} (1982) 389; Progr. Theor. Phys. {\bf 68} (1982) 1175.

\bibitem{KAM} V.I. Arnold and A. Avez, {\sl Ergodic Problems of Classical
Mechanics} (Springer: New York, 1968); A. N. Kolmogorov, Dokl. Akad. Nauk.
SSSR, {\bf 98}, 525 (1954) (An English version can be found in {\it %
Proceedings of the 1954 International Congress of Mathematics},
North-Holland, Amsterdam,m 1957); V.I. Arnold, Russ. Math. Surv., {\bf 18},
85-191 (1963); J. Moser, Nachr. Akad. Wiss. Goettingen Math. Phys., {\bf K1}
1 (1962).

\bibitem{Rzheng} H. Reichl and R. Zheng: NonLinear Resonance and Chaos in 
{\sl Directions in Chaos}, (World Scientific 1987) ed. B. Hao.

\bibitem{justin} J.J. Malecki and R.B. Mann, to appear.
\end{thebibliography}
\end{document}